\documentclass[preprint]{aastex}
\usepackage{subfig}
\usepackage{epsfig}
\newcommand{\noprint}[1]{}

\shorttitle{VLM Binary Masses}
\shortauthors{Konopacky et al.}
\begin{document}

\title{High Precision Dynamical Masses of Very Low Mass Binaries}
\author{Q.M. Konopacky\altaffilmark{1,2},
  A.M. Ghez\altaffilmark{1,3},
  T.S. Barman\altaffilmark{4}, E.L. Rice\altaffilmark{5},
  J.I. Bailey, III\altaffilmark{6},
  R.J. White\altaffilmark{7}, I.S. McLean\altaffilmark{1}, G. Duch\^{e}ne\altaffilmark{8,9}}
\altaffiltext{1}{UCLA Division of Astronomy and Astrophysics, Los Angeles, CA 90095-1562
; ghez, mclean@astro.ucla.edu}
\altaffiltext{2}{Current address: Lawrence Livermore National
  Laboratory, 7000 East Avenue, Livermore, CA 94550, konopacky1@llnl.gov}
\altaffiltext{3}{Institute of Geophysics and Planetary Physics, University of
California, Los Angeles, CA 90095-1565}
\altaffiltext{4}{Lowell Observatory, 1400 W. Mars Hill Rd.,
  Flagstaff, AZ 86001; barman@lowell.edu}
\altaffiltext{5}{American Museum of Natural History, Central
  Park West at 79th Street, New York, NY 10024-5192}
\altaffiltext{6}{Department of Astronomy, University of
  Michigan, 500 Church Street, Ann Arbor, MI 48105, baileyji@umich.edu}
\altaffiltext{7}{ Department of Physics and Astronomy, Georgia
  State University, Atlanta, GA 30303, white@chara.gsu.edu}
\altaffiltext{8}{Astronomy Department, UC Berkeley, 601
  Campbell Hall, Berkeley, CA 94720-3411, USA; gduchene@berkeley.edu}
\altaffiltext{9}{Universit\'e Joseph Fourier - Grenoble 1 /
  CNRS, Laboratoire d'Astrophysique de Grenoble (LAOG) UMR 5571, BP 53,
F-38041, Grenoble Cedex 9, France}
\keywords{stars: binaries, visual; stars: low-mass, brown
  dwarfs; stars: individual (2MASSW J0746425+200032, 2MASS
  J08503593+1057156, 2MASS J09201223+3517429, 2MASS
  J14263161+1557012, 2MASS J15344984-2952274, 2MASS
  J17281150+3948593, 2MASS J17501291+4424043, 2MASS
  J18470342+5522433, 2MASS J21402931+1625183, 2MASS
  J22062280-2047058, HD130948B, LHS 2397a, LP 349-20, LP
  415-20), stars: fundamental parameters}

\begin{abstract}
We present the results of a 3 year monitoring program of a
sample of very low mass (VLM) field binaries using both
astrometric and spectroscopic data obtained in conjunction
with the laser guide star adaptive optics system on the
W.M. Keck II 10 m telescope.  Among the 24 systems studied,
fifteen have undergone sufficient orbital motion, allowing us
to derive their relative orbital parameters and hence their
total system mass.  These measurements triple
the number of mass measurements for VLM objects, and include
the most precise mass measurement to date ($<$2$\%$).  Among
the 11 systems with both astrometric and spectroscopic
measurements, six have sufficient radial velocity variations
to allow us to obtain individual component masses.  This is
the first derivation of the component masses for five of these systems.
Altogether, the orbital solutions of these low mass systems
show a correlation between eccentricity and orbital period, consistent with their higher mass counterparts.
In our primary analysis, we find that there are systematic
discrepancies between our dynamical mass measurements and the
predictions of theoretical evolutionary models (TUCSON and
LYON) with both models either underpredicting or
overpredicting the most precisely determined dynamical masses.
These discrepancies are a function of spectral
type, with late M through mid L systems tending to have their
masses underpredicted, while one T type system has its mass
overpredicted.  These discrepancies imply that either the
temperatures predicted by evolutionary and atmosphere models
are inconsistent for an object of a given mass, or the
mass-radius relationship or cooling timescales predicted by
the evolutionary models are incorrect.  If these spectral type
trends are correct and hold into the planetary mass regime, the
implication is that the masses of directly imaged extrasolar
planets are overpredicted by the evolutionary models.

\end{abstract}

\section{Introduction}

Characterizing the fundamental properties of brown dwarfs is
an important step in unlocking the physics of substellar
objects.  These very cool objects
have internal and atmospheric properties that are quite
similar to gas giant planets and that differ fundamentally
from those of stars, including 
partially degenerate interiors, dominant molecular opacities,
and atmospheric dust formation (Chabrier $\&$ Baraffe 2000).
Brown dwarfs also 
represent a substantial fraction of the galactic stellar
content, and are bright and numerous enough to be studied in
great detail with current technology (Kirkpatrick 2005).
Thus, these substellar objects present an ideal laboratory in
which to study the physical processes at work in very low mass
objects that both approach and overlap the planetary mass regime. 

\begin{figure*}
\epsscale{1.0}
\plotone{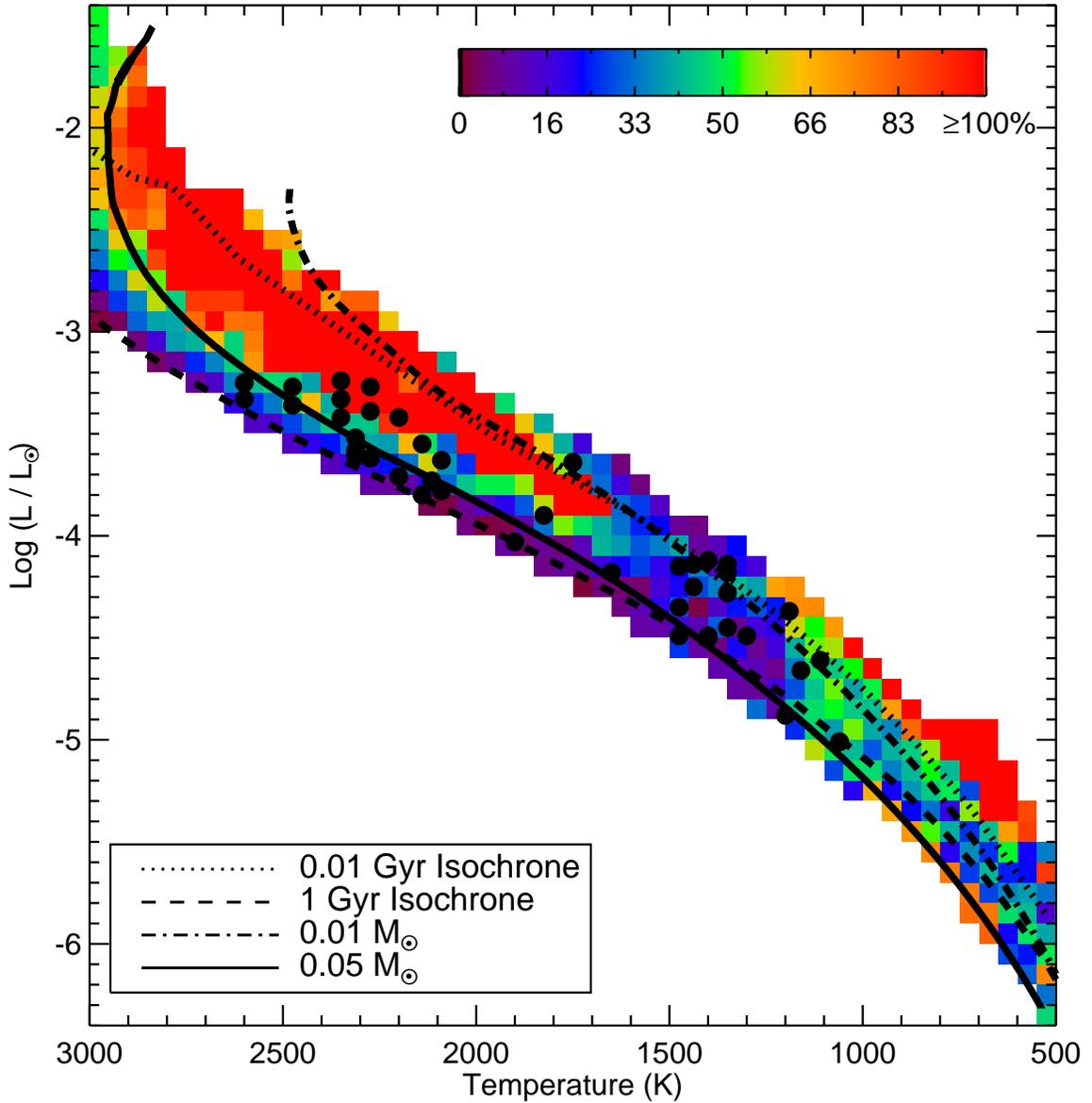}
\caption{Percent discrepancy in mass predictions between the
  Burrows et al. (1997) evolutionary and the Chabrier
  et al. (2000) evolutionary models, over the range on the H-R
  diagram with complimentary coverage.  The colors represent
  the level of the discrepancy in units of percent of the mass
  predicted by the Burrows et al. (1997) models, as shown
  by the scale bar.  For the majority of the H-R Diagram, the
  discrepancy between the model predictions is
  $\gtrsim$10$\%$, with a number of regions having
  discrepancies greater than 100$\%$.  Overplotted are two
  isochrones and lines of constant mass from the Burrows et
  al. (1997) models for points of reference.  In addition, the
  overplotted filled points show the rough location of the
  sources in our full sample.  The largest
  discrepancies are at the youngest ages, but the
  discrepancies are still substantial for older objects.}
\label{fig:mass_discrep_ave}
\end{figure*}

Mass is the most fundamental parameter in determining the
properties and evolution of a brown dwarf; unfortunately it is
also one of the most difficult to measure.  Masses of brown
dwarfs are typically inferred from the comparison of measured
luminosities and temperatures with predictions from
theoretical models.  The most commonly used models are those
of Burrows et al. (1997) and Chabrier et al. (2000).  
However, as shown in Figure \ref{fig:mass_discrep_ave}, masses obtained in this way from
different models can be discrepant, especially amongst the lowest mass objects.  These discrepancies stem from physical assumptions
about the interior and atmospheric properties of these highly
complex objects.
Examples of uncertainties in the models are atmospheric processes  
that define the transition regions between spectral types.
Specifically, this includes the formation of atmospheric clouds (M to L), the  
disappearance of clouds (L to T), and the formation of
ammonia, which makes the objects, in theory, similar
atmospherically to Jupiter (T to Y).  Additional sources of
uncertainties in the models include, but are not limited to,
the equation of state (Chabrier $\&$ Baraffe 2000), the initial conditions and accretion
history (Baraffe et al. 2009), the treatment  of convection and the possible
subsequent generation of magnetic fields, which in turn could
affect the inferred effective temperatures (Chabrier $\&$ Kuker
2006, Browning 2008).  An essential step toward properly
calibrating these models and constraining their physics is to obtain high precision
($\lesssim$10$\%$) dynamical mass measurements of brown dwarf binaries.

The advent of laser guide star adaptive optics (LGS AO) on large  
ground-based telescopes has dramatically increased the number of VLM  
objects for which high precision dynamical masses can be obtained.   
Prior to AO, only one binary brown dwarf had sufficiently  precise  
mass measurements to test the models and this was the case of
2MASS J05352184–0546085, an  
eclipsing binary
in Orion, which provided constraints at a very young age
($\sim$Myr, Stassun et al. 2006).   
Early AO, which used natural guide stars, allowed dynamical
mass estimates for two brown dwarfs (Lane et al. 2001, Zapatero
Osorio et al. 2004, Simon et al. 2006, Bouy et al. 2004).   With LGS AO, much fainter sources can be targeted,   
allowing $\sim$80\% of known brown dwarf binaries to be
observed and a much more systematic look at how the observed  
properties of brown dwarfs compare with the predictions of
models.

To capitalize on the introduction of LGS AO on 10 m class
telescopes, we initiated, in 2006, an extensive astrometric
and spectroscopic monitoring campaign of 23 very low mass
(VLM) binaries (M$_{tot}$ $\lesssim$ 0.2 M$_{\odot}$) in the
near-infrared with the Keck/LGS-AO system with the goal of
obtaining precision dynamical masses.  The astrometric aspect
of this project is similar to the work reported on three
individual targets by Liu et al. (2008) and Dupuy et
al. (2009a,b).  Our survey includes these targets as well
as others to span a wide range of late stellar and substellar
spectral types (M7.5 to T5.5) and is the first study to
include radial velocity measurements for the LGS AO targets.  Our relative
astrometric and radial velocity measurements provide estimates
of the {\it total} system mass for 15 systems.  Our absolute
radial velocity measurements, add estimates of the mass ratios
and hence {\it individual} mass estimates for 6 of these
systems.   Altogether, this work triples both the current
number of system mass measurements and individual mass 
estimates for VLM objects and, when compared to the models,
shows systematic discrepancies. 

This paper is organized as follows.  Section 2 describes the
sample selection and section 3 provides a description of the
astrometric and spectroscopic observations.  Section 4
outlines the data analysis procedures and section 5 describes
the derivation of orbital solutions.   Section 6 contains the
estimates of bolometric luminosities and effective temperatures
for the components of the binaries.   Section 7 compares the
dynamical masses to the predictions from evolutionary models
and the implications of our model comparison are discussed in
section 8.   We summarize our findings in section 9. 

\section{Sample Selection}
\label{sec:sample}
\subsection{Initial Sample}
The initial sample for this project was culled from Burgasser
et al. (2007), which listed the 68 visual, VLM binaries
known as of 2006\footnote{Though the official publication of
  Burgasser et al. (2007) in the Protostars and Planets V
  conference proceedings was in 2007, the article was
  published on astro-ph in February 2006.}.  Three cuts were
applied to this initial list. 
First, the binaries had to be observable with the Keck
telescope LGS AO system, so we imposed a declination $>$ -35
degrees requirement, which reduced the possible number of
targets to 61.  Second, the operation of the LGS AO
system requires a tip-tilt reference source of apparent R
magnitude $<$ 18 within an
arcminute of the source, and therefore  VLM binaries without a suitable
tip-tilt reference were also cut.  This lowered the total
number of observable targets to 49, 80$\%$
of the northern hemisphere sample.  

Third, we required that useful dynamical mass estimates would
likely be obtained within 3 years.  To
assess the required precision for our
dynamical mass estimates, we calculated the predicted masses for
the two most commonly used sets of evolutionary
models, those of Burrows et al. (1997) and Chabrier et
al. (2000), across the entire range of temperatures and
luminosities spanned by both models.  We calculated the
percent difference between the predictions of each model with
respect to the prediction of Burrows et al. (1997).  The results of this assessment
are shown in Figure \ref{fig:mass_discrep_ave}, which displays
in color the offset between the models across the H-R diagram
(with the discrepancies averaged in 50 K temperature
increments and 0.1 log(L/L$_{\odot}$ increments).  As
the figure demonstrates, we found that the
difference in the mass predictions of the two models varied
anywhere from a few percent to greater than 100$\%$.  We therefore chose
a precision goal of 10$\%$ because at this level the
majority of the model predictions could be
distinguished and because this level of precision was
reasonable to expect given our observing strategy.

To implement our third cut, a series of Monte Carlo simulations were
performed.  In these simulations, the total system mass
for each target was assumed based on the estimated spectral
types of the binary components from the original discovery
papers, using the Chabrier et al. (2000) models, and
held constant for all runs.  Additionally, the
semi-major axis of the orbit was chosen by sampling from the range of
possible values between 1/2 and two times the original separation measurement.  From these 
assumptions, a period was calculated,
and T$_o$ (time of periapse passage) was randomly selected
from the range allowed by this period.  All other orbital
parameters for an astrometric orbit, which include e (eccentricity), i (inclination),
$\Omega$ (longitude of the ascending node), and $\omega$
(argument of periapsis), were randomly selected from among
the complete range of possible values of each parameter.

Using these simulated orbits, it was possible
to generate simulated sets of ``astrometric datapoints''
corresponding to the likely times of measurement.  We planned
on two observing campaigns per year, one in June and one in
December.  These dates were chosen to coincide with the two
times per year that NIRSPEC is offered behind the LGS AO
system at Keck (see section \ref{obs}).  The sky coordinates of each binary then determined
whether we simulated one or two astrometric and radial
velocity data points per
year.  These simulated
measurements were chosen to correspond to appropriate
observing dates.  All synthetic data points were
combined with already existing measurements, the number of
which varied from source to source.  While the majority of
sources initially had
only one previous astrometric measurement, a few had up to
six.  Synthetic astrometric datapoints were also assigned uncertainties
based on the average uncertainty normally obtained for
short-exposure measurements of binary stars using the Keck AO system with
NIRC2 ($\sigma$ $\sim$ 1 mas).  Although the average
uncertainty in relative radial 
velocity measurements with NIRSPEC+AO (NIRSPAO) was not known at the time,
other observations with NIRSPEC suggested using a conservative
uncertainty of about 1 km/s.  All datapoints were then used to run the orbital
solution fitter, which uses the Thiele-Innes method (e.g.,
Hilditch 2001), minimizing the $\chi^2$ between the model and the
measurements (see Ghez et al. 2008).  A chi-squared cut of 10
was imposed to account for the fact that in some simulated
orbital solutions we could not generate astrometric measurements
corresponding to real data points in those systems with
multiple measurements.  In this way, we were able to utilize
more information than simply separation and estimated mass to
calculate likelihood of accurate mass measurement in a
system.  A total of 1000 simulated orbital solutions were created for each system.

From each of these simulations, the predicted uncertainty in
dynamical mass could be determined.  
All those systems for which
66$\%$ of the simulations yielded precisions of
10$\%$ or better in mass were put in the final sample.  This
generated a sample of 21 targets that we began monitoring in
the spring of 2006.  These sources are listed in Table
\ref{tab:initialSample}.  Figure \ref{fig:sample} shows the results of our
simulations, plotting the percent of solutions with precise
mass estimates versus the initial binary separation.  The
spectral type of the primary component is denoted by symbol shape,
and sources included in our initial sample are colored red.
The variation in percent of solutions with separation stems
from the variation in the estimated masses of the components
and the number of previous measurements at the start of our
monitoring program.

\subsection{Sample Refinement}

Upon commencement of the monitoring campaign, it became clear
that sample refinement and adjustment of observing priorities
was required.  Three sources had tip/tilt stars that did not
allow for successful observation (2MASS
0423-04, GJ 417B, and 2MASS 1217-03).  It is possible that
some of these tip/tilt stars were actually resolved galaxies.
In addition, 2MASS 1217-03 was later reobserved with HST and
found to be unresolved, making it unlikely to be a
binary (Burgasser et al. 2006).  Therefore, we monitored 19 of
the 22 initially identified brown dwarfs.

\begin{figure*}
\epsscale{1.0}
\plotone{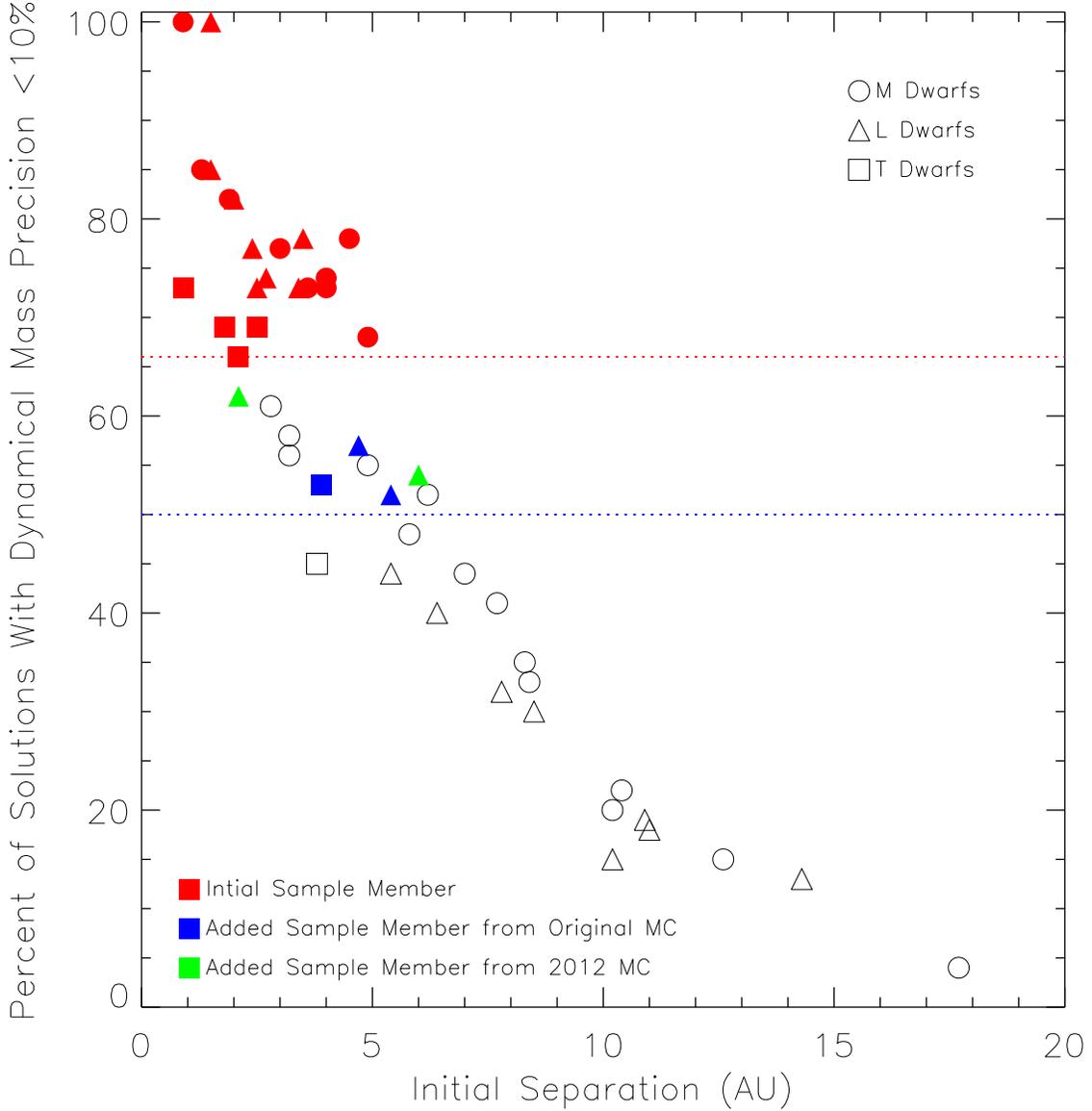}
\caption{The percent of
  solutions in our Monte Carlo simulations that yielded a mass with $\lesssim$10$\%$ precision versus the initial
separation of the binary.
Sources included in our sample are denoted in red, with the red
dotted line showing our cutoff of 66$\%$.  Additional sample
members are denoted in blue, and were chosen because they had
either L or T spectral types and because they had a
probability of $>$50$\%$ of yielding a precise mass in our
initial simulations (with an increased probability for high
precision masses by 2012).  Sources which were not included in
the original Monte Carlo simulation because of their later
discovery epoch, but that have a high likelihood of yielding a
precise mass by 2012, are shown in green.  The symbol type denotes the spectral type
of the primary component.}
\label{fig:sample}
\end{figure*}

Additionally, a few targets were added to the monitoring
program as it progressed.  First, it was recognized that some
sources did not make the cut because of the 3-year timescale
constraint, but with a slightly longer period of monitoring
could have their masses derived to a high level of accuracy.
In particular, the timescale cut introduced an obvious bias to
sources with higher predicted masses, or earlier spectral types.
Therefore, we added three objects included in Burgasser et al.
(2007) to the NIRC2 monitoring program to
provide initial epochs of data for future mass determination.
These three objects are shown on Figure \ref{fig:sample} in blue.  All three
were of spectral type L or T (we did not add additional M
dwarfs to our sample because of the large number of M dwarfs
included in our initial sample).  All three had a $>$50$\%$
probability of a precise dynamical mass estimate in our
initial Monte Carlos.
These added sources are noted in Table \ref{tab:initialSample}.
Two additional sources were added to the sample that were
discovered by Reid et al. (2006) after the initial publication of Burgasser et
al. (2007).  For these sources, we have calculated the
likelihood that they will yield precise mass estimates by
2012.  These sources are denoted in Figure \ref{fig:sample} in
green to keep them distinct from the sources from our initial
simulations, as we calculated their likelihood of yielding a
good mass estimate by 2012 instead of 2009.  We found that both sources had a $>$50$\%$
chance of yielding a precise mass estimate by 2012, and therefore added these two
sources to our astrometric program.  They are also
listed in Table \ref{tab:initialSample}. 

\section{Observations}\label{obs}
\subsection{Astrometric Data}\label{astroobs}

Targets in our sample were observed astrometrically beginning in May of 2006.
Observations were conducted twice a year between 2006 May and
2009 June UT using the
Keck II 10 m telescope with the facility LGS AO system
(Wizinowich et al. 2006; van Dam et al. 2006) and the
near-infrared camera, NIRC2 (PI K. Matthews).  The AO system, which is also used for obtaining
radial velocities (see next section),
uses the sodium laser spot (V$\sim$10.5) as the primary correction
source for all but two systems.  Tip/tilt references are
listed in Table \ref{tab:astrolog}.  NIRC2 has a plate scale of 9.963 $\pm$
0.005 mas pixel$^{-1}$ and columns that are at a PA of 0.13
$\pm$ 0.02$^o$ relative to North (Ghez et al. 2008).  The
observing sequence for each object depended upon the
brightness of the target, whether observations in multiple
filters had been previously made, and whether the target was
actually resolved into two components at that epoch.  If the
binary was not resolved, we could only obtain an upper limit on
the separation, which does not require a full observing sequence
to estimate.  We
generally tried to take at least nine individual exposures on
each target, though sometimes due to time constraints fewer
exposures were taken.   Table \ref{tab:astrolog}
gives the log of all imaging and photometric observations, listing when
each target was observed, the filters through which it was
observed and the exposure time and number of images taken in
each filter, and the tip/tilt reference source used for each
target.  In many cases, the brown dwarf targets were bright
enough to serve as their own tip/tilt reference, even though
they are not bright enough for natural guide star observations.

With only a few exceptions, all data used for astrometry were
taken through the K-prime ($\lambda_o$ = 2.124 $\mu$m, $\Delta\lambda$ = 0.351 $\mu$m) band pass
filter.  Data in both the J Band ($\lambda_o$ = 1.248 $\mu$m,
$\Delta\lambda$ = 0.163 $\mu$m) and H band ($\lambda_o$ = 1.633 $\mu$m,
$\Delta\lambda$ = 0.296 $\mu$m) were also taken at some point for most
targets to provide a complete set of spatially resolved,
near-infrared photometry.  The images were generally taken in a three position, 
2$\farcs$5 x 2$\farcs$5 dither box,
with three exposures per position (avoiding the lower left
quadrant of NIRC2, which has significantly higher noise than
the other three), which allowed sky frames to
be generated from the images themselves.  In addition, the wide
dither box insures the incorporation of known residual
distortion (Ghez et al. 2008, Yelda et al. 2009) in the camera
into our final astrometric uncertainties.  

\subsection{Radial Velocity Data}\label{rvobs}

Eleven objects ($m_K$ $\lesssim$ 12) in our astrometric sample were also observed using the
NIR spectrograph NIRSPEC on Keck II (McLean et al. 2000) in
conjunction with the LGS AO system (NIRSPAO).  
We used the instrument in its high spectral resolution mode, selecting a
slit 0$\farcs$041 in width and 2$\farcs$26 in length in AO
mode.  We elected to observe in the K band, with a particular
interest in the densely populated CO band head region
($\sim$2.3 $\mu$m, Order 33), necessitating an echelle angle of 63 degrees and
a cross-disperser angle of 35.65 degrees.  The resolution in
this setup is R$\sim$23000, as determined by the width of
unresolved OH sky lines, and the wavelengths covered are 2.044
- 2.075 $\mu$m (order 37), 2.100-2.133 $\mu$m (order 36),
2.160 - 2.193 $\mu$m (order 35), 2.224 - 2.256 $\mu$m (order
34), 2.291 - 2.325 $\mu$m (order 33), and 2.362 - 2.382 $\mu$m
(order 32), with some portions of the K band beyond the edges
of the detector.  For this work, all analysis was done using
only Order 33, the order containing the CO bandhead, and all
data presented come from this order.  

The camera was rotated such that both components of each binary fell
simultaneously on the high resolution slit, which is
at an angle of 105.9$^{o}$ with respect to north.  Typical observations
consisted of four spectra of both components, each with 1200 second
integration times, taken in an ABBA dither pattern along
the length of the slit.  In a few cases, more than four
spectra were taken or a slightly different integration time
was used, depending on the brightness of the object.  Table
\ref{tab:bd_rvlog} gives the log of our spectroscopic observations,
listing the targets observed, the date of observation, the
number of spectra, and the integration time for each
spectrum.  We successfully obtained spatially
resolved spectra for sources separated by $\gtrsim$60 mas in
all epochs.  Each target observation was accompanied by the
observation of a nearby A0V star to measure the telluric
absorption in the target spectra.  

\section{Data Analysis}\label{sec:dataan}
\subsection{Astrometric Data Analysis}\label{astroan}

The NIRC2 data were initially processed using standard data
reduction techniques for near-infrared images. 
Frames at differing dither positions were subtracted from each
other to remove sky background, followed by the
removal of bad pixels, division by a flat field, and
correction for optical distortion with a model provided in the
pre-ship review
document\footnote{http://www2.keck.hawaii.edu/inst/nirc2/preship$\_$testing.pdf}
using standard IRAF and IDL routines.  The binaries were then shifted to a common location in
all frames and the images were median combined.  
Astrometry and flux ratios were obtained using the
IDL package StarFinder (Diolaiti et al. 2000).  An empirical
point-spread function (PSF) is required by the StarFinder fitting
algorithm.  In the case of two sources, a suitable PSF star
falls within the field of view of the NIRC2 observation of the
source.  However, in the majority of cases, no such PSF source is in the field of
view.  For these observations, we use either an image of a
single star taken near
in time to the images of the binary, or if no suitable single
star was imaged, we use an idealized Keck PSF degraded to the
calculated strehl ratio of the observation for PSF fitting.
For this last case, the PSF is generated by first convolving
the idealized PSF with a
Gaussian, such that the core is broadened to the appropriate
FWHM.  Next, a simulated ``halo'' is generated by adding a
Gaussian with FWHM of 0$\farcs$5 (average near-infrared seeing halo at
Keck), normalized such that the resulting strehl ratio matches
the observations.  Internal statistical measurement errors were calculated
by fitting the components of the binaries in all individual images that
contributed to the combined images and finding the RMS of
the values derived therein.  

\begin{figure*}
\epsscale{0.8}
\plottwo{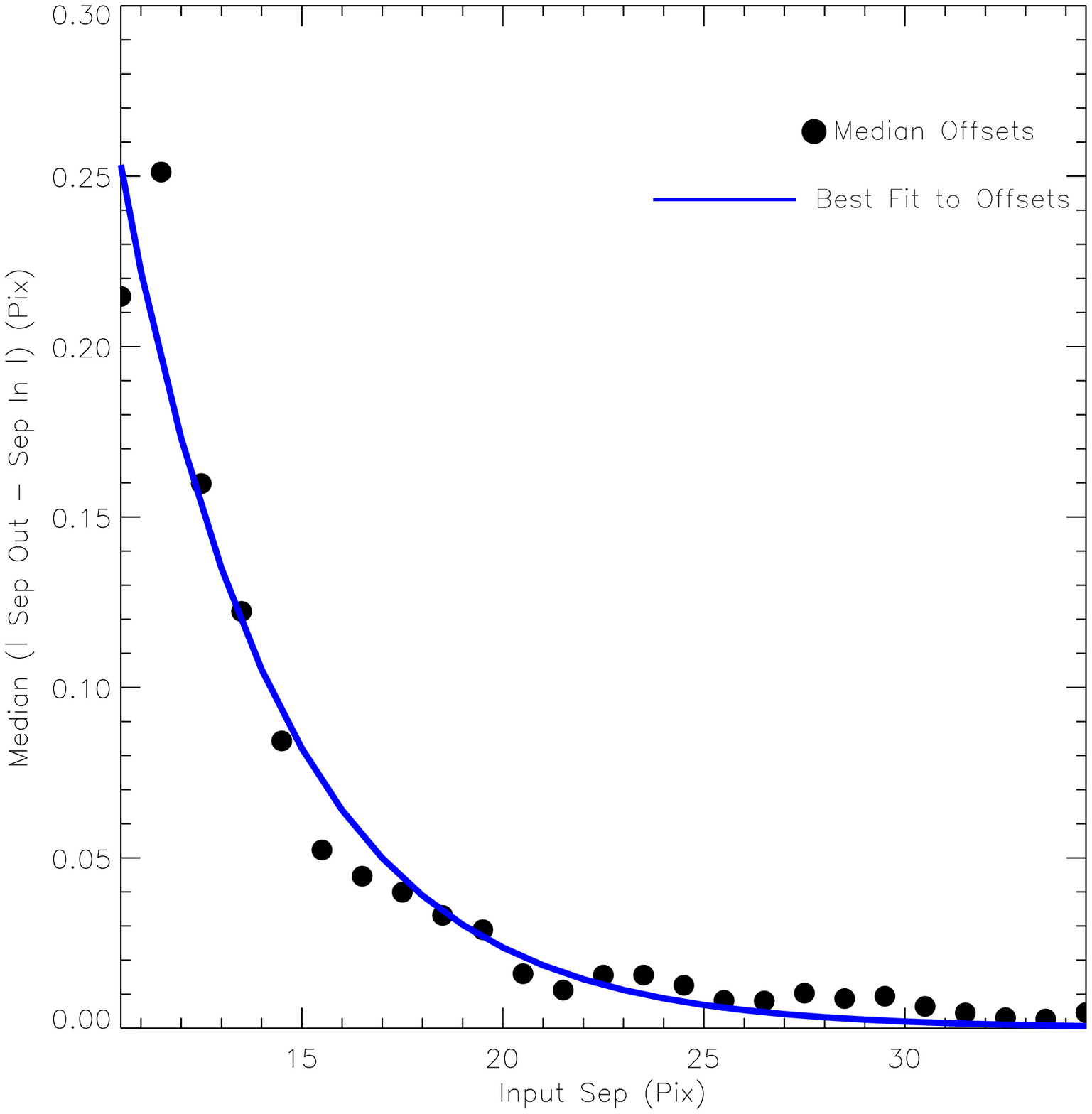}{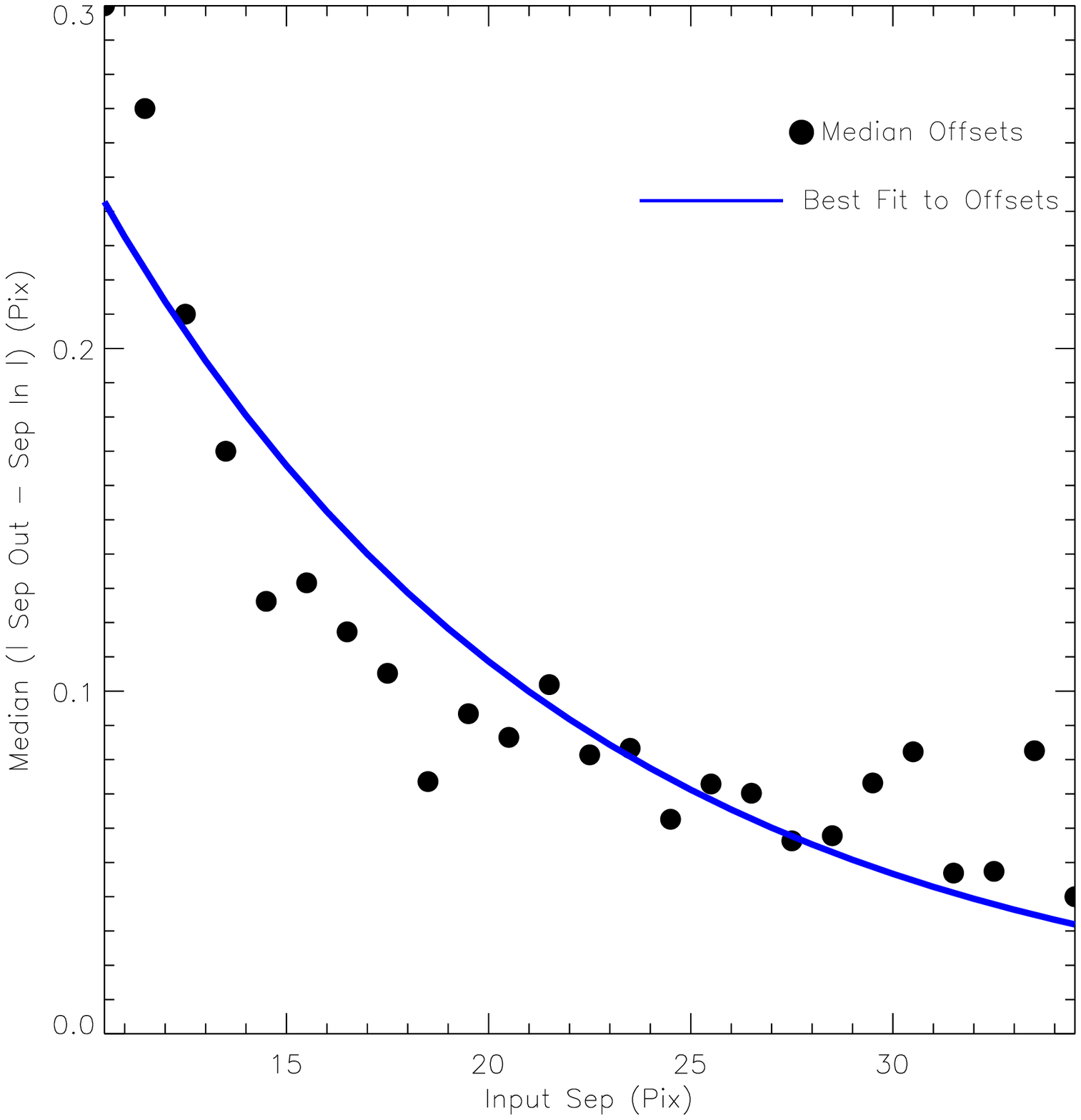}
\plottwo{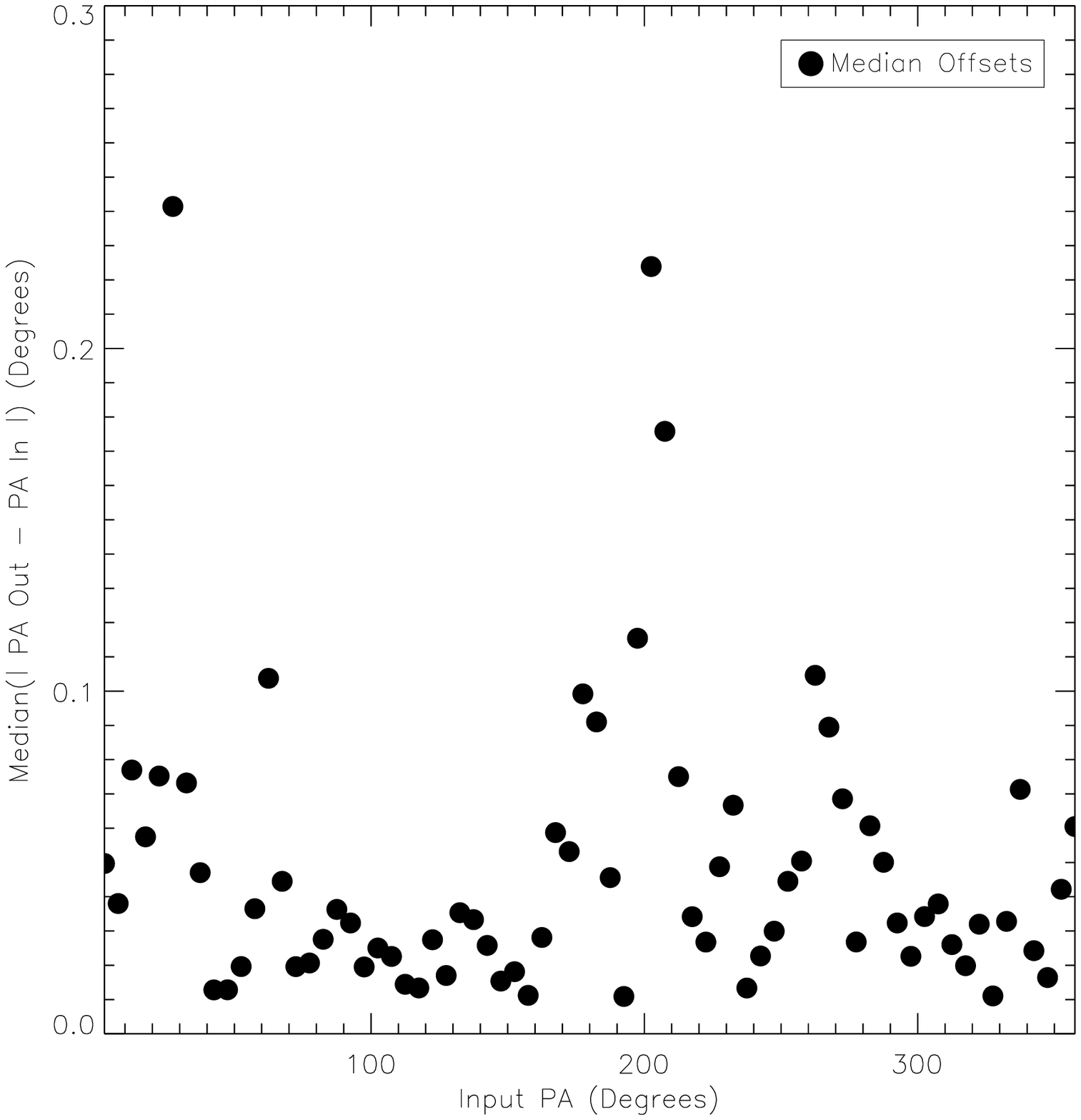}{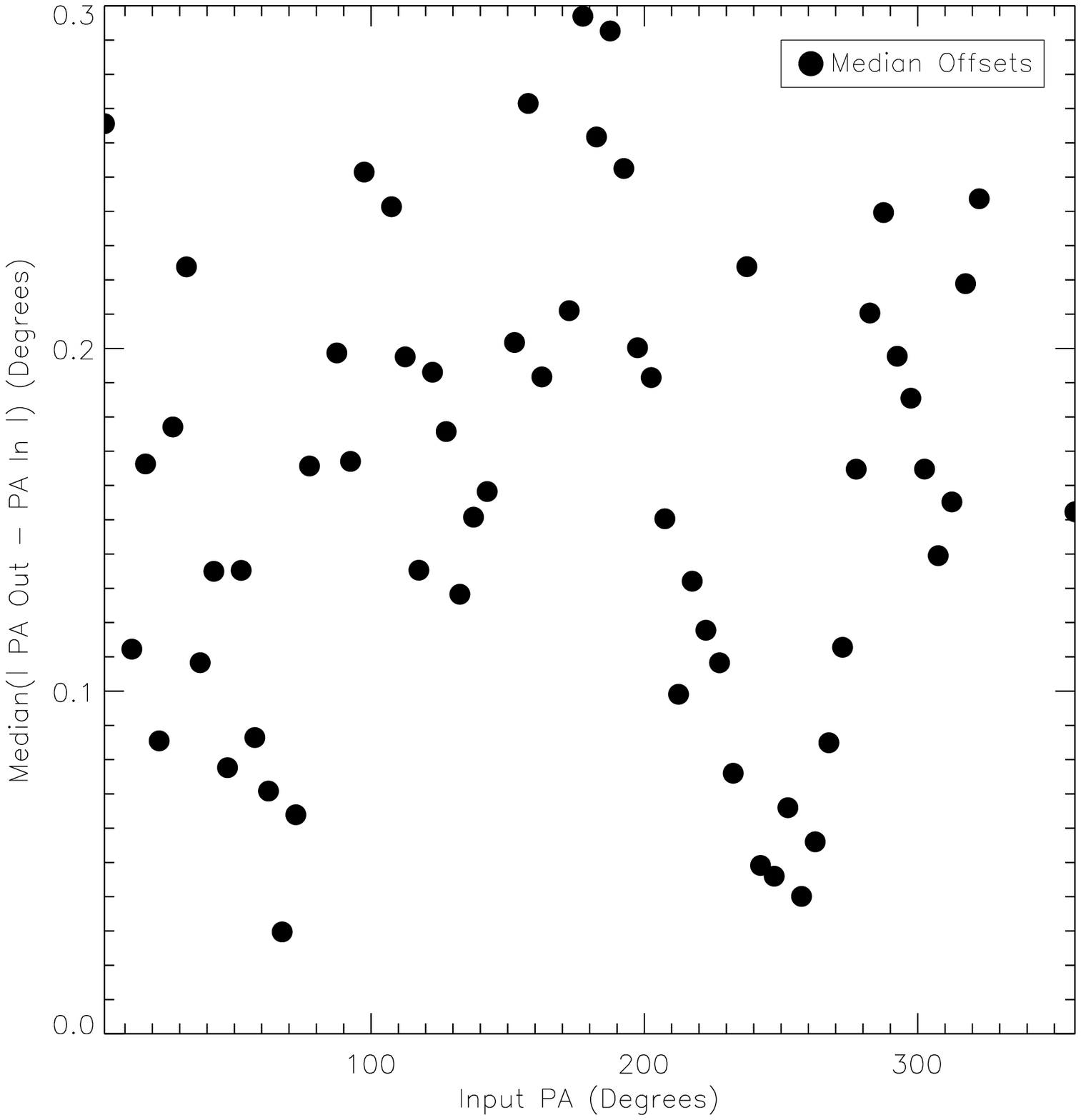}
\plottwo{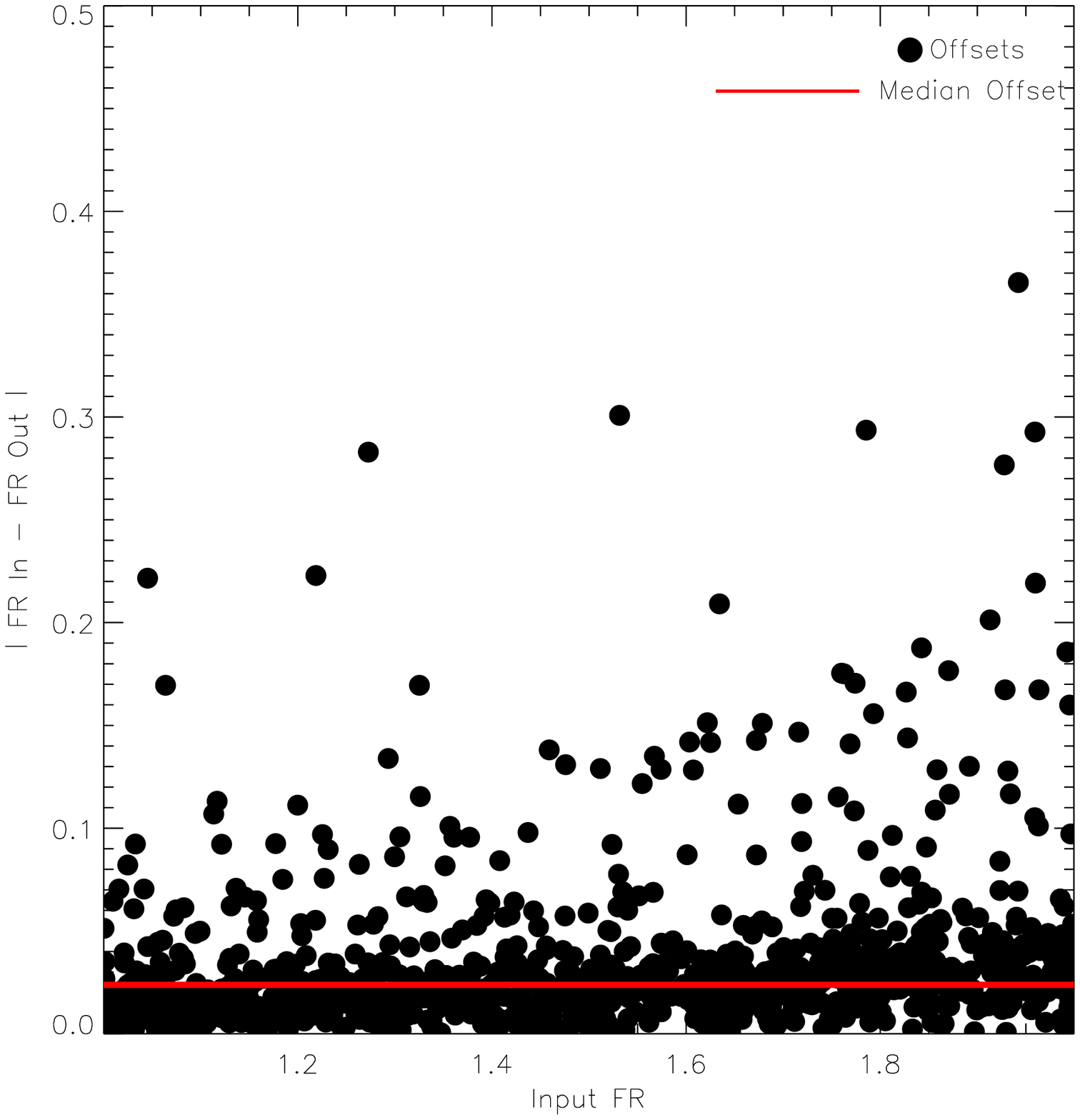}{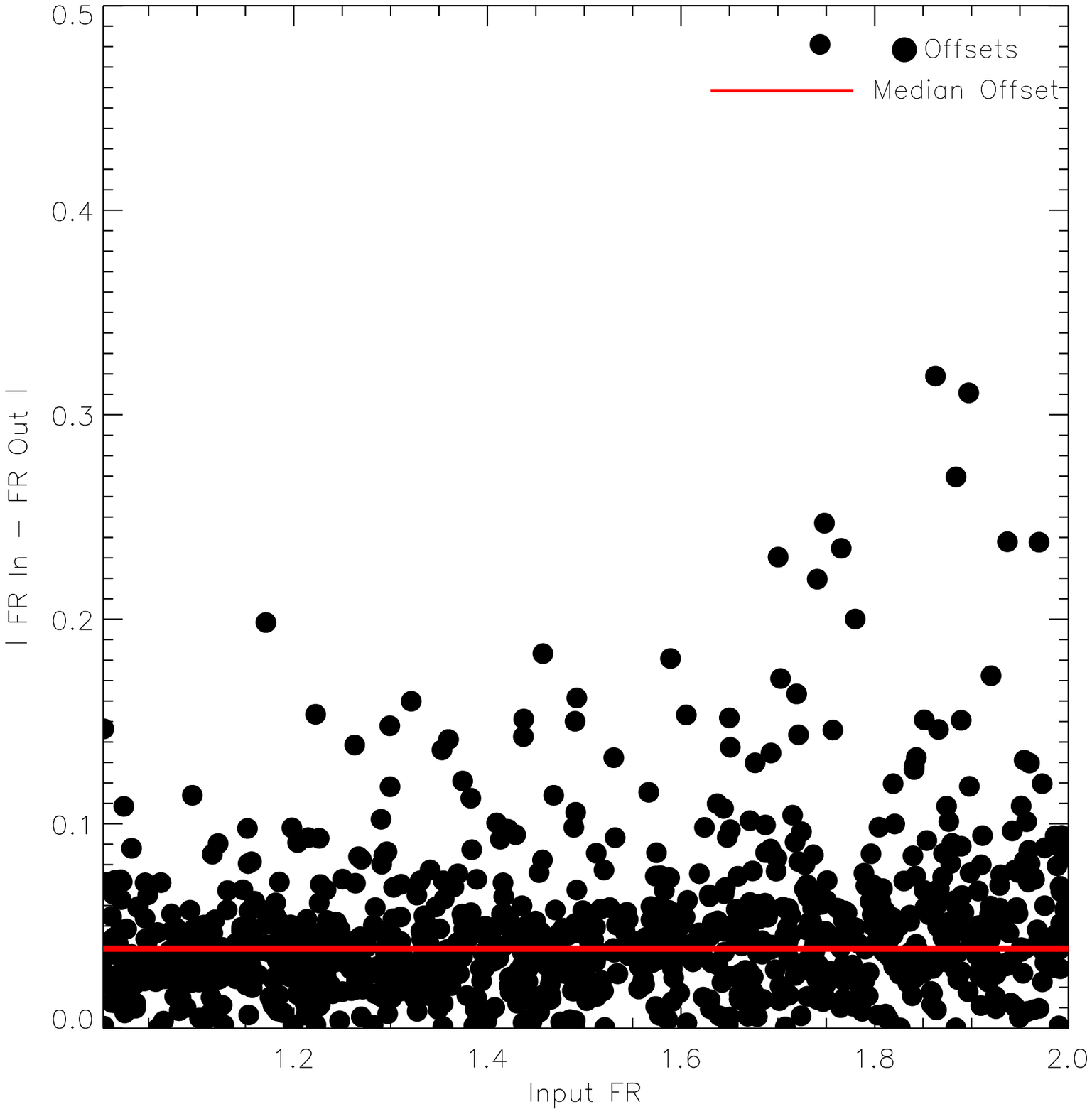}
\caption{Results of PSF systematics simulation from 2006 May
  21, using two observed PSFs (left column) and one observed,
  one simulated PSF (right column). \textbf{Top}: Median offset in fit separation from
  input separation, binned in one pixel increments.  The
  absolute value of these offsets is an exponentially
  decreasing function of separation.  The blue line shows the
  fit of an exponential function to these offsets.  We use
  this function to determine the additional uncertainty
  necessary for a source given its fit
  separation. \textbf{Middle}: Median offset in fit position
  angle (PA) from input 
  PA, binned in 5 degree increments.  Because of variable PSF
  structure, the offsets have no obvious functional form.  We
  therefore use these binned data to apply an additional
  uncertainty in PA given the PA of the binary.
  \textbf{Bottom} Measured absolute offsets in fit flux
  ratio from input flux ratio.  We use the median of all these values,
  represented by the red line, as the additional uncertainty
  in flux ratio. }
\label{psf_off}
\end{figure*}

Additional systematic uncertainties need to be accounted for when
determining the final astrometric and photometric measurements
for each binary.  First, absolute uncertainties in the plate scale
and position of north given above are accounted for in all
astrometry.  A further,
more complicated, source of uncertainty stems from using a
PSF that is not imaged simultaneously with each binary, introducing systematic uncertainties in both
astrometry and photometry.  In particular, the variability of
the AO performance over a given night generates time variable
PSF structure that can contribute to slight offsets in
astrometry and photometry.  To
estimate the additional uncertainty due to imperfect PSF
matching, we performed simulations in which 1000 artificial
binaries were generated using images of an image single sources with
separations and flux ratios spanning the range
observed for our sources.  These artificial binaries were then
fit with StarFinder using either a separate single source from the same night or
simulated source as the PSF.
This exercise was performed for every night in which
observations were taken for both PSF types.  Examples of the results of these
simulations are shown in Figure \ref{psf_off} (from the night of 2006 May
21).  We find in all
simulations that median offsets between input and fitted
separations are an exponentially decreasing function of the
separation, meaning that fits to tighter binaries were more
discrepant from the correct values than those to
wide binaries.  We also find that due to variable structure in AO PSF
halos (even after accounting for pupil rotation), the offset in fitted position angle is a 
function of
the position angle of the binary.  Finally, the median offset in fitted flux ratio with respect to
the input flux ratio was essentially constant for all separations
and position angles. Therefore, for every measurement of each
target, we compute the necessary uncertainties from imperfect
PSFs based on the relationships detailed above, taking the median values of the measured offsets as the
 magnitude of
the additional uncertainty.  The PSF uncertainties have the
greatest impact on the tightest systems or on nights when
the performance was poor (strehl ratios $\lesssim$20$\%$).  In about 25$\%$ of our
measurements, this PSF uncertainty is larger than our
statistical uncertainty.  The astrometry and relative photometry for
all sources at all epochs is given in Table \ref{tab:bd_astro}.  Those sources
that were unresolved in our observations have
upper limits on binary separation only.  Uncertainties in
Table \ref{tab:bd_astro} are listed separately for the purpose
of illustrating the relative magnitude of each source of
uncertainty, but for all further analyses, we add
them together in quadrature to give a final uncertainty.

\subsection{Spectroscopic Data Analysis}\label{sec:specan}

The basic reduction of the NIRSPAO spectra was performed with
REDSPEC, a software package designed for
NIRSPEC\footnote{http://www2.keck.hawaii.edu/inst/nirspec/redspec/index.html}.
Object frames are reduced by subtracting opposing
nods to remove sky and dark backgrounds, dividing by a flat
field, and correcting for bad pixels.  As mentioned above, for
the purposes of this work we only analyzed Order 33, which
contains the CO bandhead region.  This order is
spatially rectified by fitting the trace
of each nod of A0 calibrators with third order
polynomials, and then applying the results of those fits
across the image.  A first-order guess at the wavelength
solution for the spectra is obtained using the etalon lamps
that are part of the lamp suite of NIRSPAO (this is used as a
starting point for our derivation of the true wavelength
solution).  Order 33 has very
few OH sky lines or arc lamp lines to use for this purpose.  To obtain
the correct values of the wavelengths for the etalon lines, we
followed the method described by Figer et al. (2003).  The
wavelength regime that Order 33 encompasses was found to be
between $\sim$2.291 and $\sim$2.325 $\mu$m.  The output we
used from REDSPEC was therefore a reduced, spatially rectified
and preliminarily spectrally rectified fits image of order 33.

 As these systems are fairly tight
binaries, cross-contamination can be an issue when extracting
the spectra.  This made the simple square-box extraction
provided in REDSPEC unsuitable for these observations.  We
therefore extracted the spectra by first
fitting a Gaussian to the trace of one component of the binary
and subtracting the result of this fit from the frame to leave
only the other component.  The width of the Gaussian is
allowed to vary with wavelength, although over the narrow
wavelength range covered by order 33, the variation is
small.  Typically the binaries are separated by more than a
FWHM of this Gaussian, making the fit of the bright stars' trace unbiased by
the other.  In the few cases where the traces were separated
by less than about 7 pixels, the fitted FWHM would be
artificially widened due to the presence of the companion.  In
these cases, we fixed the FWHM of the Gaussian to that
measured for other, more widely separated sources observed on
the same night.  After the trace of one component was fitted
and subtracted from the frame, the trace of the remaining
component was then also fit with a Gaussian for extraction.  We
normalized this Gaussian such that the peak was given a value
of one and corresponded to the peak of the trace in the
spatial direction.  We then weighted the flux of each pixel by
the value of the normalized Gaussian at that pixel location,
and then added these weighted fluxes together to get our final
extracted spectrum.  We do not remove telluric absorption from our order 33 spectra
because telluric lines are used for radial velocity determination.

Radial velocities are determined from the extracted spectra relative to the
telluric absorption features, which provide a stable absolute
wavelength reference (e.g. Blake et al. 2007). Our specific
prescription is identical to that outlined in detail in Bailey 
et al. (2010); they demonstrate radial velocity precisions of 50 m/s with
NIRSPEC spectra for slowly rotating mid-M dwarfs. Here we
provide only a brief overview of this method.  Each 
extracted spectrum is modeled as a combination of a KPNO/FTS telluric spectrum 
(Livingston $\&$ Wallace 1991) and a synthetically generated
spectrum derived from the PHOENIX atmosphere models
(Hauschildt et al. 1999). The model spectrum is parameterized
to account for the wavelength solution, continuum normalization, 
instrumental profile (assumed to be Gaussian), projected
rotational velocity ($v$sin$i$)\footnote{While this method has been shown
to produce reliable radial velocities, the \textit{V sin i}'s
have known systematics that are perhaps due to an additional
degeneracy with the instrumental PSF.}
 and the radial velocity. The best fit is
determined by minimizing the variance-weighted reduced
$\chi^2$ of the difference between the model and the extracted
spectrum, once this difference has been Fourier filtered to
remove the fringing present in NIRSPEC K-band  
spectra. This fit is only done using a single order of our
NIRSPEC spectra (order 33), since it uniquely contains a rich
amount of both telluric and stellar absorption features,
sufficient for precise calibration. 

To account for any systematic effects from using a template of
a given temperature, which can impact the value of $v$sin$i$ and
potentially cause slight shifts in the derived radial velocity, we use
multiple templates spanning 300 K in temperature to determine
the systematic uncertainty in 
radial velocity
due to our synthetic template.  We find that this systematic error amounts to
approximately $\sim$0.2-0.3 km/s for most targets.  

\begin{figure*}
%\epsscale{1.0}
\epsfig{file=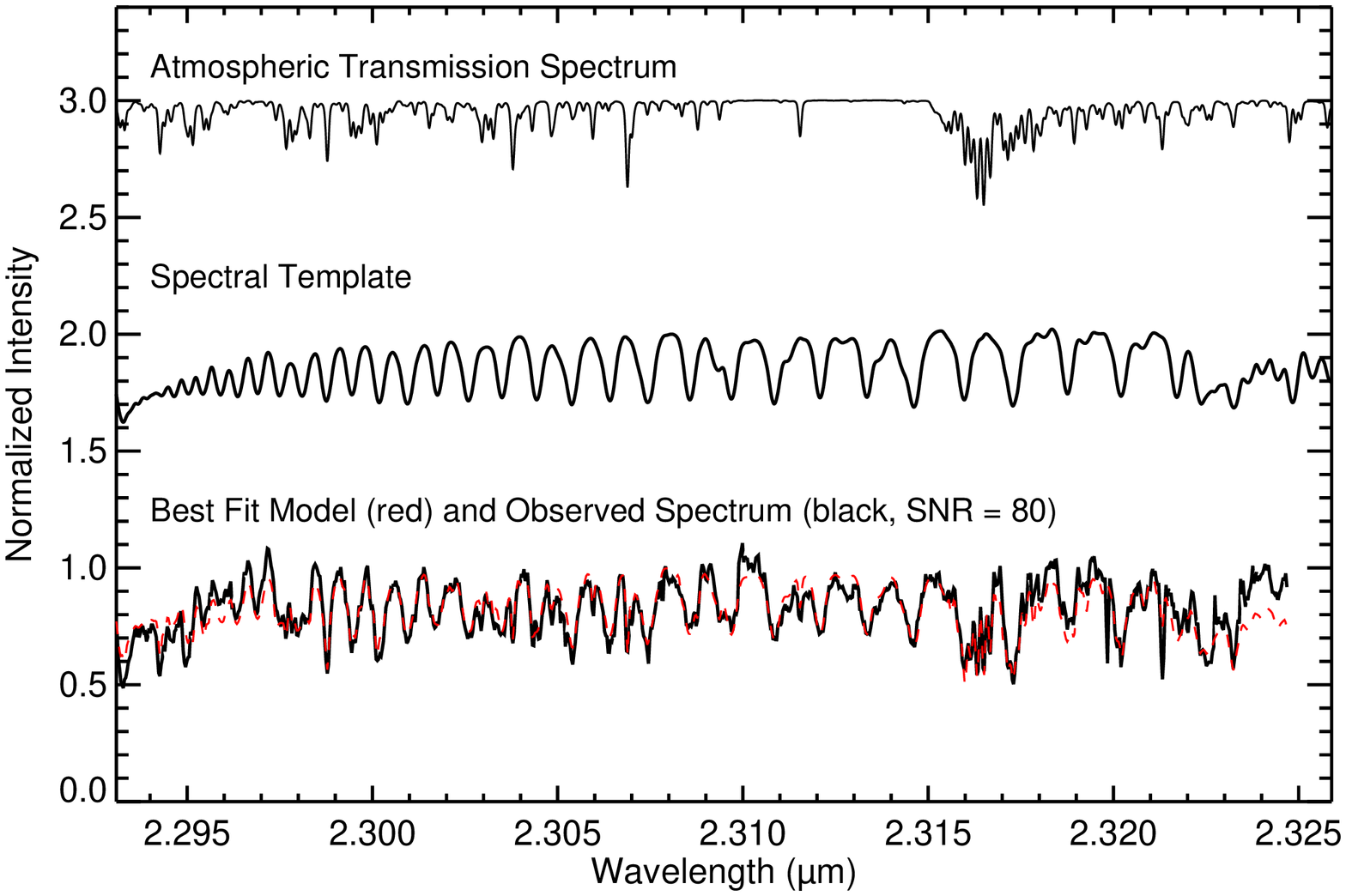,angle=90,width=\linewidth,height=3.5in}
\epsfig{file=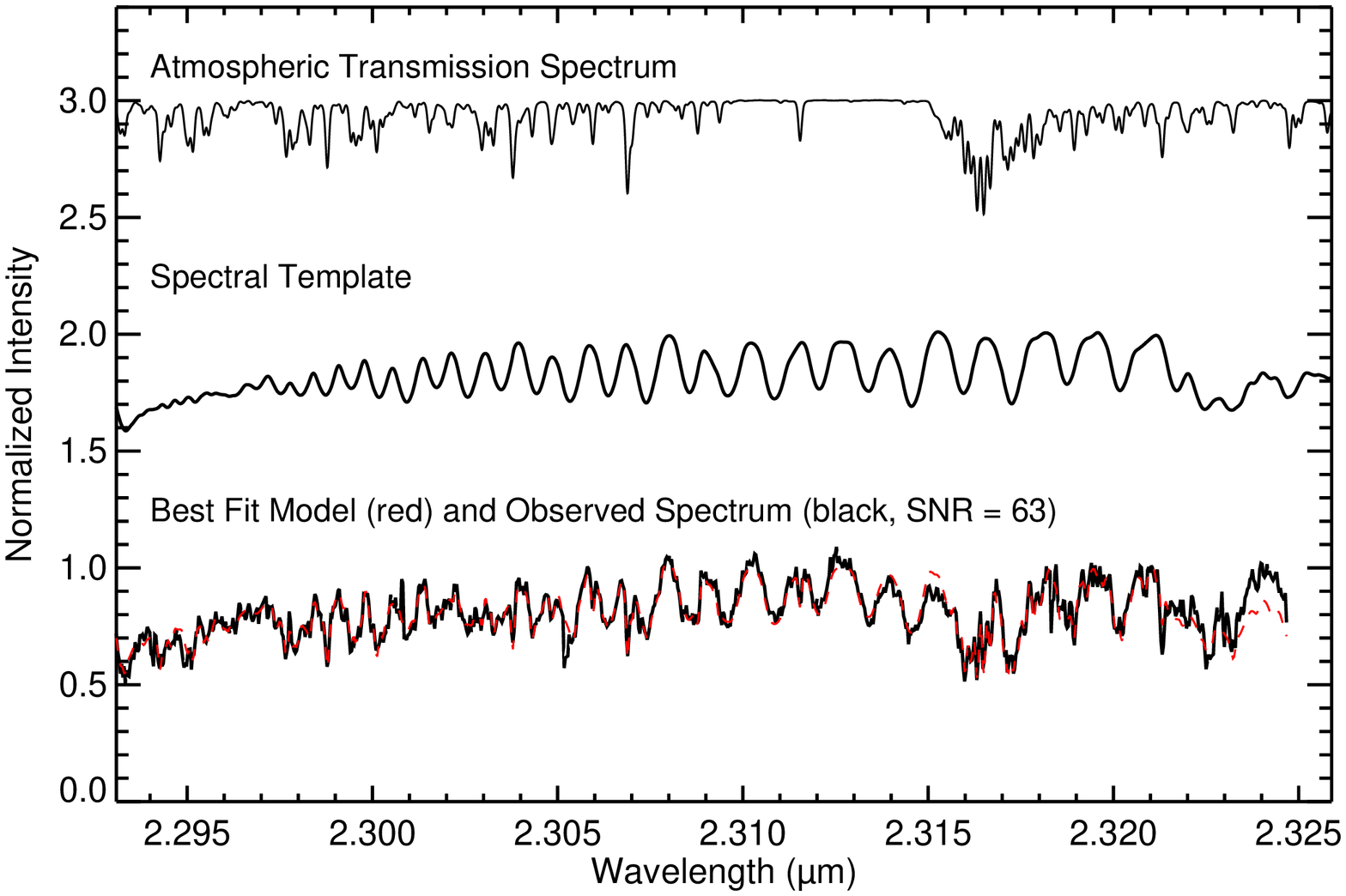,angle=90,width=\linewidth,height=3.5in}
\caption{Example of a fit for radial velocity for the components
  of 2MASS 0746+20 A (\textbf{top}) and B (\textbf{bottom}) from the night of
  2007 Dec 04.  The atmospheric transmission spectrum used for
wavelength calibration is shown, as well as the theoretical
spectral template.  On the bottom of each panel, we plot our
actual spectrum in black (note that the telluric features have
not been removed, as is necessary for the fitting) and
overplot in red the best fitting model that combines the
synthetic atmospheric and spectral templates.  Example spectra for all other
systems with NIRSPAO measurements are shown online.}
\label{2mass07464_specfit}
\end{figure*}

The measured radial velocities from this method are reported in
Table \ref{tab:bd_rvdata}.  Our uncertainties in radial
velocity range from 0.4 to 2.8 km/s, consistent with the
assumptions used in our original Monte Carlo simulations
(section \ref{sec:sample}).  In Figure \ref{2mass07464_specfit}, we show
example fits for each of the sources with spectroscopic
observations (additional examples are included in online only figures).

\section{Orbital Analysis}\label{sec:orbs}

The orbital analysis of the 24 stars monitored for this study fall  
into the following three categories:

\begin{itemize}

\item [1.]  There are 9 stars that do not have sufficient kinematic  
information to
do any orbital analysis.  Of the stars that fall into this category, 3  
were unresolved
in all epochs of our NIRC2 observations.  These sources may have  
orbits that take them below the Keck
diffraction limit during this study.  We note that the initial  
separations measurements, made with HST, were 0.17" (2MASS 0652+47),
0.057" (2MASS 1600+17), and 0.051" (2MASS 0518-28).   Another binary in this  
category (2MASS 1047+40) was resolved
in the first epoch of NIRC2 observations, but unresolved in the  
subsequent epochs.  The remaining
5 sources in this category have been resolved in all NIRC2  
observations, but have not yet shown significant astrometric curvature  
due to either their late addition to the sample (2MASS 0700+31, 2MASS  
1021-03, 2MASS 2101+17, 2MASS 2152+09) or, in the case of 2MASS 1017+13, large  
projection effects.  We are continuing to monitor the first 4 but  
have stopped observing the latter target, as updated Monte Carlo  
simulations with the new epoch of data showed that this source was no  
longer "likely" to yield a mass with the
necessary precision on a few years timescale.  The unresolved and  
resolved astrometric measurements of these 9 systems are reported for  
completeness in Table 4.

\item [2.] There are 15 stars that have enough kinematic information  
measured to solve for the system's relative orbit, from
which the total mass and eccentricity of the system can be inferred.    
All of them have at least 4 independent measurements; 4 of them have  
only astrometric data and 11 of them have multiple epochs of  
astrometry and at least one epoch of radial velocity measurements.   
The relative orbit analysis is described in \S5.1.

\item [3.] There are 6 systems, which are a subset of those in category  
[2], that have three or more radial velocity measurements
for the individual components, allowing for estimates of the system's  
absolute orbital parameters from which individual masses can be  
derived.  We have shown that radial velocity measurements are possible  
for an additional 5 stars.  Two other sources, which have been unresolved in  
NIRC2 observations, have K magnitudes that are comparably bright (K$\lesssim$  
12.0).  Further measurements of these systems are likely to also yield  
individual masses.  The absolute orbital analysis is described in \S5.2.

\end{itemize}

\subsection{Relative Orbit Model Fits}
\label{sec:astroorb}

To derive total mass estimates from relative orbital solutions
for our sources, we 
combine our relative astrometric measurements from Section
\ref{astroan}, previous astrometry 
reported in the literature, and the relative radial
velocity between the components as determined in section
\ref{sec:specan}.  As described in Ghez et al. (2008), our model
for the relative orbit always contains six free parameters:
period (P), semi-major axis (a),
eccentricity (e), time of periapse passage (T$_o$), inclination (i), position
angle of the ascending node ($\Omega$), and longitude of periapse
passage ($\omega$).  We can remove the degeneracy in the values of $\Omega$
and $\omega$ which exists
without information in this third dimension for the 11 sources
that have radial velocity information.  The radial
velocity data also allows distance to be a free parameter in
the fit for those sources without a previously-measured
parallax (5 systems).  For those systems with a parallax measurement (9
systems), we
do not allow distance to be a free parameter, but rather we
constrain it to be consistent with the parallax distance and
its uncertainties.  The uncertainties on parallax
measurements are smaller than those from fitting for distance
as a free parameter, and the values are consistent in all cases.  The distances,
either used or derived in our fits, are given in Table \ref{tab:astro_orb}.  In
the case of one system, 2MASS 0920+35, we had neither radial
velocities nor a parallax measurement, so we use instead the
photometric distance as determined from the relationship in
Cruz et al. (2003), which is based on J band photometry and
spectral type (here assumed to have an uncertainty of $\pm$2
spectral subclasses).  The
best fit orbital parameter values are found by
minimizing the total $\chi^2$, which is found by summing the $\chi^2$ of each data type ($\chi^2_{tot}$ =
$\chi^{2}_{ast}$ + $\chi^2_{rv}$; see Ghez et al. (2008) for
more details on this fitting procedure).

After the best fit is determined, the uncertainties in the
orbital parameters are found via a
Monte Carlo simulation.  First, 10,000 artificial data sets are
generated to match the observed data set in number of points,
where the value of each point (including the distance when
determined from parallax measurements) is
assigned by randomly drawing
from a Gaussian distribution centered on the best-fit model value with a
width corresponding to the uncertainty on that value.  Each of
these artificial data sets is then fit with an orbit model as
described above, and the best fit model is saved.  The
resulting distribution of orbital parameters represents the
joint probability distribution function of those parameters.
We obtain the uncertainties on each
parameter as in Ghez et al. (2008), where the distribution of
each parameter is marginalized against all others and
confidence limits are determined by integrating the resulting
one-dimensional distribution out to a
probability of 34$\%$ (one sigma) on each side of the best fitting value.  On occasion, when one or more
parameters are not well-constrained, the best fit value does
not correspond to the peak of the probability distribution.
However, in almost all cases the best fit value for a parameter is
within 1$\sigma$ of the peak.  The few fit parameters in which
this is not the case are normally represented by bifurcated or
poorly constrained flat distributions (see for example the distributions of e and
$\omega$ for 2MASS 1847+55 AB, online version of Figure \ref{fig:astrhist}).

The resulting best-fit orbital parameters and their
uncertainties are given in Table \ref{tab:astro_orb}.  We find
the orbital solutions for 15 of the systems in our sample.  The orbital solutions are 
shown, along with both the astrometric and relative
radial velocity data points, in Figures
\ref{fig:2mass07464_orb}-\ref{fig:lhs2397a_orb}.  The dotted
blue lines represent the 1$\sigma$ range of separations and relative
radial velocities allowed at a given time based on the orbital
solutions from the Monte Carlo.  The
distributions of orbital parameters for three sources are shown
in Figures
\ref{fig:2mass0746_astrhist}-\ref{fig:2mass21402_astrhist},
chosen to be representative of the full sample.  The rest of
the distributions are shown online.  The
shaded regions on the histograms show the 1$\sigma$ ranges of
each parameter.  If the distances were sampled from previous
parallax measurements, they are denoted with a red histogram.
These figures are alphabetically ordered based on the sources' names.

\subsection{Absolute Orbit Model Fits}

For 6 systems in our sample, sufficient absolute radial velocity
measurements (at least 3) have been made, in conjunction with their relative
orbits, to derive the first estimates of
their absolute orbits, and hence the individual masses of
the binary components.  Common parameters between
absolute and relative orbits, namely the P, T$_{o}$,
e, and $\omega$ make it possible to only
have to fit two free parameters: the semiamplitudes of the
velocity curve for the primary (K$_{Primary}$) and
the systemic velocity ($\gamma$).  K$_{Secondary}$ is derived
from the constraint that K$_{Primary}$ + K$_{Secondary}$ = 2$\pi$ a
sin\textit{i} / P (1 - e$^{2}$)$^{1/2}$.   

To first obtain the best fit
solution for these parameters, we use our radial velocities
from Table \ref{tab:bd_rvdata} and fix the values of P, a, T$_{o}$,
e, i, and $\omega$ to the values obtained in the relative
orbit fitting to perform a least-squares minimization
between the equations for the spectroscopic orbit of each
component and our data.  We fully map $\chi^2$ space (where in
this case $\chi^2_{tot}$ = $\chi^{2}_{Primary}$ +
$\chi^2_{Secondary}$) by first sampling randomly 100,000 times
from a uniform distribution of K$_{Primary}$ and $\gamma$ that
are wide enough to allow 
mass ratios between 1 and 5 (where M$_{primary}$ /
M$_{Secondary}$ = K$_{Secondary}$ / K$_{Primary}$) for all
sources except LHS 2397a AB, for which we allow for mass
ratios between 1 and 10.  To determine the uncertainties on our fit parameters, we
again perform a Monte Carlo simulation.  We use the
distributions of P, a, T$_{o}$, e, i, and $\omega$ derived from our
astrometric orbit Monte Carlo as inputs into the fits to
account for the uncertainty in these parameters.  We also then
resample our radial velocity measurements to generate
10000 artificial data sets such that the value of each point is
assigned by randomly drawing
from a Gaussian distribution centered on the true value with a
width corresponding to the uncertainty on that value (as was
done with the astrometric data).  We then find the best fit
solution for each of these data sets (coupled with the sampled
parameters from the astrometric fits).  As with the
astrometric orbit, we find the uncertainties
by marginalizing the resulting distribution of
each parameter against all others and integrating the resulting
one-dimensional distribution out to a
probability of 34$\%$ on each side of the best fitting value.      

\begin{figure*}
\epsscale{0.9}
\plotone{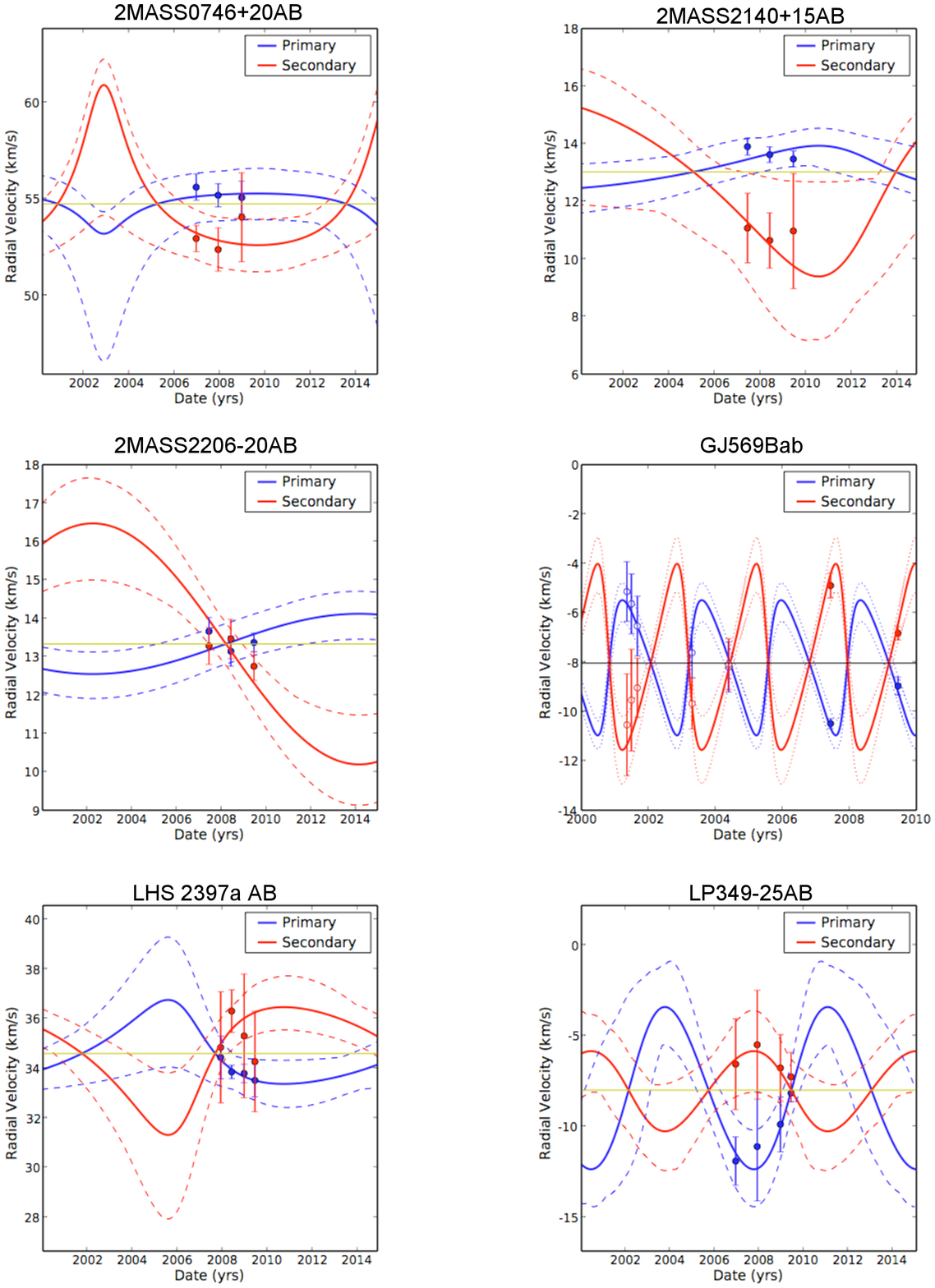}
\caption{Best fit absolute orbits for 6 systems in our
  sample.  Absolute radial velocity data points overplotted with the best
fit orbits for both components.  Radial velocity data from the
literature for GJ 569Bab is taken from Zapatero Osorio et
al. (2004) and Simon et al. (2006).  The green line represents the
best fit systemic velocity.  The dotted lines represent the
1$\sigma$ allowed ranges of radial velocity at a given time.}
\label{fig:2mass21402_specorb}
\end{figure*}

The resulting best-fit orbital parameters for the absolute motion and their
uncertainties are given in Table \ref{tab:spec_orb}.  The
absolute orbital solutions are 
shown with the absolute radial velocity datapoints in Figure
\ref{fig:2mass21402_specorb} and
the distributions of orbital parameters for LHS 2397a AB, as a
representative example, is shown
in Figure \ref{fig:lhs2397a_spechist}.  All other
distributions are shown in the online version of the figure.  By
combining our mass ratio distribution derived with these data
with the total system mass derived in Section \ref{sec:astroorb}, we have
computed the first direct measurements of the individual
masses of the components for 5 of these 6 systems.  These
individual masses are given in Table \ref{tab:spec_orb}.  

\subsection{Eccentricity Distribution}
\label{sec:ecc}

\begin{figure*}
\epsscale{1.0}
\plottwo{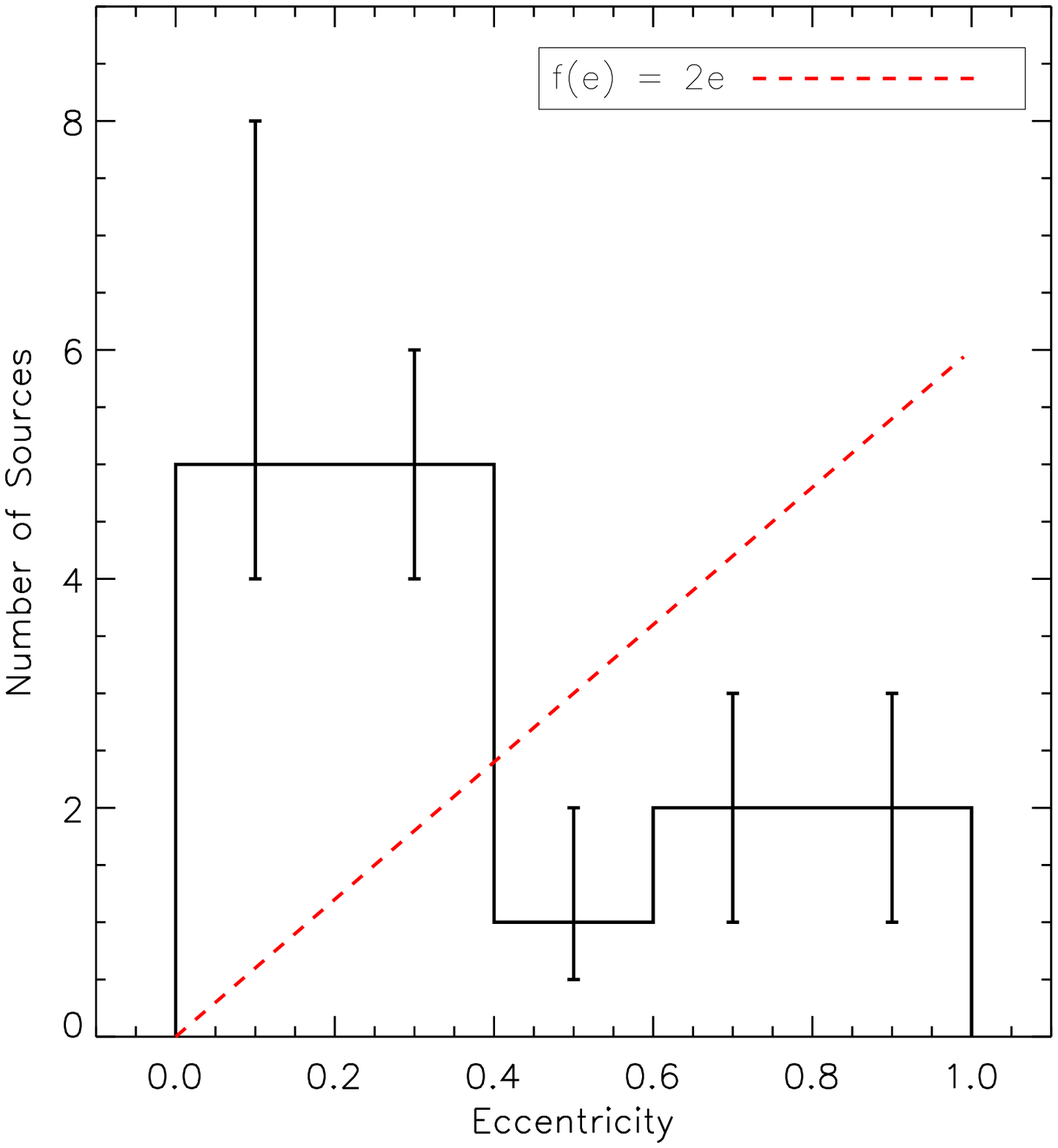}{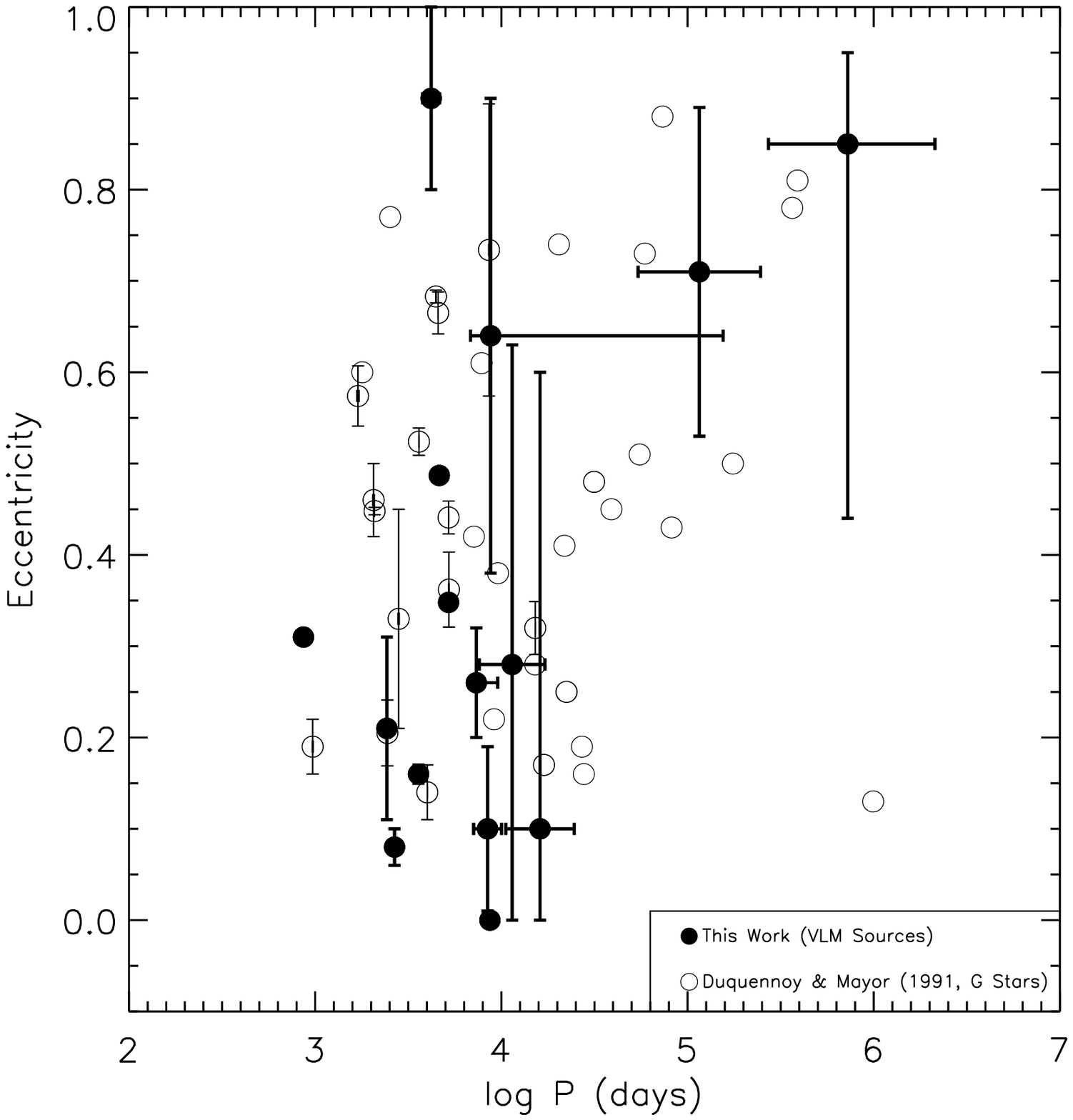}
\caption{\textbf{Left:} The eccentricity
distribution of our sample based on the eccentricity
distributions for each source from the relative orbit Monte
Carlos.  Overplotted is the relation for field solar-like
stars from Duquennoy $\&$ Mayor (1991), where f(e) = 2e
(normalized to 15 systems).   \textbf{Right:} Eccentricity as
a function of period for the sources in our sample (filled
circles).  Overplotted are the systems from Duquennoy $\&$
Mayor (1991) with periods greater than 1000 days (open circles).  As in Duquennoy $\&$
Mayor (1991), eccentricity tends to increase with period.}
\label{fig:ecc_dist}
\end{figure*}

Using the distributions from our Monte Carlo simulations in
section \ref{sec:astroorb}, we can begin to examine the bulk
orbital properties of our sample of VLM objects.  In
particular, we can determine the distribution of
eccentricities for our sample (9 of which are constrained to
better than 30$\%$), which may shed light on the
formation of VLM binaries.  To determine our eccentricity
distribution, we performed a 
Monte Carlo analysis in which we randomly sampled one value of eccentricity per
source from the distributions in Section
\ref{sec:astroorb}.  In each trial, the total number of
sources per bin was calculated in bins of width 0.1 over the range of
values from 0 to 1.  We performed 10,000 of these trials,
which gave a distribution for each bin of the number of
expected sources.  This distribution provided a predicted
number of sources in each bin along with an uncertainty.  We
then combined distributions such that 
the final bin width was 0.2.  The resulting distribution for
our 15 sources is shown in the left panel of Figure \ref{fig:ecc_dist}.  

Though we have a small sample, the eccentricities of the
binaries in our sample appear to follow a rough trend with orbital period.  In the right panel of Figure
\ref{fig:ecc_dist}, we plot the eccentricity of our sources as
a function of period.  In addition, we have overplotted on Figure
\ref{fig:ecc_dist} the periods and eccentricities from Duquennoy $\&$ Mayor
(1991) for solar-like field stars with periods $>$ 1000 days
(all sources in our sample meet this criteria except GJ 569B, which has a period
of 865 days and an eccentricity of 0.31).  Duquennoy
$\&$ Mayor (1991) also found that eccentricity appeared to be a
function of period, with longer period systems tending towards
higher eccentricities (albeit with fairly large scatter past
the tidal circularization period of $\sim$10 days).  A
2-dimensional K-S test between 
our distribution and the distribution from Duquennoy $\&$
Mayor (1991) shows that the two samples have an 11$\%$ chance
of being drawn from the same distribution (therefore being
consistent to within 1.6$\sigma$), suggesting the
distributions are statistically consistent (again with the
caveat that we have a much smaller sample than those authors).  Thus, although
our sample appears to have a slight overabundance of moderate
to low eccentricities compared to the expected distribution of
eccentricities if a population is dynamically relaxed of
f(e)$\sim$2e (shown as a red line on the left panel of Figure
\ref{fig:ecc_dist}), the similar trend in our sources with
period to Duquennoy $\&$ Mayor (1991) suggests that the VLM
objects may ultimately have a similar eccentricity distribution.

\subsection{Individual System Remarks}
\label{sec:notes}

\subsubsection{2MASS 0746+20AB}

2MASS 0746+20 AB originally had its total system mass derived
by Bouy et al. (2004).  Those authors found a total mass of
0.146$^{+0.016}_{-0.006}$ M$_{\odot}$.  Our new astrometric
and radial velocity data has allowed us to improve this total mass
estimate by a factor of 4 to 0.151 $\pm$ 0.003 M$_{\odot}$, or to a precision
of 2$\%$.  This measurement represents the most precise mass
estimate for a VLM binary yet determined.  Our
individual mass estimates are the first for this system and
the first for a binary L dwarf.

Though we have achieved superb precision in total mass for
this system, the uncertainties in our radial velocity
measurements are large compared to the current difference
between in the velocities of the components (as can be seen in
Figure \ref{fig:2mass21402_specorb}).  This has limited the precision we can
currently achieve for the individual component masses, meaning
cannot yet resolve the debate on whether the secondary is a
brown dwarf or a low mass star (Gizis $\&$ Reid 2006).  However, our measurements can shed some
light on the result by Berger et al. (2009), in which a radius
for the primary component of the system was estimated using
periodic radio emission from the system.  Based on the
assumptions that the emission was coming from the primary,
that the previously reported, spatially unresolved \textit{V sin i}
measurements reflected the \textit{V sin i} ($\sim$25 km/s) of the primary, and that the
rotation axis of the primary was aligned with the binary
inclination, Berger et al. (2009) derive a radius of 0.76
$\pm$ 0.10 R$_{Jup}$ for this system.  These authors note that
this radius is about 30$\%$ smaller than expected based upon
the models.  Though we do not report definitive \textit{V sin
  i} measurements for each component from \textit{spatially
  resolved} spectra, providing an assumption free value for
this type of analysis, preliminary work comparing all K band
spectra across multiple orders for this system to high
resolution atmosphere models suggests that the rotational
velocity of the secondary component may be $\sim$35 km/s.
If it were the case that the radio emission were coming from
this component of the binary instead of the primary (which
does seem to have \textit{V sin i}$\sim$ 25 km/s), this would
increase the derived radius to something more in line with
predictions.  We leave the full analysis of our spectroscopy
to derive quantities such as spatially resolved \textit{V sin i} for a
future publication.  

\subsubsection{2MASS 0850+10AB}

2MASS 0850+10 AB has two independent measurements of its
distance via parallax.  The first was from Dahn et al. (2002,
25.6 $\pm$ 2.5 pc) and the second was from Vrba et al. (2004,
38.1 $\pm$ 7.3 pc).  These values are about 1.5$\sigma$
discrepant from each other, a fact noted by Vrba et
al. (2004).  We performed a full Monte Carlo analysis as
described in section \ref{sec:astroorb} using both estimates
for distance.  Since the current uncertainties in the period and
semimajor axis for this system are large, the
impact on the total mass estimate of choosing one distance over the other is negligible.
We choose to present the values of mass as derived
from the Vrba et al. (2004) distance estimate here because the
larger uncertainties on this value make it the marginally more
conservative choice.

\subsubsection{2MASS 0920+35AB}

2MASS 0920+35 AB was discovered to be binary by Reid et
al. (2001) using HST.  A follow-up
monitoring campaign of the system was performed by Bouy et
al. (2008) using both HST and the VLT in conjunction with
their facility AO system.  In each of the five observations
performed by Bouy et al. (2008), the system was unresolved
(2002 Oct, 2003 Mar, 2005 Oct, 2006 Apr).
These authors postulated that the binary was therefore perhaps
on a highly inclined orbit with a period of roughly 7.2
years.  When our monitoring of the system began in 2006, the
system was again resolved, and remained resolved for all of
our measurements until our most recent in 2009 June.  We
therefore utilize both the resolved and unresolved
measurements to perform our orbit fits.  First, we fit the
resolved astrometric measurements for relative orbital parameter solutions as
described in Section \ref{sec:astroorb}.  We then took the
output orbital solutions for those trials and calculated the
predicted separation of the binary at each of the epochs in
which it was unresolved - if the predicted separation was
above the detection limits given by Bouy et al. (2008) or our
2009 June 10 measurement, it was thrown out.  These unresolved
measurements therefore provided tighter constraints on the
orbital parameters for this system.  

The results of the Monte
Carlo simulation for this system are shown in Figure
\ref{fig:2mass09201_astrhist}.  As shown in this figure,
the resulting distribution of periods has a strong
bifurcation, whereby $\sim$45$\%$ of the solutions favor an
orbital period of $\sim$3.3 years and a very high eccentricity,
and 55$\%$ favor the best fit solution of $\sim$6.7 years and more
modest eccentricities.
Since these solutions are nearly equally preferred but quite
distinct, we display the best fit of both solution families in
Figure \ref{fig:2mass07464_orb}.  The two solution sets cause the
current mass uncertainty to be fairly high.  The mass
distributions for the two sets overlap, creating the
continuous distribution seen in Figure
\ref{fig:2mass07464_orb}.  The long tail out to masses greater
than 1 M$_{\odot}$ is generated by the short period solution
set.  An additional astrometric measurement before mid-2010 should distinguish
between the two sets, as it will not be resolved for periods
of $\sim$6.7 years but it will be resolved for periods of
$\sim$3.3 years.  Further, we have found the inclination of
this system to be nearly edge on, meaning it has a
non-negligible chance of being an eclipsing system (see
section \ref{sec:disc}).

\subsubsection{2MASS 1534-29AB}

The first derivation of the orbit of 2MASS 1534-29AB was
performed by Liu et al. (2008), where they calculated a total
system mass of 0.056 $\pm$ 0.003 M$_{\odot}$.  By combining
our astrometry with that reported by Liu et al. (2008), we
find a slightly higher, but consistent, total system mass of
0.060 $\pm$ 0.004 M$_{\odot}$.  We note that if we perform our
analysis on only the astrometry given in Liu et al. (2008), we
obtain a mass of 0.056 $\pm$ 0.004 M$_{\odot}$, consistent
with their values.  

\subsubsection{2MASS 2140+16AB and 2MASS 2206-20AB}

We have acquired sufficient radial velocity data to make the
first calculations of the absolute orbits of these systems.
However, the uncertainty in the radial velocities is comparable
to the difference between the values.  Because of this, the best
fit is typically the one that minimizes K$_{Primary}$, which
in turn maximizes K$_{Secondary}$.  This leads to
relatively high predicted mass ratios.  Though there is some
spread in the value of mass ratio, as shown in the online
version of Figure
\ref{fig:lhs2397a_spechist}, the mass ratio is quite peaked
at this high value.  This leads not only to mass values for
the secondary that are likely too low given their approximate
spectral types, but also uncertainties that are too small for the
secondary given the uncertainty in the mass of the primary.
For these two systems, we therefore extend the uncertainty in
the secondary mass by combining in quadrature the uncertainty
in the total system mass and the uncertainty in the mass of
the primary component.  We have noted that we have taken this
approach in Table \ref{tab:spec_orb}, and have shaded the
histograms in Figure \ref{fig:lhs2397a_spechist} to reflect our chosen
uncertainties.  Though these first estimates of individual
mass are fairly uncertain, they will improve with continued
monitoring of these systems.

\subsubsection{GJ 569Bab}
\label{sec:gj569b_disc}

The first derivation of the relative orbit of GJ 569Bab was
performed by Lane et al. (2001), and was followed with
improvements by Zapatero Osorio et al. (2004) and Simon et
al. (2006).  The work of Zapatero Osorio et al. (2004) and
Simon et al. (2006) also contained spatially resolved, high
resolution spectroscopic measurements for this system, which
is one of two targets in our sample that is an NGS AO target.
Zapatero Osorio et al. (2004) derived the  
first estimate of the individual component masses of this
system using their J band spectroscopic measurements.  Simon
et al. (2006) made their radial velocity measurements in the H
band and noted that their derived center of mass velocity
(-8.50 $\pm$ 0.30 km/s) is
discrepant from that of Zapatero Osorio et al. (2004, -11.52
$\pm$ 0.45 km/s) by $\sim$3 km/s.  Simon
et al. (2006) postulate that this stems from the choice of
lines used to make their measurements.  Zapatero Osorio et
al. (2004) use the K I doublet location referenced to
laboratory wavelengths while Simon et al. (2006) perform
cross-correlation of their full order 48 ($\lambda$ = 1.58 -
1.60 $\mu$m) and 49 ($\lambda$ = 1.55 - 1.57 $\mu$m) spectra with
spectral templates.  Simon et al. (2006) also note that the
relative radial velocities are consistent with what is
predicted based on astrometry.

\begin{figure*}
\epsscale{1.0}
\plotone{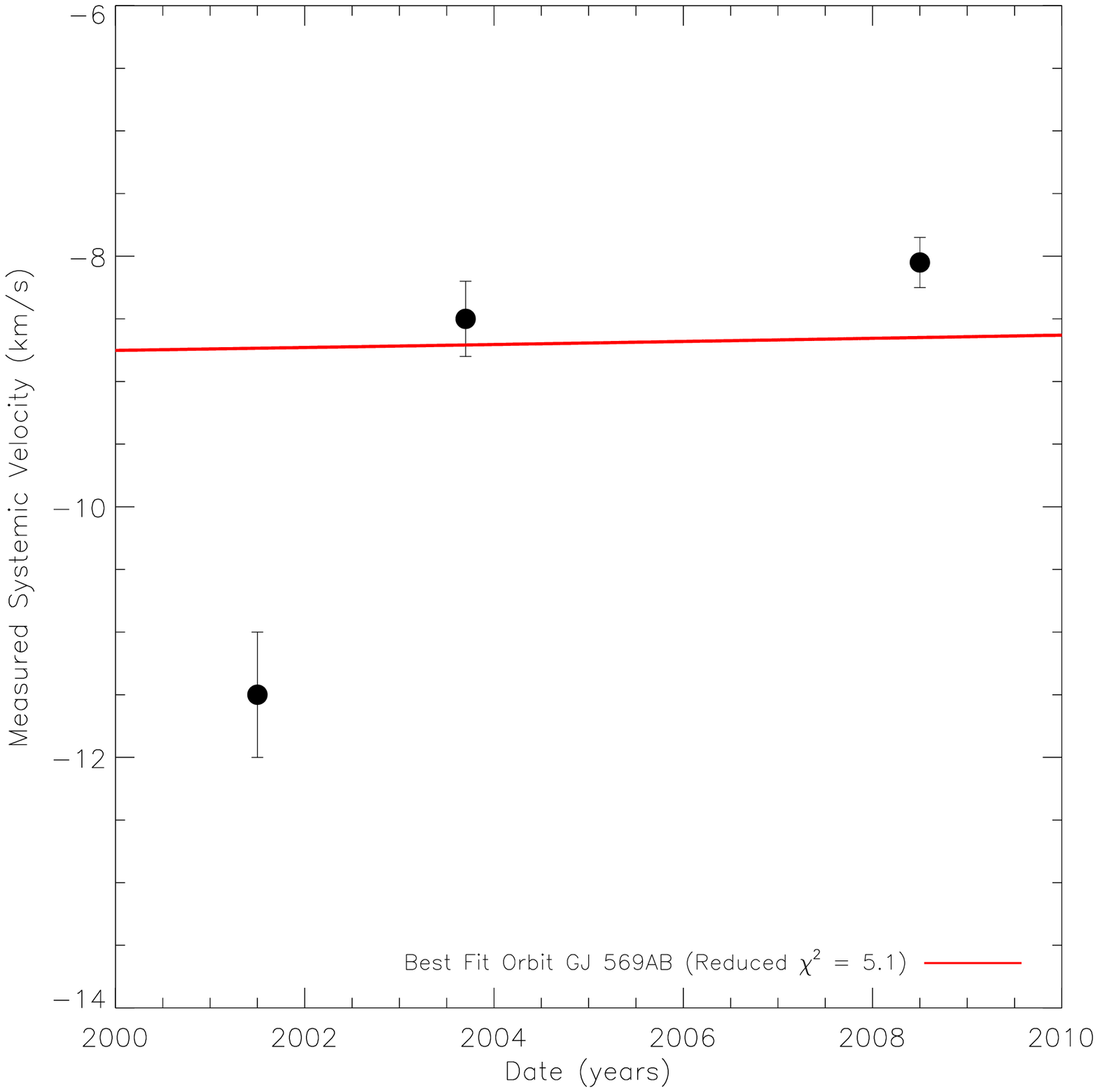}
\caption{The
  measured systemic velocities from Zapatero-Osorio et
  al. (2004), Simon et al. (2006), and this study, as a
  function of the median time of observation.  Because this
  source is a wide companion to GJ 569A, an M star, it is
  expected to undergo some change in velocity due to its orbit
around GJ 569A.  Overplotted in red is the best fit orbit to
all astrometric data for GJ569AB and these three radial
velocity measurements.  The Zapatero Osorio et al. (2004) data
point lies clearly off the best fit, which has a reduced
$\chi^{2}$ of 5.1 due to the offset of this data point, and is
pulled by this data point to a very high eccentricity (0.9).  Thus,
the differences between the Zapatero Osorio et al. (2004)
systemic velocities and those from this study and Simon et
al. (2006) are likely not due to orbital motion.}
\label{fig:gj569b_veldrift}
\end{figure*}

We now note that our spectra, measured in
the K band and fit for radial velocity as described in Section
\ref{sec:specan}, appear to be systematically offset from both
the measurements of Zapatero Osorio et al. (2004), but
consistent to within 1.25$\sigma$ of Simon et
al. (2006).  We find a center of mass velocity using just our
data points of -8.05 $\pm$ 0.20 km/s, which is the most consistent
of the three sets of measurements with that of the M2V
primary of this tertiary system (-7.2 $\pm$ 0.2 km/s).  We
find, as shown in Figure \ref{fig:2mass22062_orb}, that
the relative velocities are consistent with what is expected
for the relative orbit.  Thus the velocity differences truly
seem to be driven by an offset in their absolute value.  It is
possible that these velocity offsets between Zapatero Osorio
et al. (2004) and all other measurements may be related to the orbit of this binary
around GJ 569A.  To examine whether this is the case, we fit a
rough relative orbital solution to all astrometric data in the
literature for GJ569AB (Forrest et al. 1988, Mart{\'{\i}}n et
al. 2000, Lane et al. 2001, Simon et al. 2006) and the relative velocities
between the components as determined by the difference between
the velocity of GJ 569A and each of the three measurements for
GJ569B.  We assume for this exercise that the total system
mass is $\sim$0.5 M$_{\odot}$.  We find that the best fit
orbit that can be obtained for all data has a reduced
$\chi^{2}$ of 5.1, driven entirely by the velocity point from
Zapatero Osorio et al. (2004).  This is demonstrated in Figure
\ref{fig:gj569b_veldrift}, where we plot the predicted
velocity from the best fit orbit with each systemic radial
velocity measurement.  The Zapatero Osorio et al (2004) lies
far from the best fit, while at the same time driving it to
require a very high eccentricity (0.9) and a time of periapse
passage close to the time of the measurements.  Thus, we
conclude that the differences in systemic velocity are likely
due to systematics in absolute radial velocity calibration as
described by Simon et al. (2006) and not orbital motion.  

In order to use all the radial velocity measurements to
calculate individual masses, we opt to shift all data points
from Zapatero-Osorio et al. (2004) and Simon et al. (2006)
such that their center of mass velocity is consistent with ours.  We also
increase the uncertainties in these values such that they
incorporate the uncertainties in our value of systemic
velocity and in the systemic velocity derived in each work,
which we combine in quadrature.  We then use
these shifted velocities in conjunction with our measurements
to derive the absolute orbit, which is shown in Figure
\ref{fig:2mass21402_specorb}.  The application of this offset
results in a very nice fit with a reduced $\chi^{2}$ of 0.56.
We find a mass ratio of 1.4 $\pm$ 0.3, which is lower than the
value of 5.25 found by Simon et al. (2006).  Those authors
postulated that since the mass of the primary appeared to be
so much higher than that of the secondary, that the primary
may be a binary itself (something potentially suggested by the
wider lines seen in GJ 569Ba).  Our values of primary and
secondary mass suggest that the sources actually have fairly
similar masses of 0.073 $\pm$ 0.008 M$_{\odot}$ and 0.053
$\pm$ 0.006 M$_{\odot}$.  We cannot, however, definitively
rule out that GJ 569Ba is comprised of two components, as
suggested Simon et al. (2006), although this possibility is
more unlikely given that we find the mass of GJ 569Ba to be
lower than those authors found.    

\subsubsection{HD 130948BC}

The first derivation of the relative orbit of HD 130948BC was performed
by Dupuy et al. (2009a), where they calculated a total system
mass of 0.109 $\pm$ 0.003 M$_{\odot}$.  By combining our
astrometry with that reported by Dupuy et al. (2009a), we find
an identical, but slightly more precise, total system mass of
0.109 $\pm$ 0.002 M$_{\odot}$.  Although we only have one
radial velocity measurement for this system, which is
insufficient to calculate individual masses for the
components, our radial velocity measurement allows us to
resolve the degeneracy in the values of $\omega$ and $\Omega$.

\subsubsection{LHS 2397a AB}

The first derivation of the relative orbit of LHS 2397a AB was
performed by Dupuy et al. (2009b), where they calculated a
total system mass of 0.146$^{+0.015}_{-0.013}$ M$_{\odot}$.
Combining our astrometry with that reported by Dupuy et
al. (2009b), we also find a consistent total mass of
0.144$^{+0.013}_{-0.012}$ M$_{\odot}$.  Performing our analysis
on just the astrometry given in Dupuy et al. (2009b), we find a
slightly different, but consistent, mass of
0.150$^{+0.014}_{-0.013}$.  Dupuy et al. (2009b) also use their
results in conjunction with a bolometric luminosity and the
evolutionary models (both Burrows et al. (1997) and Chabrier
et al. (2000)) to derive the individual component masses.
Here we derive the first individual mass estimates
free of assumptions, which
allows for a direct comparison to the models.  We find
component masses of 0.09 $\pm$ 0.06 M$_{\odot}$ for the primary and 0.06 $\pm$
0.05 M$_{\odot}$ for the secondary.  The well-mapped velocity
curve of the primary allows for this absolute orbit to be
relatively well-defined with a comparable
number of radial velocity measurements to other sources that
do not yet have well-defined absolute orbits.

\section{Bolometric Luminosity and Effective Temperature Derivation}
\label{sec:tefflbol}

In order to compare the predictions of theoretical
evolutionary models to our dynamical mass measurements,
estimates of both the effective temperature and bolometric
luminosity are required.  With input of these parameters, the
evolutionary models can be used to derive a mass and an age
for a source.  Thus, we must derive these parameters for all
binary components.  

Our method for deriving both of these quantities relies on the
spatially resolved photometry we have obtained with our
imaging data.  In our NIRC2 data, we have measured the flux
ratio of the binary components in the J, H, and K' bands,
given in Table \ref{tab:bd_astro}.  We convert these flux ratios into
individual apparent magnitudes using the unresolved photometry for
these sources from 2MASS (Cutri et al. 2003).  The apparent magnitudes can
then be converted into absolute magnitudes using the distances
from Table \ref{tab:astro_orb}.  We also find the absolute magnitudes for each
system in all other photometric bands for which spatially
resolved measurements exist.  The majority of these
measurements were made in the optical with HST.  The absolute photometry
for all sources is given in Table \ref{tab:phot}.  

The determination of effective temperature for these sources
is complex.  Generally speaking, spectral type is
not as accurate a proxy for the temperature of brown dwarfs as it
is amongst hydrogen burning stars, with derived temperatures
spanning several hundred Kelvin for different sources of the
same spectral type (Leggett et al. 2002, Golimowski et
al. 2004, Cushing et al. 2008).  We therefore opt to perform
spectral synthesis modeling using atmospheric models on
each source individually, which given sufficient wavelength coverage allows
for lower temperature uncertainties for most objects than
would be achieved by using a temperature vs. spectral type
relationship.  Though this introduces a
model assumption into our comparison of these sources
to evolutionary models, we can use our mass estimates to
determine the consistency of the atmospheric and evolutionary
models with each other.
For our sources of late M to L spectral types, we
derive effective temperature using the DUSTY form of the
PHOENIX atmosphere models (Allard et al. 2001).  These
models, in which all refractory elements are assumed to form
dust grains and create thick dust clouds, have
been shown to reproduce well the colors and spectra of these types
of objects. 

 Updated opacities and grain size
distributions, which are used in our analysis, have improved the correspondence of these
models to observations (Barman et al. in prep, Rice et
al. 2009).  Among the 30 individual components in our dynamical
mass sample, 21 have previously-determined late M to early L spectral types for
which the DUSTY models are appropriate.  For the two sources in our
sample of mid-T spectral type, we
use the COND version of the PHOENIX atmosphere models, which
have been shown to reproduce the colors and spectra of T
dwarfs well.  In these models, all refractory elements have
been removed from the atmosphere through an unspecified ``rain
out process'', resulting in dust free atmospheres and blue
near infrared colors.  The treatment of L/T transition sources
is discussed below.  Due to the close proximity of these
sources, we assume that extinction is negligible.

\begin{figure*}
\epsscale{1.2}
\plottwo{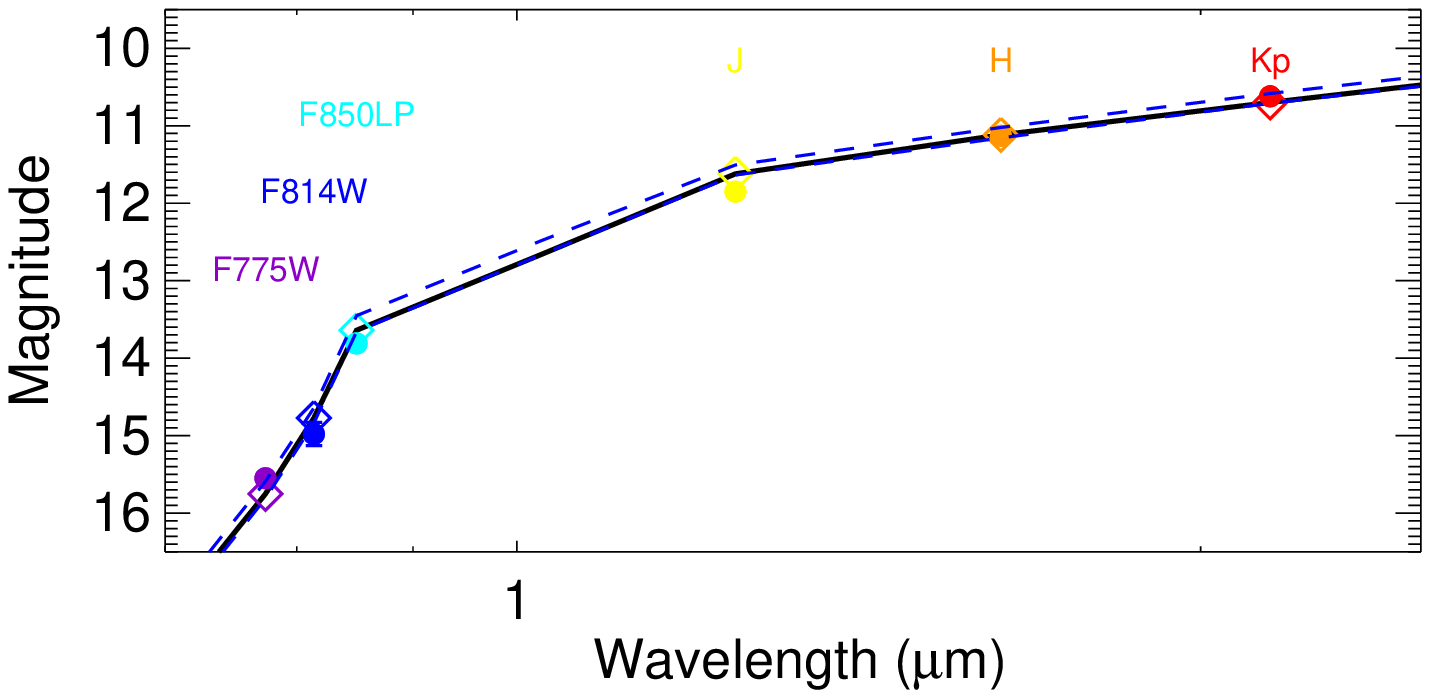}{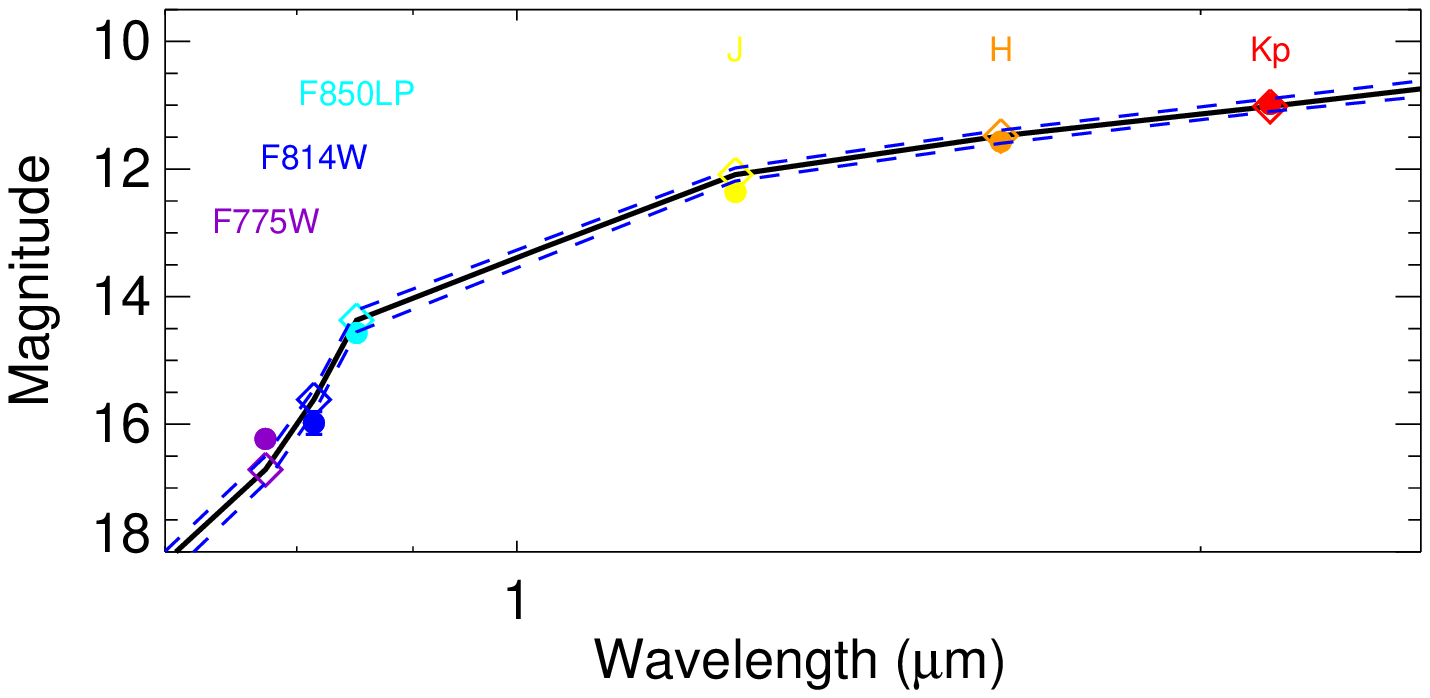}
\plottwo{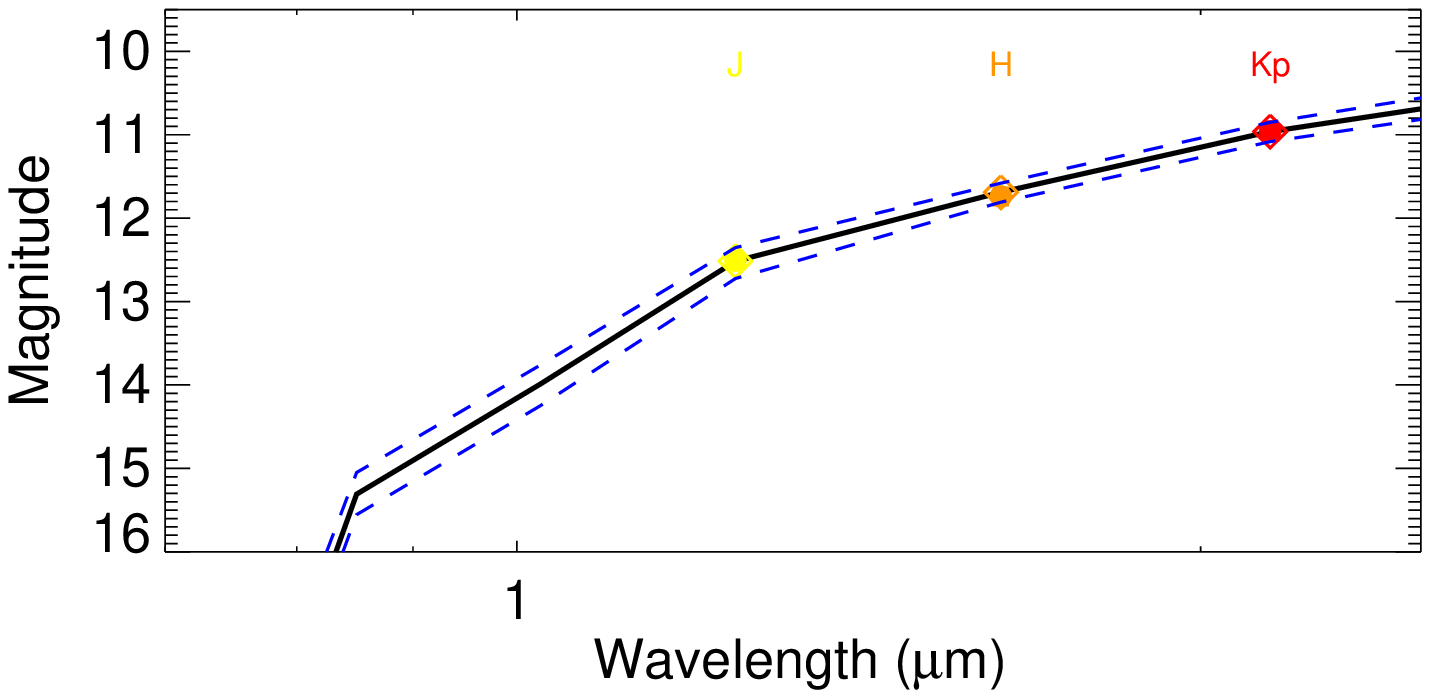}{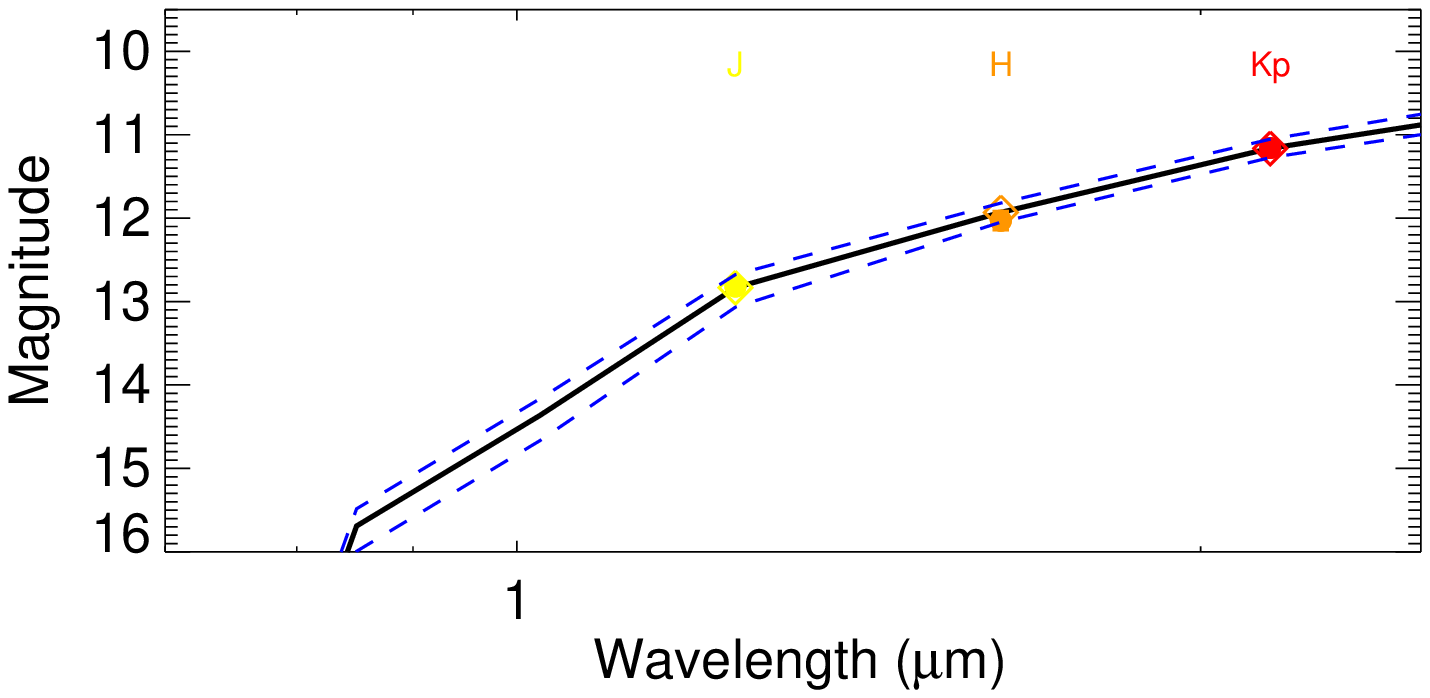}
\caption{Photometry and best fit SEDs for 2MASS 0746+20A
  (top left, 2205 K) and 2MASS 0746+20B (top right, 2060 K).
  These are examples of fits in which optical data is
  available.  Photometry and best fit SEDs are also shown for
  HD 130948B (bottom left, 1840 K) and HD 130948C (bottom right,
  1790 K), representing fits without optical data.   Photometric
  measurements are shown as filled circles, and
  best fit photometry from the DUSTY atmosphere models are
  show as open diamonds.  The full best fit SED (generated by
  interpolating between the best fit photometry from the models) is overplotted
  in black, and the 1$\sigma$ allowed ranges of magnitudes are
  shown as dashed blue lines.  Best fit SEDs for all other
  sources are shown online.}
\label{fig:2mass07464_temp}
\end{figure*}

\begin{figure*}
\epsscale{1.0}
\plottwo{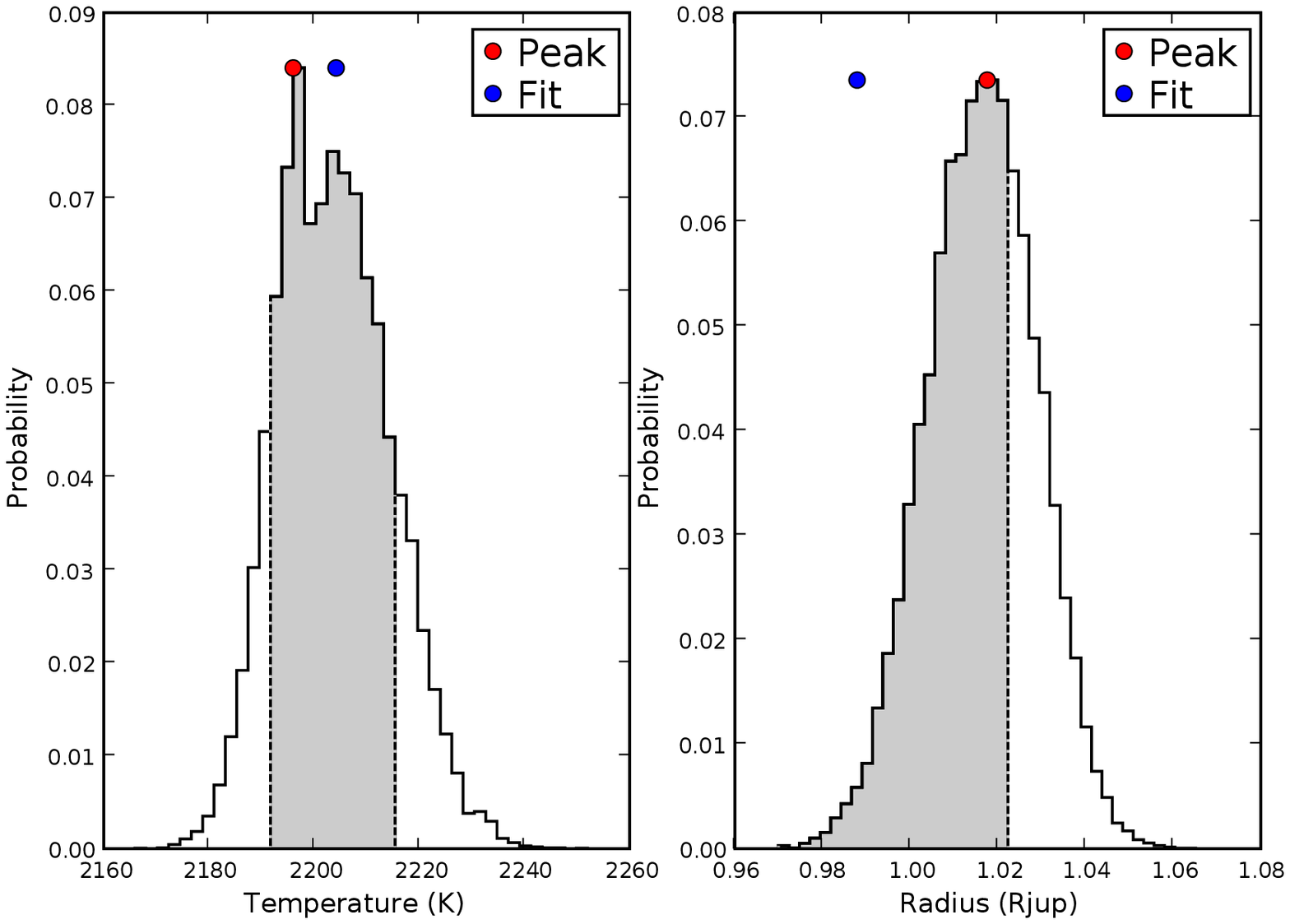}{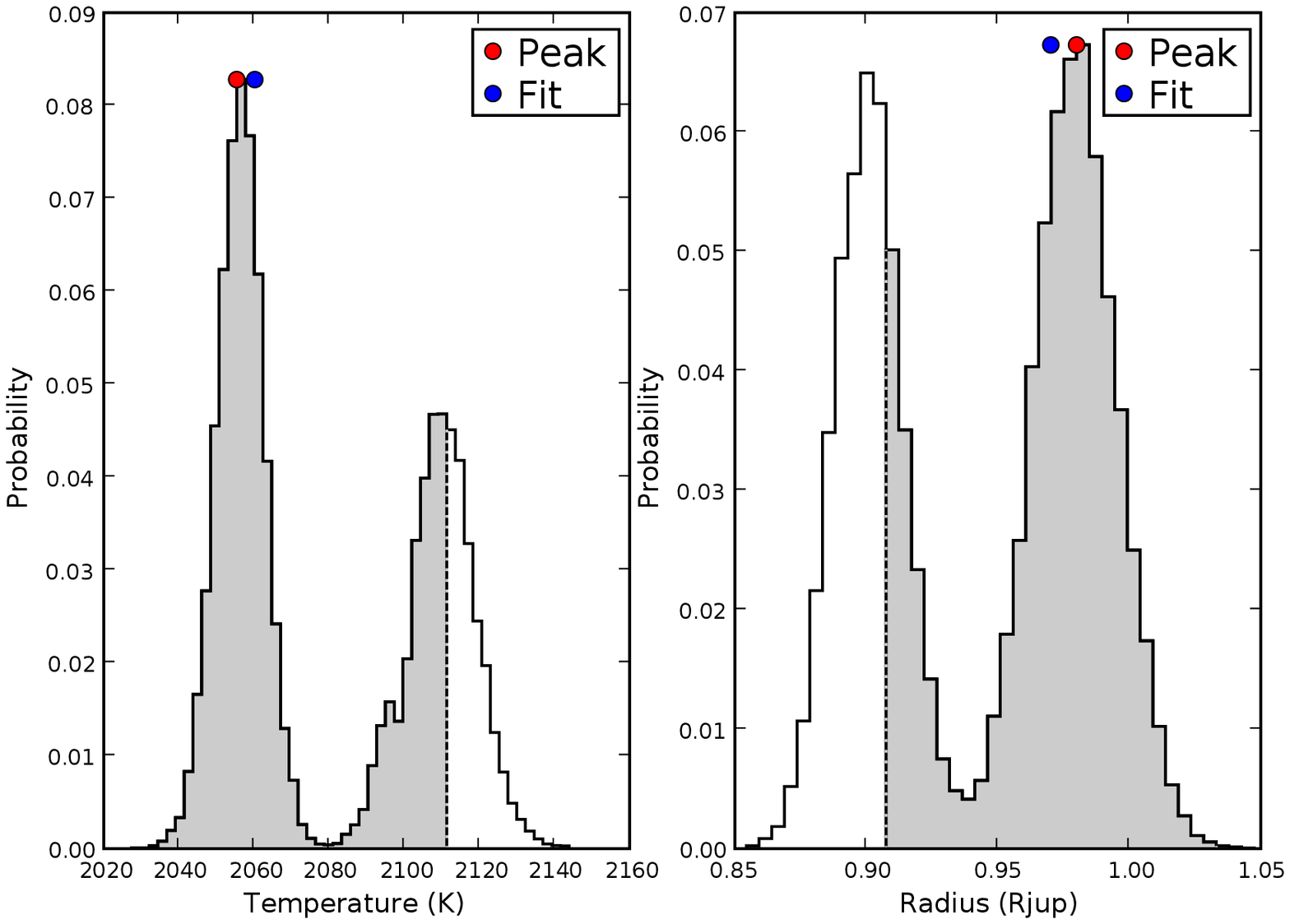}
\plottwo{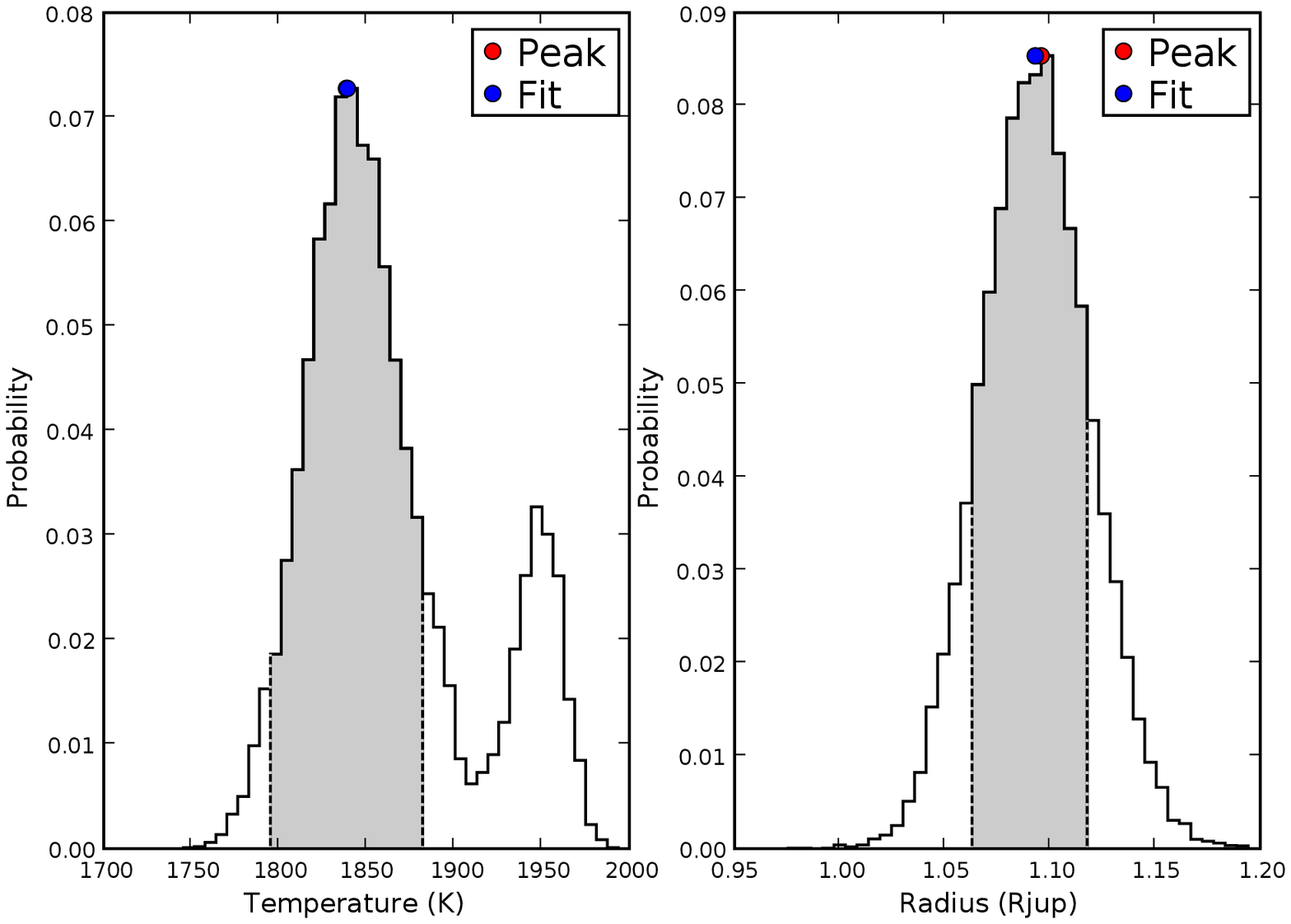}{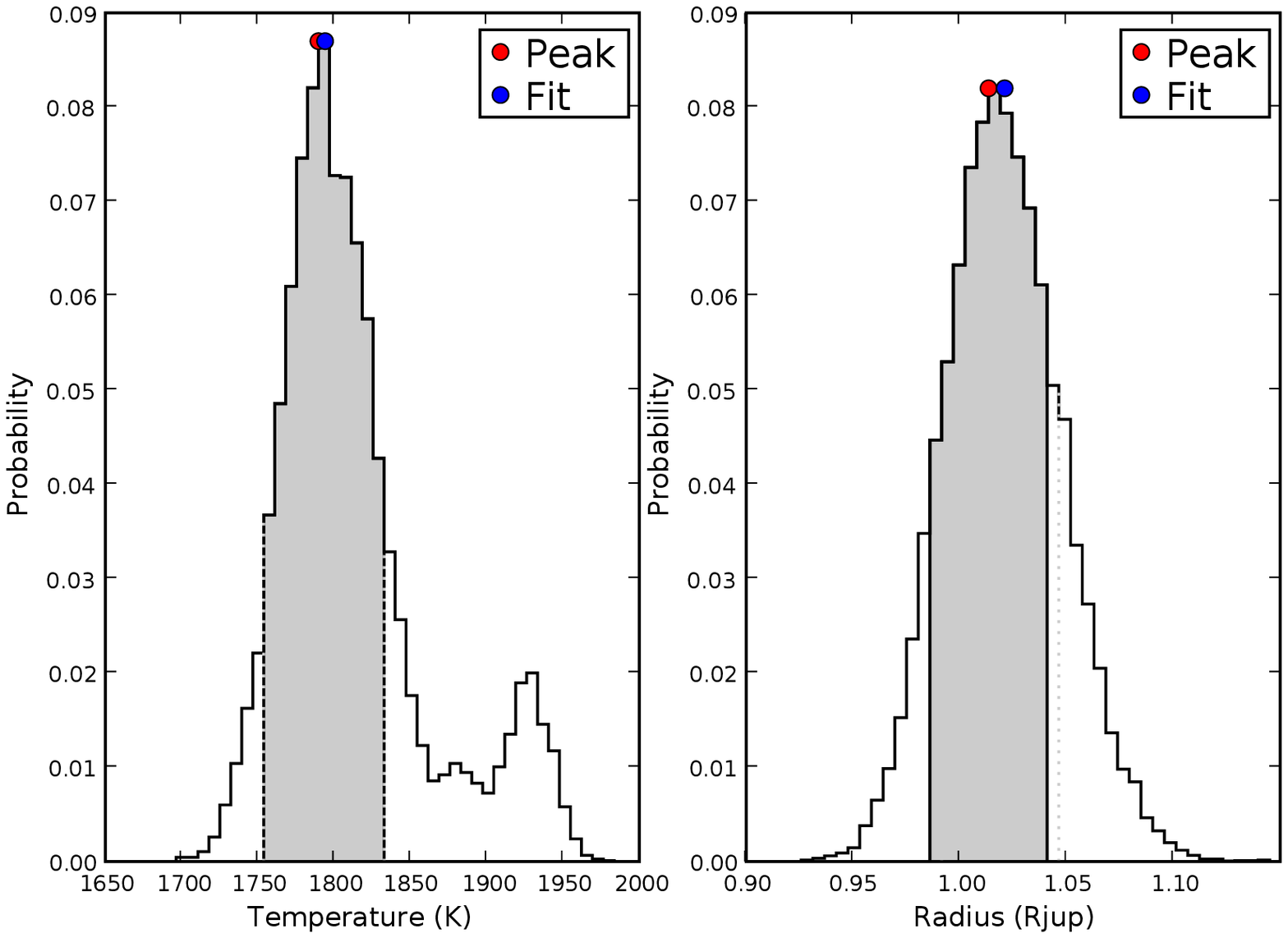}
\caption{One dimensional PDFs
  of temperature and radius from the spectral synthesis
  modeling for 2MASS 0746+20A and B (top) and HD 130948B and C
  (bottom).  PDFs for all other sources are shown online.}
\label{fig:2mass07464_temphist}
\end{figure*}

Since temperature can be most effectively constrained by
comparing synthetic atmosphere data over a broad
range in wavelengths, we elect to use our spatially resolved
photometry to perform the spectral synthesis modeling.  The
wavelength and bandpass 
information for each of our photometric measurements in Table
\ref{tab:phot} were used in conjunction with the PHOENIX
models to generate a grid of synthetic photometry for
objects with T$_{Eff}$ = 1400 - 4500 K for DUSTY and T$_{Eff}$
= 300 - 3000 K for COND, with log g = 4.0 - 5.5.  This range
of surface gravity should be appropriate for all sources in
our sample (McGovern et al. 2004, Rice et al. 2009).  
We then use this grid to fit the measured photometry for each
source, allowing for interpolation between finite grid
points.  Since the grid contains surface flux densities, the
model values must be scaled by (radius / distance)$^{2}$.
Since the distance is known, the radius becomes a simple
scaling parameter to fit simultaneously with gravity and
effective temperature.  Uncertainties in the derived
temperature and radius are then calculated via Monte
Carlo simulation, in which 20000 new photometric data points are
generated by sampling from a Gaussian distribution centered on
each apparent magnitude with a width given by the uncertainty in each
magnitude.  The apparent magnitudes are then converted to
absolute magnitudes using a distance sampled from a Gaussian
distribution centered on the values given in Table
\ref{tab:astro_orb}.  These datapoints are then fit in the same manner, and
confidence limits on effective temperature and radius are then calculated
by integrating our resulting one-dimensional distribution out to a probability of 34$\%$ on each side of
the best fitting value.  The best fit
SEDs from the atmosphere models are shown overplotted on the
photometry for each source in Figure
\ref{fig:2mass07464_temp}.  The one
dimensional PDFs for temperature and radius are show in
Figure \ref{fig:2mass07464_temphist}.  Although surface gravity is also
allowed to vary, we do not have sufficient photometric
precision to distinguish between values of surface gravity for
these field binaries, and the distributions of surface gravity
are essentially flat.  

\begin{figure*}
\epsscale{1.0}
\plotone{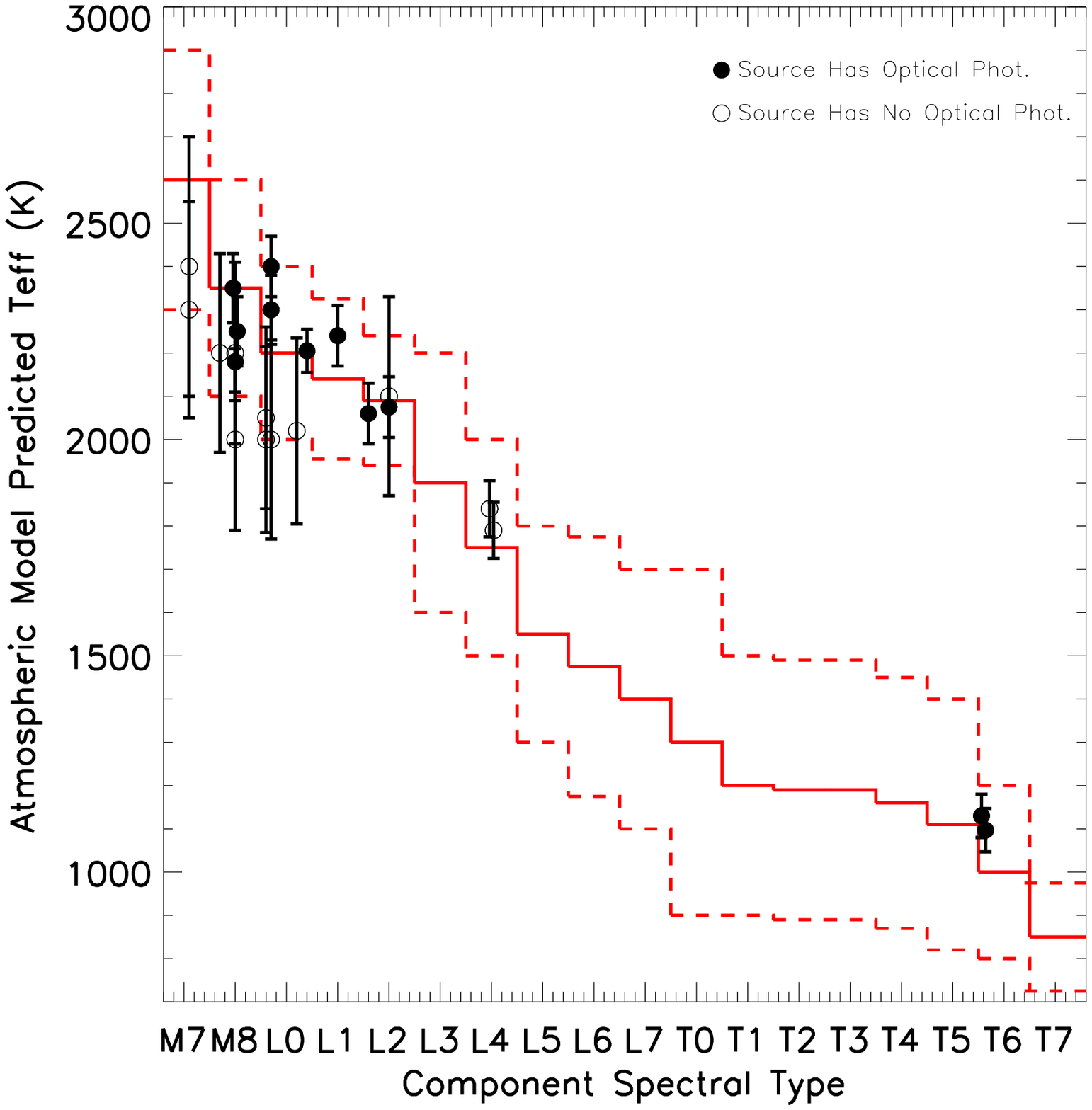}
\caption{The fitted effective temperatures from the
  atmospheric models are plotted as a function of spectral
  type.  Overplotted in red is an effective
  temperature/spectral type relationship derived from the
  results of Golimowski et al. (2004), Cushing et al. (2008),
  and Luhman et al. (2003).  In most cases where we have
  optical photometry in addition to near infrared photometry, the uncertainties
  in our derived temperatures are smaller temperature than
  those predicted by the temperature/spectral type relationship.}
\label{fig:teff_spty}
\end{figure*}

In addition to the temperature uncertainties resulting from
our photometric uncertainties, there are several systematic
uncertainties in temperature that must be accounted for.
First, the intrinsic uncertainty in the models themselves is
estimated to be on the order of 50 K.  We therefore combine
this uncertainty in quadrature with the uncertainties from our
Monte Carlo simulations for all sources.  In addition, the
lack of optical photometry for sources with spectral types
earlier than L2 tends to bias the derived temperatures towards
cooler values than is calculated for sources with optical
photometry.  The systematic offset is on average $\sim$200 K.
Therefore, for those systems in this spectral type range with
no optical photometry, we add in quadrature an additional
uncertainty of 200 K.  
In Figure \ref{fig:teff_spty}, we plot our
derived effective temperatures as a function of spectral
type.  We also combine the results of Golimowski et
al. (2004), Cushing et al. (2008), and Luhman et al. (2003) to
illustrate previous measures of temperature versus spectral
type.  This relationship is plotted in red on Figure
\ref{fig:teff_spty}, along with error bars representing the range
of allowed values by these works.  This comparison
demonstrates that in the cases where our photometry is
well constrained and spans a broad range of wavelengths, the
temperatures we derive using atmospheric 
modeling have lower uncertainties than we would be able
to obtain using spectral type.

\begin{figure*}
\epsscale{1.0}
\plotone{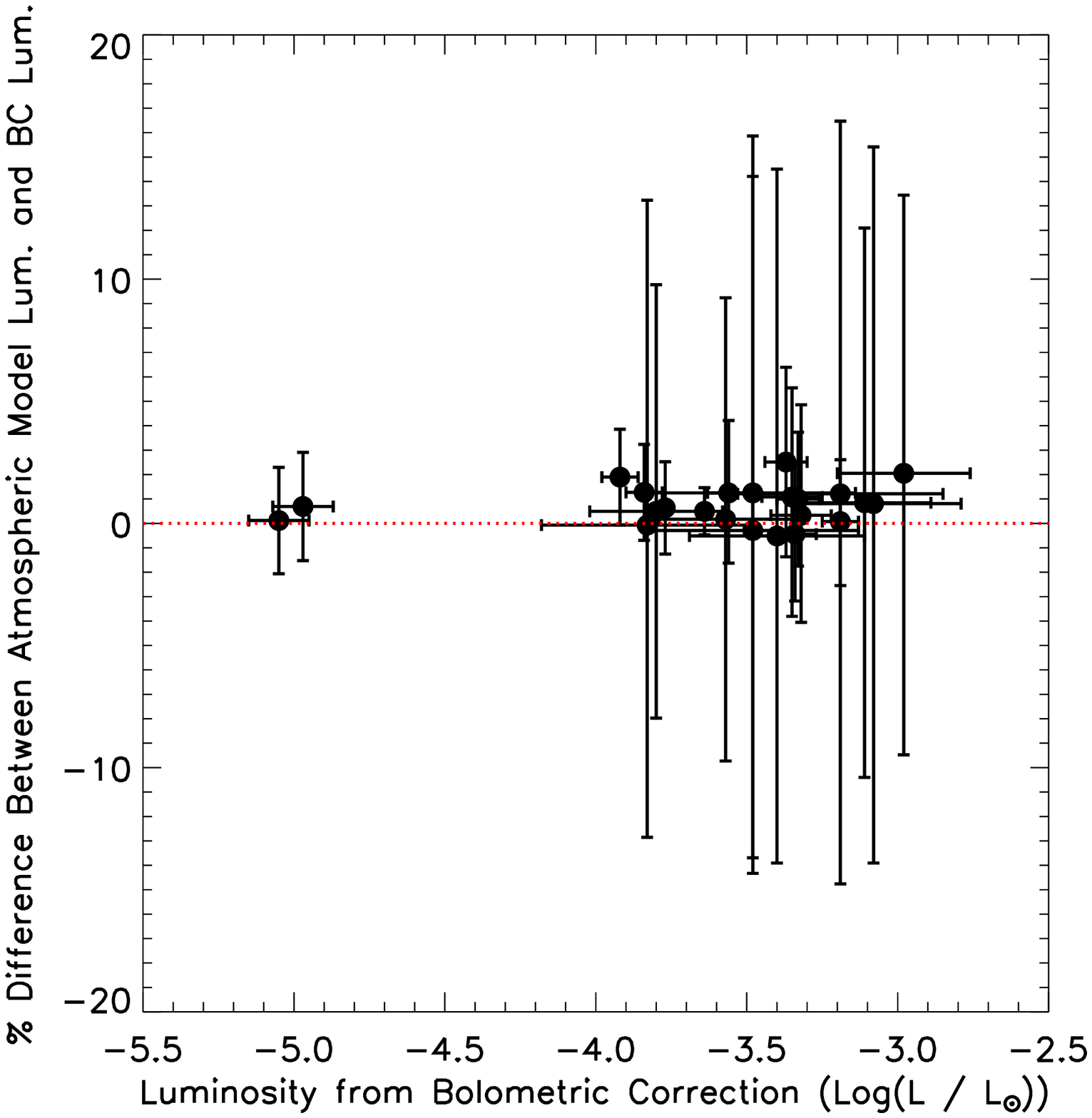}
\caption{The luminosities
  implied from our atmospheric model fits for T$_{Eff}$ and
  radius versus the luminosities derived from the bolometric
  corrections in Golimowski et al. (2004).  The red line
  represents 1:1 correspondence.  All values are consistent
  with each other.  We use the luminosities from bolometric
  corrections for further analysis because they are completely
independent of models.}
\label{fig:lum_comp}
\end{figure*}

Because we have a derived temperature and radius, we can also
calculate the PHOENIX model predicted bolometric luminosity.
However, this would also generate a model-dependence in our
value of luminosity.  Instead, we elect to determine
bolometric luminosity using the K band
bolometric corrections provided by Golimowski et al. (2004).
These corrections are a function of spectral type and were
derived using sources with photometric measurements over a
broad range of wavelengths, integrating under their SEDs.  The only assumption required to use these
corrections is that spectral type is a good proxy for K-band
bolometric corrections.  In contrast to predicted effective temperature,
the change in the K band bolometric correction with spectral
type is quite gradual with lower scatter.  In addition, Liu et
al. (2008) and Dupuy et al. (2009b) showed that
by deriving bolometric luminosities from the SEDs of four
sources, they obtain values fully consistent with 
those they would have obtained using the bolometric
corrections of Golimowski et al. (2004).  To be conservative,
we assume an uncertainty in the
spectral type of each source of $\pm$2 spectral subclasses to
determine our uncertainty in bolometric correction.  Even with
this assumption, the bolometric correction
uncertainty is never the limiting factor in our bolometric
luminosity uncertainty.  Generally, the uncertainty is dominated by
the distance uncertainty.  Our estimates of bolometric
luminosity from using these bolometric corrections are
given in Table \ref{tab:phot}.  To demonstrate the
correspondence between the luminosities calculated in this way
and the luminosities predicted by the atmosphere models, we
plot the percent difference between the luminosities derived
in each method in Figure \ref{fig:lum_comp}.  The measured
scatter around perfect correspondence of these values is
smaller than or consistent with our uncertainties.  We
therefore feel confident that our model-independent 
estimates of bolometric luminosity are appropriate for these
sources and consistent with our other methodology.  

In principle, our high
resolution spectroscopy can also be used to calculate
effective temperature.
However, the narrow wavelength coverage in the near infrared
provides relatively loose constraints on temperature, with
many temperatures being allowed by our K band spectra.  We
have, however, performed a few comparisons of our K band
spectra to the same models we use for the photometric fitting
and find that the results are consistent, with the photometry
providing lower uncertainties than the spectroscopy alone.
Ultimately, the best temperatures would be derived by fitting
a combination of the photometry and the spectroscopy.
However, such fitting has known challenges associated with how
data are weighted (Cushing et al. 2008, Rice et al. 2009).  In the future, we hope to
perform fitting of this kind, combining all spectral data.     

\begin{figure*}
\epsscale{1.0}
\plotone{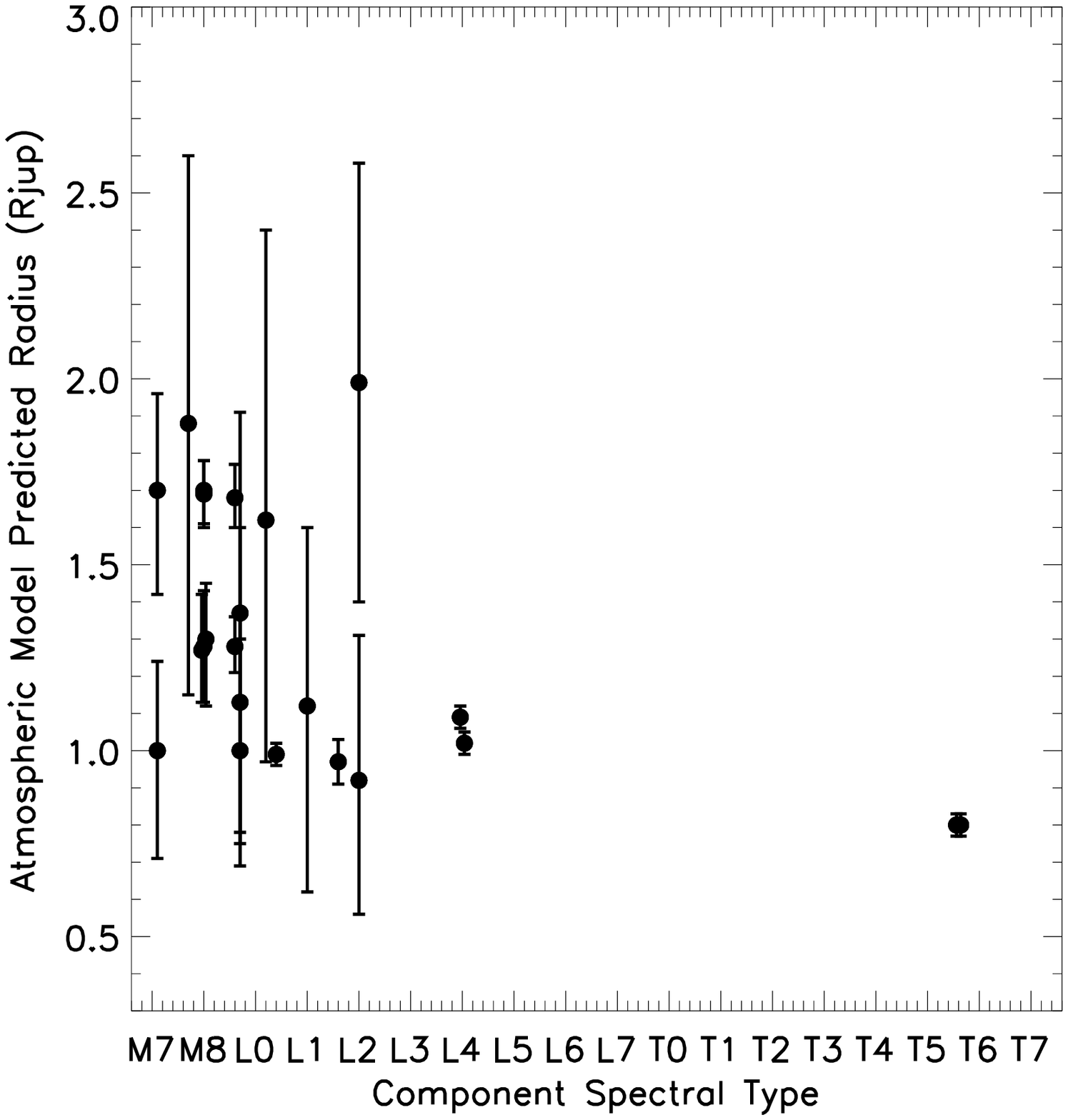}
\caption{The fitted radii from the
  atmospheric models are plotted as a function of spectral
  type.  The values are consistently in the range
  expected for VLM objects of between 0.5 and 2 R$_{Jup}$.
  This result justifies our choice of assuming a radius
  of 1.0 $\pm$ 0.3 R$_{Jup}$ for the L/T transition objects
  that cannot be fit by the atmospheric models.}
\label{fig:rad_spty}
\end{figure*}

For the seven sources in our sample in the L/T transitions
region, we must take a different approach to obtaining
effective temperatures.  The DUSTY and the COND models can be
thought of as boundary conditions to the processes occurring
in brown dwarf atmospheres, meaning each atmosphere is either
fully dusty or completely dust free.  There is no transitional
dust phases represented in the current versions of these
models.  
Though we attempted to fit sources in this region via the method described above
with both models, we obtained very high temperatures
($\gtrsim$1900 K) and
unphysically small radii ($\lesssim$0.5 R$_{Jup}$).  Therefore averaging the predictions
of the two models does not work.  For these sources, we elect to use the bolometric luminosity of the source
and assume a radius with a large uncertainty, chosen to
conservatively span the values derived in our atmospheric
model fitting  (1.0 $\pm$ 0.3
R$_{Jup}$).  A radius in this range is also what is expected for
these objects theoretically.  In Figure \ref{fig:rad_spty},
we plot the radii from our fits as a function of spectral
type.  Although there is a lot of scatter in this relationship
due to the mixed ages in our sample, the large uncertainty we
have assumed for radii at the L/T transition region should
account for this variation.  The
result of assuming a radius is higher temperature
uncertainties for these objects.  All derived 
temperatures and radii are given in Table \ref{tab:phot}.

\section{Comparisons to the Predictions of Evolutionary Models}\label{sec:compmodel}

The derived temperatures and bolometric
luminosities are used to determine the model-predicted mass for each source
in our sample.  We consider both the Chabrier et al. (2000)
evolutionary models, called DUSTY and COND, and the Burrows et
al. (1997) evolutionary models (TUCSON).  The DUSTY and COND
evolutionary models are named as such because they use the 
boundary condition between the interior and the atmosphere provided by the DUSTY and COND atmosphere
models, respectively.  Thus the evolutionary and atmosphere
models are not strictly independent.  In
comparing to the Chabrier et al. (2000) evolutionary models,
we are consistently testing model predictions because we have used
the same atmospheric models in our analysis.  Comparisons to
the Burrows et al. (1997) models require a caveat, as we do
not have access to the 
atmospheric models employed by those authors.  However, we still perform the
comparison to test the correspondence of these models to our
measurements, as the effect of the atmospheric model boundary condition
should only have a minor impact on the evolutionary
predictions (Chabrier $\&$ Baraffe 2004).

To do this comparison, we first interpolate over the
surface defined by the grids of temperature, luminosity, mass,
and age provided by the evolutionary models using spline
interpolation.  Then, the temperature and luminosity
point on the interpolated surface closest to our input
value of temperature and luminosity is determined.  For the sources from
late-M to mid-L, we calculate the predictions of the DUSTY
version of the Chabrier et al. (2000) models, while for the T
dwarfs we use the COND version on these models.  For the L/T
transition objects, we calculate the predictions of both sets of
Chabrier et al. (2000) models.  The Burrows et al. (1997,
TUCSON) models do not assume a different atmospheric treatment
for different spectral types and assume that dust species
have condensed out of the atmosphere across the entire
substellar regime.  We therefore compare the predictions of
these models to all objects in our sample.  These comparisons provide the
predicted mass and age for each source.  

To determine the uncertainties in each model
prediction of mass, we sample from temperatures and luminosities
defined by the uncertainties in each for each sources,
accounting for the correlations between bolometric luminosity and
temperature (Konopacky et al. 2007)\footnote{In contrast to
  Konopacky et al. (2007), we are not obtaining temperatures
  based upon the color of our systems. The atmospheric model
  fits are only linked to bolometric luminosity through the K
  band magnitude, and thus the correlations between the two
  parameters are very weak in this study}.  The
range of masses and ages predicted from this sampling,
marginalized against the other parameters, provides the
uncertainties.  The values of mass predicted by each model are
provided in Table \ref{tab:models}.  

\begin{figure*}
%\epsscale{0.5}
\epsfig{file=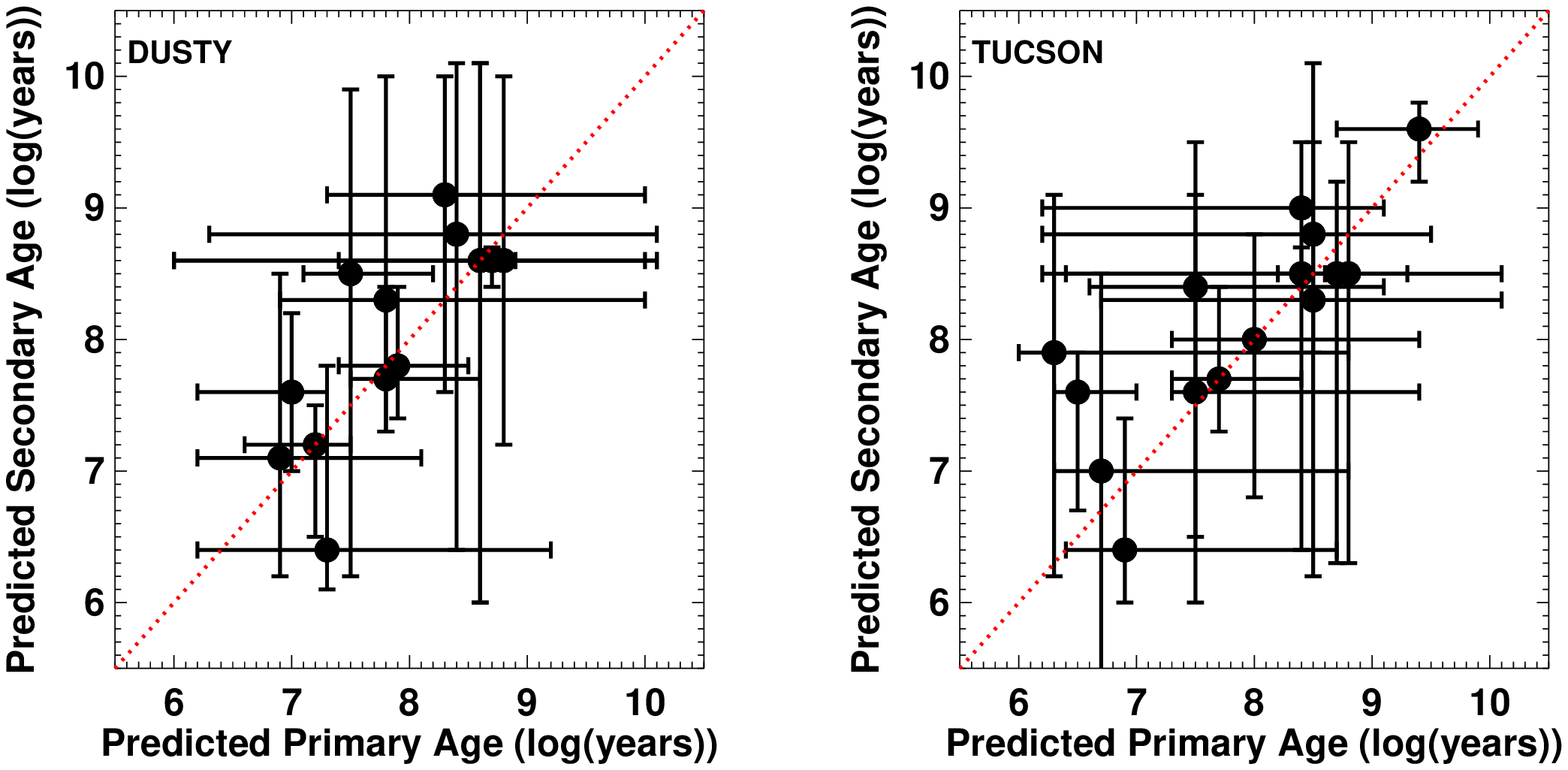,angle=90,width=\linewidth}
\caption{The predicted ages for secondary components versus primary
  components by the DUSTY (\textbf{left}) and TUCSON
  (\textbf{right}) models.  The line of 1:1 correspondence is
  plotted in red.  Within the uncertainties, all binary
  components are predicted to be coeval for all models.}
\label{fig:age_comp}
\end{figure*}

The majority of the
sources in the sample have little to no age information -
hence, we look for whether the models predict that the components are
coeval as opposed to correct age prediction by the
models.  For the two sources with age information, HD 130948
BC ($\sim$500 Myr, Gaidos 1998) 
and GJ 569Bab ($\sim$100 Myr, Simon et al. 2006), the uncertainties on these
ages are such that both models predict ages for these systems
that are consistent with these values.
For all sources in the sample, all binary components are
consistent with being coeval within
the uncertainties by both models.  Figure \ref{fig:age_comp} shows the
predicted ages of the binary components in the DUSTY and the
TUCSON models plotted versus each other.  A line of 1:1
correspondence is overplotted.  The relatively large uncertainty in
age estimates stems from the fact that the model
isochrones become more closely packed with increasing age.
Because of this fact, empirical age estimates for field
objects provide relatively weak constraints on the models.
Thus, it is not surprising that both models predict all
binary components are coeval.  Stronger constraints on the
ages predicted by the evolutionary models are likely to be
made using younger sources, for which isochrones are less dense.
    
\begin{figure*}
\epsscale{1.0}
\plottwo{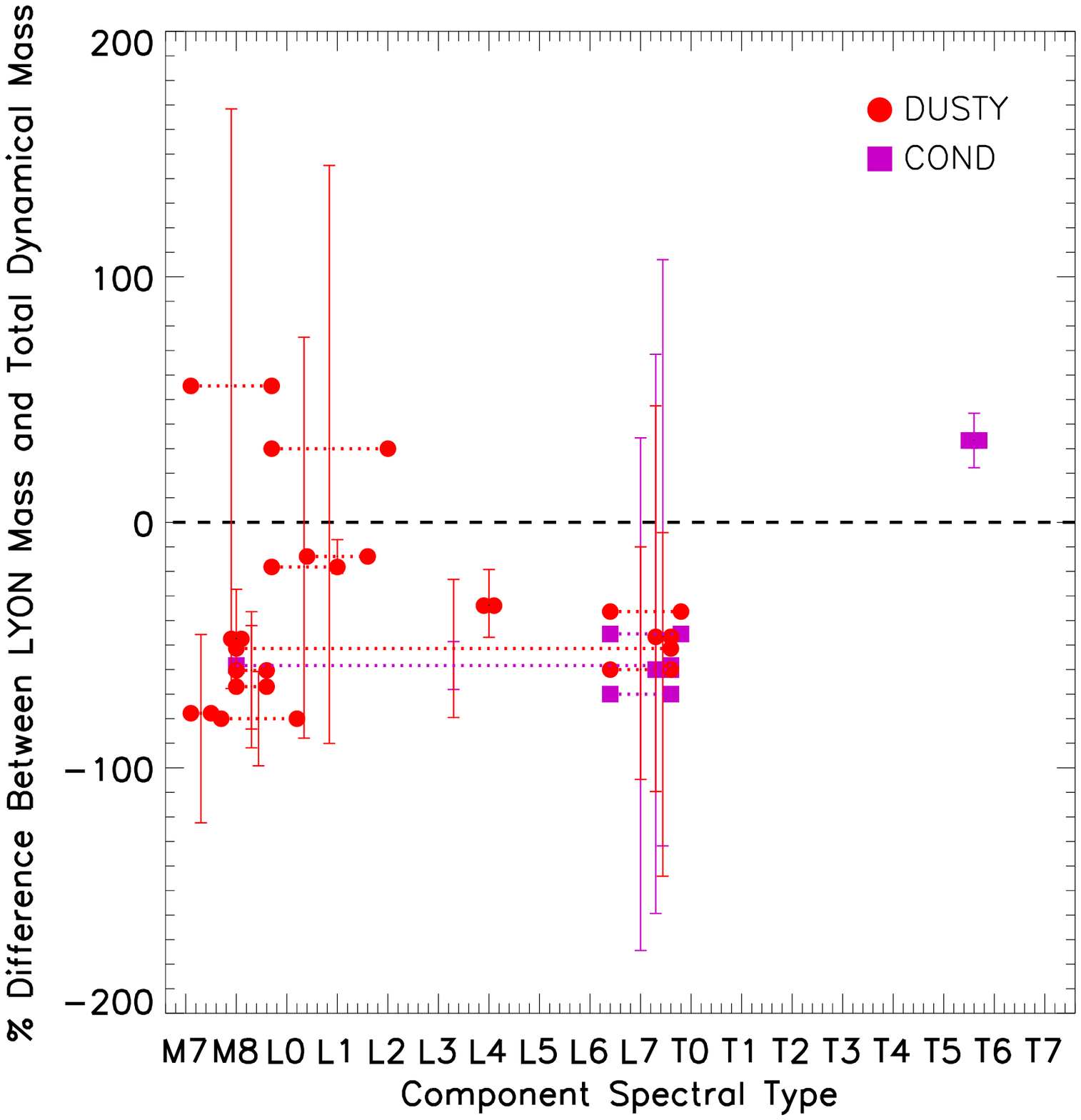}{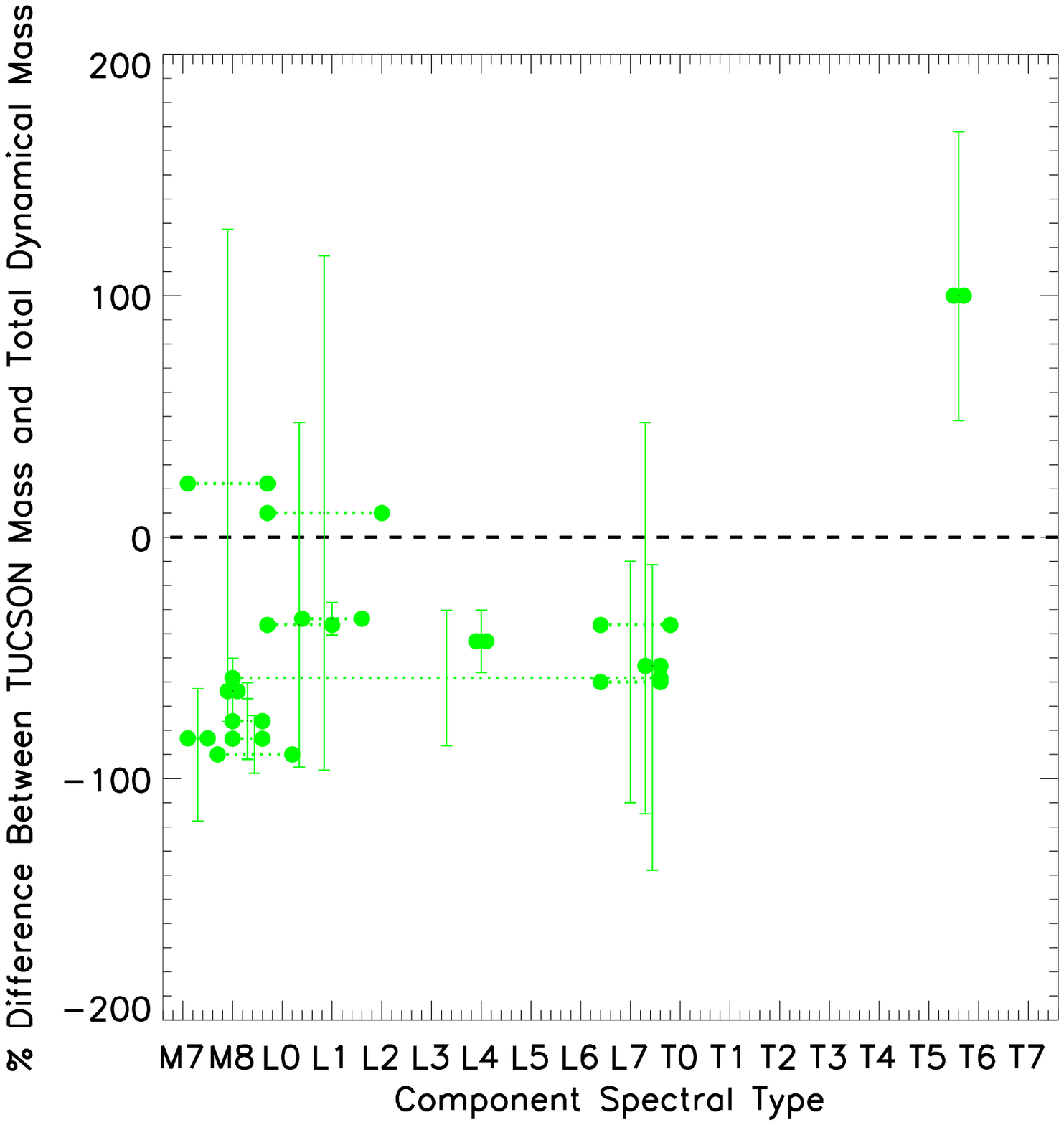}
\caption{\textbf{Left:} The percent difference between the
  predictions of the Lyon (both DUSTY and COND, Chabrier et al. 2000) models and
  our total dynamical masses as a function of spectral
  type.  Each system is denoted by the spectral type of its
  components, which are connected with a horizontal bar.  We
  find that 7 of the 14 systems we have compared to the DUSTY models
  have their masses underpredicted by these models.  These
  systems all have primary component spectral types earlier
  than L4.  We find that one T dwarf system we compared
  to the COND models has its mass overpredicted by the
  models.  All sources with primary component spectral types
  in the L/T transition region have mass predictions that are
  consistent with the total dynamical mass.  \textbf{Right:} The percent difference between the
  predictions of the TUCSON (Burrows et al. 1997) models and
  our total dynamical masses as a function of spectral type.
  We note that while we have used different atmospheric
  models to derive effective temperature than is employed in
  the Burrows et al. (1997) models, the effect of the
  atmospheric model is thought to be minor.  We have compared
  all 15 systems to these models.  We find that 7 systems have
their masses underpredicted by these models, all of which have
primary component spectral types earlier than L4.  We find
that one mid-T system has its mass overpredicted by the
models.  All sources with primary component spectral types
  in the L/T transition region have mass predictions that are
  consistent with the total dynamical mass.}
\label{fig:dusty_spty}
\end{figure*}

Since the highest precision measurements are currently in total system
mass, the model predictions can be compared most effectively to
these measurements.  To do this, we add the model masses derived for
each component together and add their uncertainties in
quadrature.  The combined mass predictions are also given in
Table \ref{tab:models}.  For 7
systems, all models underpredict the total system
mass by greater than 1$\sigma$.  These 7 systems have the
smallest uncertainties in 
dynamical mass and primary component spectral types earlier
than L4.  For 7 other systems, 
all models considered predict masses that are consistent with
the dynamical mass within 1$\sigma$.
These systems all have mass uncertainties over 60$\%$, and
also generally higher temperature uncertainties.  Finally,
the mass of one system is overpredicted by both models by
greater than 1$\sigma$ and is
the only system with a mid-T spectral type.  All systems in
our sample with mass precisions better than 60$\%$ are
therefore discrepant with the models by more than 1$\sigma$.
To illustrate the apparent dependence in the direction of the
mass discrepancy 
with spectral type, we have plotted the percent difference between the
model prediction and the total dynamical mass for each
model.  These plots are shown in Figure \ref{fig:dusty_spty}.
While the significance of the discrepancy is only on the order
of 2$\sigma$ (at most) for each individual target, the fact
that there is a systematic trend (as a function of spectral
type) suggests that the problem is more profound. Indeed, the
hypothesis that the evolutionary models predict the correct
masses for all systems can be tested by computing the
associated reduced-$\chi^2$ ($\frac{1}{(N-1)}~
\displaystyle\sum^N\frac{(M_{dyna}-M_{model})^2}{\sigma_{dyna}^2+\sigma_{model}^2}$). The 
probability of obtaining the observed value of $\chi^2=4.81$
with 15 measurements is $\sim$5x$10^{-9}$. In other words, the
significance level of the discrepancy between empirical and
modeled systems masses is very high if one considers the
entire sample as a whole.  

\begin{figure*}
\epsscale{1.0}
\plottwo{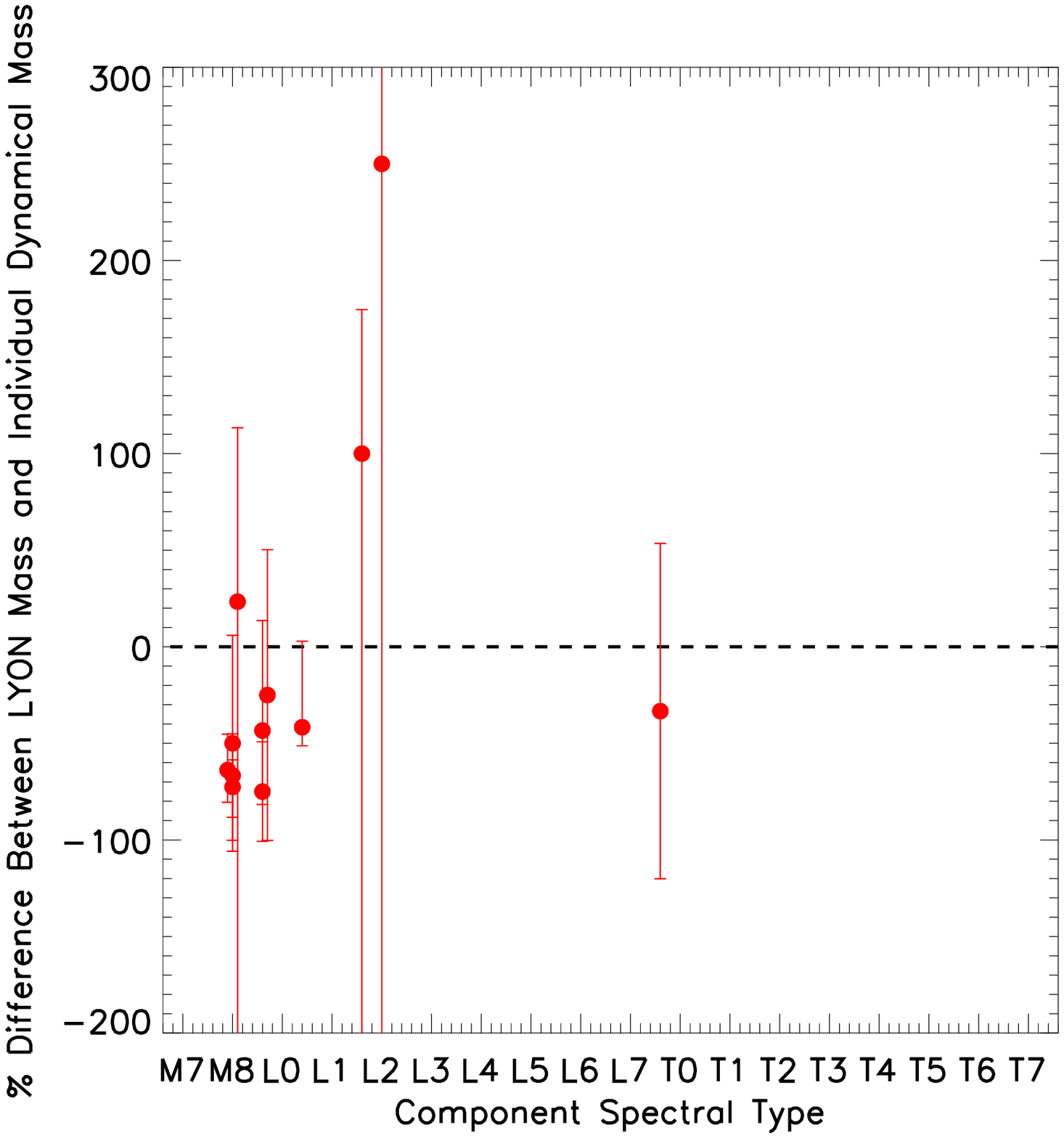}{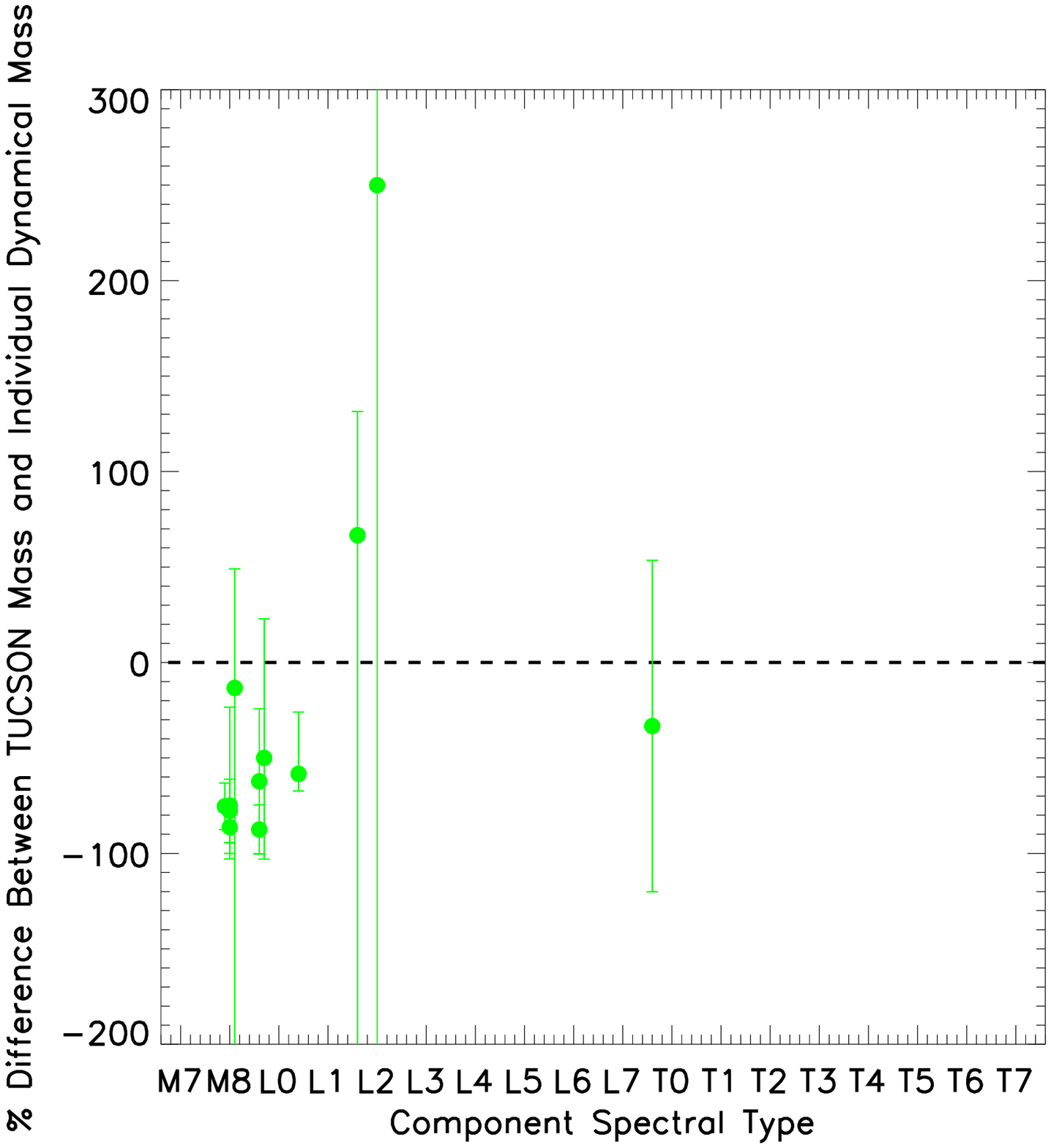}
\caption{\textbf{Left:}  The percent difference between the
  predictions of the DUSTY (Chabrier et al. 2000) models and
  our individual dynamical masses as a function of spectral
  type.  We compare our 12
  individual mass measurements to these models, and find that
  five sources have their masses underpredicted by these
  models.  All five sources have spectral types of M8 -
  M9. \textbf{Right:}  The percent difference between the
  predictions of the TUCSON (Burrows et al. 1997) models and
  our individual dynamical masses as a function of spectral
  type. 
  We note that while we have used different atmospheric
  models to derive effective temperature than is employed in
  the Burrows et al. (1997) models, the effect of the
  atmospheric model is thought to be minor. We compare our 12
  individual mass measurements to these models, and find that
  five sources have their masses underpredicted by these
  models.  All five sources have spectral types of M8 - M9.}
\label{fig:dusty_indiv}
\end{figure*}

We can test the predictions of the models a bit
further by considering our
handful of individual mass measurements.  Although our individual mass
measurements do not yet have the high precision we have
achieved in total mass, we can already see for the most
precise cases that the discrepancy holds.  That is, for the
primary components of 2MASS 2206-20AB, GJ 569Bab, and LHS 2397a
AB (which have the highest precision in component mass), the
models underpredict the mass.  These three systems all have
approximate spectral types of M8.  We also see that the
secondary component LP 349-25 AB has its
mass underpredicted by the models, and the TUCSON models
underpredict the mass of GJ569Bb.  These systems are both
of approximate spectral type M9.  For further illustration of these points, we again plot the
percent difference between the masses predicted by the models and the
dynamical masses, this time plotting the individual component
mass.  These plots are shown in Figures \ref{fig:dusty_indiv}\footnote{Since only one system has a spectral type later
  than L5, we do not perform this exercise for the COND
  models}.  Although the uncertainties are larger,
the trends we saw amongst total system mass holds.  These
figures also demonstrate the 
power of using individual component masses to perform model
comparisons, allowing for the investigation of where
discrepancies lie without assumptions (this is particularly
apparent in the case of LHS 2397a AB, which has an M8 primary
and an L7.5 secondary).  In addition, individual component
masses effectively double the sample of sources that can be
used for comparison (here, we have compared 12 sources,
already approaching the 15 we can do with total system masses).  Emphasis in the future will be placed
on obtaining more precise individual mass estimates
for these systems to see if these trends persist.   

We note that discrepancies between the evolutionary and
atmosphere models have been noted before for three of our
systems.  The systems HD 130948BC, LHS 2397a AB, and 2MASS
1534-29 AB had their relative orbits derived by Dupuy et
al. (2009a), Dupuy et al. (2009b), and Liu 
et al. (2008).  These authors use spectral synthesis by Cushing et
al. (2008) as a proxy for performing atmospheric fitting on
the sources, and see that there is an offset between the
evolutionary and atmosphere models, though they make the
comparison only in terms of temperature (using their
luminosities and total system masses with a mass ratio
assumption to derive an evolutionary model
predicted temperature).  Thus, though their
approach is different, they arrive at similar conclusions.

\section{Discussion}\label{sec:disc}

\begin{figure*}
\epsscale{1.0}
\plotone{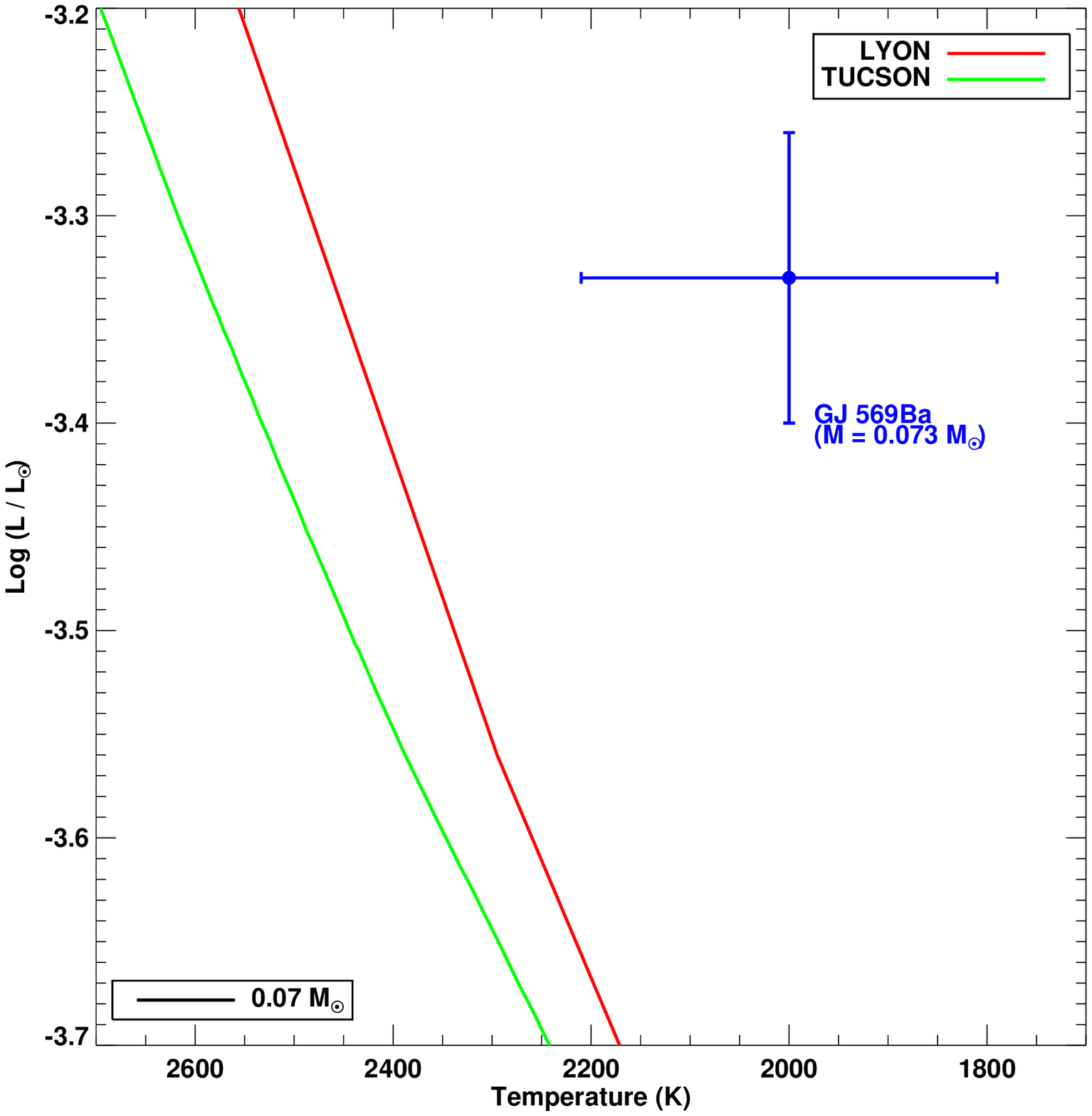}
\caption{Location of GJ 569Ba on
  the H-R diagram given our derived temperature and
  luminosity.  Since this system had a mass of 0.073 $\pm$
  0.008 M$_{\odot}$, it should lie close to the line of
  constant mass for a 0.07 M$_{\odot}$ object in the evolutionary
  models.  The location of this line for both LYON and TUCSON
  are also plotted.  As with all discrepant sources in our
  sample of spectral type M or L, the source lies above and to
the right of these lines, implying either the temperature is
too high in the evolutionary models, the radius is too small
in the evolutionary models, or the temperature is too low in
the atmosphere models for these sources.}
\label{fig:gj569bb_hrdiag}
\end{figure*}

We have found systematic discrepancies between our measured
dynamical masses and the predicted masses 
from theoretical evolutionary models, where overall the M
and L dwarfs have higher dynamical masses than predicted and one T
dwarf has a lower dynamical mass than predicted by
evolutionary models.  We determined the mass predicted by each evolutionary
model using our measured
parameters of luminosity and temperature, which are related to
each other through the canonical equation L
=4$\pi$R$^{2}\sigma$T$^{4}$. Our observed bolometric
luminosity is the most constrained of these parameters and
does not rely on evolutionary or atmospheric models; therefore, it is the least
likely parameter to contribute to disagreement.  Instead, the
radius and temperature are the most
likely cause of the discrepancy between the predicted
evolutionary model masses and our dynamical masses, either
those predicted by the evolutionary model or those from our
atmospheric model fits.  In this section, we
explore temperature and radius and discuss other assumptions
used with both atmospheric and evolutionary models which may
give rise to differences between our measured masses and
predicted modeled mass.  For reference, we show in Figure
\ref{fig:gj569bb_hrdiag} the location of GJ 569Ba of the H-R Diagram, whose
individual mass measurement was underpredicted by the
evolutionary models.  We show the location of the line of
constant mass for a 0.07 M$_{\odot}$ source as given in both
the LYON and TUCSON models, which should align with the position GJ 569Ba if
there was no discrepancy.  The direction of the offset between
these lines and the position of GJ 56Ba is representative of
the direction of the offset for all discrepant systems of M
and L spectral types.  Though we cannot make a
corresponding plot for the case of our overpredicted T dwarf
system, for which we do not have individual component masses,
the direction of the offset is opposite that of GJ 569Ba.

We first consider the case in which the driver for the
discrepancy is primarily the evolutionary models, which begin
with mass and age
as input parameters and then predict quantities of radius,
temperature, and, in turn, the luminosity.  For the sources in
the late M through mid L
spectral types that are discrepant, the mass tracks that agree
with our dynamical masses lie at
higher temperatures and/or lower luminosities than our input
values.  To bring these sources into agreement would require
either a decrease in the evolutionary model-predicted temperature for these
sources of $\sim$100-300 K or an increase in the radii by
a factor of $\sim$1.3-2.0.  Meanwhile, for the discrepant T
dwarfs, the correct mass lines lie at lower temperatures
and/or higher luminosities than our input values.  To bring
them into agreement would require an increase in the
evolutionary model predicted temperatures by $\sim$100 Kelvin,
or an decrease in the radii by a factor of $\sim$1.5.  

There are a number of implications for the physics of the
evolutionary models.  If the radii are off, this
suggests that the mass-radius relationship in these
models might be off.  The predicted radius for a
source is driven almost entirely by the assumed
equation of state, with a very minor dependence on the assumed
atmosphere (Chabrier et al. 1997).  An
update of the equation of state from that given by Saumon et
al. (1995), which is the equation of state used by both
Burrows et al. (1997) and Chabrier et al. (2000), could
potentially modify the mass-radius relationship, although
improvements with using new experiments are unlikely to have a
major impact in the brown dwarf mass regime.

  Meanwhile, if the required change is in the
predicted effective temperature, this implies that
adjustments need to be made to the efficiency of interior
energy transport, or an offset is needed in the interior/atmospheric
boundary condition.   Magnetic activity, which is not included
in the models, could possibly inhibit the efficiency of
convection, lowering the effective temperatures of these
objects (Chabrier $\&$ Baraffe 2000).  This may be important
for the discrepant sources of 
spectral type M or L, which have lower temperatures than
predicted by the evolutionary models.  Several studies have
shown that cooler temperatures are measured for low mass
eclipsing binaries than predicted by evolutionary models
(e.g., Stassun et al. 2007, Morales et al. 2008), and the lack of accounting for
activity is thought to be a likely culprit for this
discrepancy.  

The other case we consider is that in which the discrepancy is
caused by the
temperatures and radii predicted by the PHOENIX atmosphere
models.  In our spectral synthesis modeling, these two
parameters are linked through the bolometric luminosity.
Because our luminosity is a fixed, model-independent quantity, the
parameter that matters in this case is
temperature, because the radius is effectively set by the
measurement of L$_{Bol}$ and enters only as a scaling
factor for the SEDs which are shaped by temperature.
Therefore, if the discrepancy is caused by the atmosphere
models, it is through the temperature prediction.  In this
case, the temperatures predicted would be too low in the case
of the M and L dwarfs by $\sim$100-300 K, and too high in the case of the T
dwarf by about $\sim$100 K.  A change in temperature would cause
the atmosphere model-predicted radius to change as well, but
in a way that again maintains correct L$_{Bol}$.

The PHOENIX atmosphere models we have used are thought to represent the limiting
cases in terms of atmospheric dust treatment.  If the temperatures predicted
for the discrepant M and L dwarfs are too low, it implies that
the dust clouds are too opaque, trapping too much radiation.
For the T dwarfs, removal of all refractory elements from the
atmosphere may have resulted in a drop in opacity that allows
too much radiation to escape, causing a higher than predicted
temperature.  Recent work by Helling et al. (2008) has
shown that the treatment of dust clouds in atmospheric models
has a dramatic effect on the output photometry.  
Though Helling et al. (2008) only compared two test cases, one
at 1800 K and one at 1000 K, a rough comparison between the
colors of our discrepant sources and those test cases show
that models with thinner dust clouds and uniform grain sizes may bring the
temperatures into alignment with what is predicted by the
evolutionary models.  

Therefore, there are a number of scenarios in which a slight
change in the input physics to either the evolutionary models
or the atmosphere models (possibly both) could generate
agreement between our dynamical masses and the model predicted masses.
We also note that if the discrepancies between the models and
the dynamical masses continue to follow the same trend, the
implication for pushing into the planetary mass regime is that,
like the T dwarfs, the masses of planets would be
overestimated by the evolutionary models.  For instance, in
the case of the directly-imaged extrasolar system, HR 8799,
relatively high masses of 7, 10, and 10 M$_{Jup}$ have been
derived using evolutionary models (Marois et al. 2008).  These higher masses have
generated some difficulty in terms of allowing for systemic
stability over long timescales (Go{\'z}dziewski \& Migaszewski
2009, Fabrycky \& Murray-Clay 2008).  The decrease of these
masses by only a few tens of percent
would imply the system was much more stable over long
timescales.  Thus, this work may have important implications
for the masses derived for directly imaged extrasolar planets
using evolutionary models.  

\begin{figure*}
\epsscale{1.0}
\plotone{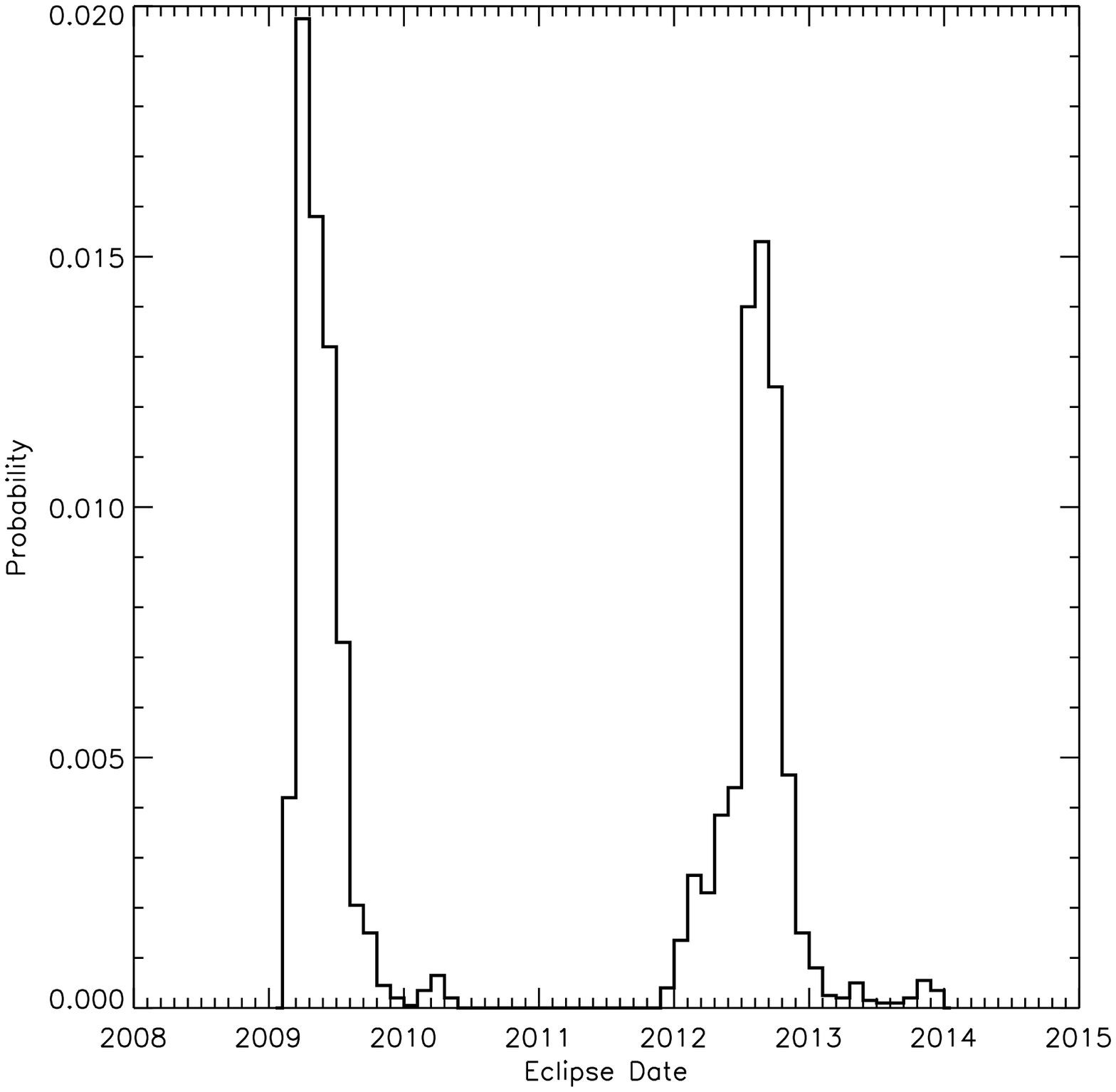}
\caption{The probability of eclipse as a function
  of date of occurence for 2MASS 0920+35 AB.  Overall, the
  system has a 6.8$\%$ chance of being an eclipsing system,
  with the most likely date of eclipse having occurred in April
of 2009.  The next most likely date of an eclipse is in mid-2012.}
\label{fig:eclipse}
\end{figure*}

In order to place further constraints on these models, more
measurements of brown dwarf radii are required in addition to mass.  Thus far,
only one eclipsing binary brown dwarf 
has been reported (Stassun et al. 2006), providing the only
empirical measurement of a brown dwarf radius (for a very
young system in Orion).  In our sample, one source in our
sample, 2MASS 0920+35 AB 
(which has components in the L/T transition region), is
on a highly inclined orbit with an inclination of 88.6 $\pm$
1.2$^{o}$.  Assuming that the components have a radius of 1
R$_{Jup}$, the system will be an eclipsing system if it has an
inclination between 89.89$^{o}$ and 90.15$^{o}$.  Based on our full
relative orbital solution distribution, we find the system has
an 6.8$\%$ probability of eclipsing.  If we consider only those solutions
with a $\sim$6.5 year period, which is the best fit period, we
find the system has a only a 
3.1$\%$ chance of eclipsing.  If we instead only consider those
solutions with a $\sim$3.5 year period, the system has an
11.3$\%$ chance of eclipsing.  In Figure \ref{fig:eclipse}, we plot the
total probability distribution of eclipse dates, considering
both periods.  The highest
probability of eclipse occurred in April of 2009.  The next
most likely date of eclipse is in mid-2012.  The duration of
the eclipse would most likely be between 2 and 4 hours.  If this system does
eclipse, it will provide for a direct measurement of its
radius, allowing for a very powerful test of models at the L/T
transition region.

\section{Conclusions}\label{sec:con}

We have calculated relative orbital solutions for 15 very low
mass binary systems, using a combination of astrometric and
radial velocity data obtained with the Keck Observatory LGS AO
system.  For 10 of these systems, this is the
first derivation of the relative orbits, one of which gives
the most precise mass yet measured for a brown dwarf binary.  We have also
calculated the absolute orbital solutions for 6 systems, 5 of
which are the first for those systems, representing the first
individual component masses for several L dwarfs. 

The masses we have
calculated based on these orbital solutions and our derived
temperatures have allowed us to
perform the first comprehensive comparison of a sample of VLM
objects to theoretical evolutionary models.  
 All systems with mass precision better
than 60$\%$ show discrepancies with the predictions of
evolutionary models.
We find that for
6 systems, their total system masses are underpredicted by
both evolutionary models considered.  In these systems, 11 of
the 12 components have spectral types earlier than L4.  We
find that one binary T dwarf has its total system mass
overpredicted by the evolutionary models.  We postulate that
for those systems in
which we see a discrepancy, the possible cause is either an
incorrect radius prediction by the evolutionary models, an
incorrect temperature prediction by the evolutionary models,
or an incorrect temperature prediction by the atmospheric
models.  

Future work that would illuminate the apparent mass
discrepancies include (1) improving the precision of the
dynamical masses, with a particular emphasis on obtaining more
individual masses across a broader range of spectral types and (2)
obtaining radius measurements for our sources, which we can
potentially pursue through the calculation of surface gravity
(which, with individual mass, provides a means for calculating
radius), or measure directly if 2MASS 0920+35AB is eclipsing.
Such measurements
will allow us to test the evolutionary and atmospheric models
independently.  

\acknowledgements

The authors thank observing assistants Joel Aycock, Heather
Hershley, Carolyn Parker, Gary Puniwai, Julie Rivera, Chuck Sorenson,
Terry Stickel, and Cynthia Wilburn and support astronomers Randy
Campbell, Al Conrad, Jim Lyke, and Hien Tran for
their help in obtaining the observations.  We thank Jessica
Lu, Tuan Do, Sylvana Yelda, Marshall Perrin, and Will Clarkson
for helpful discussions of this work.  We also thank an
anonymous referee for helpful suggestions for the improvement
of this document.  Support for this
work was provided by the NASA Astrobiology Institute, the
Packard Foundation, and the NSF Science
\& Technology Center for AO, managed by UCSC
(AST-9876783).  Portions of this work performed under the
auspices of the U.S. Department of Energy by Lawrence
Livermore National Laboratory under Contract
DE-AC52-07NA27344.  QMK acknowledges support from the NASA Graduate
Student Research Program (NNG05-GM05H) through JPL and the
UCLA Dissertation Year Fellowship Program.  This
publication makes use of data products from the Two Micron All
Sky Survey, which is a joint project of the University of
Massachusetts and the Infrared Processing and Analysis
Center/California Institute of Technology, funded by the
National Aeronautics and Space Administration and the National
Science Foundation.  The W.M. Keck Observatory is operated as
a scientific partnership among the California Institute of
Technology, the University of California and the National
Aeronautics and Space Administration. The Observatory was made
possible by the generous financial support of the W.M. Keck
Foundation.  The authors also wish to recognize and
acknowledge the very significant cultural role and reverence
that the summit of Mauna Kea has always had within the
indigenous Hawaiian community.  We are most fortunate to have
the opportunity to conduct observations from this mountain.

%%tables

\begin{singlespace} 
\begin{deluxetable}{lccccc} 
\tabletypesize{\scriptsize} 
\tablewidth{0pt} 
%\rotate
\tablecaption{VLM Binary Sample} 
\tablehead{ 
  \colhead{Source Name} & \colhead{RA} & \colhead{Dec} &
  \colhead{Estimated} & \colhead{Discovery} &
  \colhead{2MASS}\\
  \colhead{} & \colhead{(J2000)} & \colhead{(J2000)} &
  \colhead{Sp Types\tablenotemark{a}} & \colhead{Reference} & \colhead{K Band Mag.}
}
\startdata 
LP 349-25AB & 00 27 55.93 & +22 19 32.8 & M8+M9 & 13 & 9.569 $\pm$ 0.017\\
LP 415-20AB & 04 21 49.0 & +19 29 10 & M7+M9.5 & 8 & 11.668 $\pm$ 0.020\\
2MASS J05185995-2828372AB\tablenotemark{b} & 05 18 59.95 & -28 28 37.2 & L6+T4& 2 & 14.162 $\pm$ 0.072\\
2MASS J06523073+4710348AB\tablenotemark{b} & 06 52 30.7 & +47 10 34 & L3.5+L6.5 & 3 & 11.694 $\pm$ 0.020\\
2MASS J07003664+3157266 & 07 00 36.64 & +31 57 26.60 & L3.5+L6 &3&11.317 $\pm$ 0.023 \\
2MASS J07464256+2000321AB & 07 46 42.5 & +20 00 32 & L0+L1.5 & 4 & 10.468 $\pm$ 0.022\\
2MASS J08503593+1057156 & 08 50 35.9 & +10 57 16 & L6+L8 & 4 & 14.473 $\pm$ 0.066 \\
2MASS J09201223+3517429AB & 09 20 12.2 & +35 17 42 & L6.5+T2 & 4 &  13.979 $\pm$ 0.061\\
2MASS J10170754+1308398AB\tablenotemark{c} & 10 17 07.5 & +13 08 39.1 & L2+L2 & 5 & 12.710 $\pm$ 0.023\\
2MASS J10210969-0304197 & 10 21 09.69 & -03 04 20.10 & T1+T5 & 15 &15.126 $\pm$ 0.173 \\
2MASS J10471265+4026437AB & 10 47 12.65 & +40 26 43.7 & M8+L0 & 6 & 10.399 $\pm$ 0.018\\
GJ 569b AB  & 14 54 29.0 & +16 06 05 & M8.5+M9 & 14 &$\sim$9.8 \\
LHS 2397a AB & 11 21 49.25 & -13 13 08.4 & M8+L7.5 & 12 & 10.735 $\pm$ 0.023 \\
2MASS J14263161+1557012AB & 14 26 31.62 & +15 57 01.3 & M8.5+L1 & 6 & 11.731 $\pm$ 0.018 \\
HD 130948 BC & 14 50 15.81 & +23 54 42.6 & L4+L4 & 10 & $\sim$11.0\\
2MASS J15344984-2952274AB & 15 34 49.8 & -29 52 27 & T5.5+T5.5 & 7 & 14.843 $\pm$ 0.114 \\
2MASS J1600054+170832AB\tablenotemark{b} & 16 00 05.4 & +17 08 32 & L1+L3 & 5 & 14.678 $\pm$ 0.114 \\
2MASS J17281150+3948593AB & 17 28 11.50 & +39 48 59.3 & L7+L8 & 5 & 13.909 $\pm$ 0.048\\
2MASS J17501291+4424043AB & 17 50 12.91 & +44 24 04.3 & M7.5+l0 & 8 & 11.768 $\pm$ 0.017\\
2MASS J18470342+5522433AB & 18 47 03.42 & +55 22 43.3 & M7+M7.5 & 9 & 10.901 $\pm$ 0.020 \\
2MASS J21011544+1756586 & 21 01 15.4 & +17 56 58 & L7+L8 & 5 & 14.892 $\pm$ 0.116 \\
2MASS J21402931+1625183AB & 21 40 29.32 & +16 25 18.3 & M8.5+L2 & 6 & 11.826 $\pm$ 0.031 \\
2MASS J21522609+0937575 & 21 52 26 & +09 37 57 & L6+L6 & 3 & 13.343 $\pm$ 0.034  \\
2MASS J22062280-2047058AB & 22 06 22.80 & -20 47 05.9 & M8+M8 & 6 & 11.315 $\pm$ 0.027\\
\enddata
\tablenotetext{a}{From discovery reference}
\tablenotetext{b}{In all observations of these sources, the
  binary was never resolved.  We report upper limits to the
  separations of these binaries, but no orbital solutions can
  be derived}
\tablenotetext{c}{Source cut from sample due to additional
  astrometry showing that it was not likely to yield a mass to
  a precision of better than 10$\%$ in the required timeframe}
\tablecomments{References - (1) Burgasser et al. 2005 (2) Cruz
et al. 2004 (3) Reid et al. 2006(4) Reid et al. 2001 (5) Bouy et
al. 2003 (6) Close et al. 2003 (7) Burgasser et al. 2003 (8)
Siegler et al. 2003 (9) Siegler et al. 2005 (10) Potter et
al. (2002) (11) Koerner et al. 1999 (12) Freed et al. 2003
(13) Forveille et al. 2005 (14) Martin et al. 2000 (15)
Burgasser et al. 2006}
\label{tab:initialSample}
\end{deluxetable} 
\end{singlespace}

\begin{singlespace}
\begin{deluxetable}{lccccc}
\tabletypesize{\scriptsize}
\tablecolumns{6}
\tablewidth{0pc}
\tablecaption{Log of NIRC2 LGS AO Observations}
\tablehead{
\colhead{Target} &  \colhead{Date of} & \colhead{Tip/Tilt} & \colhead{Filter} &
\colhead{Exposure Time} & \colhead{No. of} \\
\colhead{Name} &\colhead{Observation (UT)} &  \colhead{Reference} & 
\colhead{} & \colhead{(sec x coadds)} &
\colhead{Frames}
}
\startdata
2MASS 0518-28AB   & 2006 Nov 27 & USNO-B1.0 0615-0055823 & Kp & 30x4 & 18 \\
                  & 2007 Dec 02 &  & Kp & 30x4 & 9 \\
                  & 2008 Dec 18 &  & Kp & 20x4 & 5 \\
2MASS 0652+47AB   & 2006 Nov 27 & USNO-B1.0 1371-0206444 & Kp & 8x12 & 6 \\
                  & 2007 Dec 02 &  & Kp & 5x12 & 9 \\
2MASS 0746+20AB   & 2006 Nov 27 & source & Kp & 2x30 & 9 \\
                  & 2007 Dec 01 &  & Kp & 2x30 & 8 \\
                  & 2007 Dec 01 &  & J       & 4x15 & 9 \\
                  & 2008 Dec 18 &  & Kp & 2x30 & 8 \\
                  & 2008 Dec 18 &  & H       & 2x30 & 6 \\
2MASS 0850+10AB   & 2007 Dec 02 & USNO-B1.0 1009-0165240 & Kp & 30x4 & 9 \\
                  & 2008 Dec 18 &  & Kp & 10x1 & 5 \\
2MASS 0920+35AB   & 2006 Nov 27 & USNO-B1.0 1252-0171182 & Kp & 30x4 & 7 \\
                  & 2007 Dec 02 &  & Kp & 30x4 & 4 \\
                  & 2007 Dec 02 &  & J       & 30x4 & 2 \\
                  & 2008 May 30 &  & Kp & 30x4 & 6 \\
                  & 2008 Oct 21 &  & H       & 30x4 & 6 \\
                  & 2008 Dec 18 &  & H       & 10x10 & 7 \\
                  & 2009 Jun 10 &  & H       & 10x5 & 6 \\
2MASS 1017+13AB   & 2006 Nov 27 & USNO-B1.0 1031-0208442 & Kp & 13x12 & 3 \\
2MASS 1047+40AB   & 2006 Jun 21 & source & Kp & 1x60 & 9 \\
                  & 2006 Nov 27 &  & Kp & 2x30 & 12 \\
                  & 2007 Dec 02 &  & Kp & 2x30 & 6 \\
                  & 2008 Dec 18 &  & Kp & 1x30 & 9 \\
2MASS 1426+15AB   & 2006 Jun 20 & USNO B1.0-1059-0232527 & Kp & 10x12 & 3 \\ 
                  & 2008 May 30 &  & Kp & 10x12 & 8 \\
                  & 2008 May 30 &  & J       & 15x5 & 5 \\
                  & 2009 May 02 &  & Kp & 5x12 & 9 \\
                  & 2009 May 02 &  & H       & 5x12 & 6 \\
2MASS 1534-29AB   & 2006 Jun 20 & USNO-B1.0 0601-0344997 & J   & 30x4 & 9 \\ 
                  & 2008 May 30 &  & Kp & 40x2 & 7 \\
                  & 2008 May 30 &  & H       & 40x2 & 6 \\
                  & 2008 May 30 &  & J       & 40x1 & 3 \\
                  & 2009 May 04 &  & H       & 30x4 & 6 \\
2MASS 1600+17AB   & 2007 May 20 & USNO-B1.0 1071-0293881 & Kp & 30x4 & 9 \\
                  & 2008 May 30 &  & Kp & 10x1 & 2 \\
2MASS 1728+39AB   & 2007 May 20 & USNO-A2.0 1275-09377115 & Kp & 30x4 & 5 \\
                  & 2008 May 30 &  & Kp & 30x2 & 4 \\
                  & 2008 May 30 &  & J       & 60x2 & 5 \\
                  & 2009 May 03 &  & Kp & 30x2 & 7 \\
                  & 2009 Jun 11 &  & H       & 30x4 & 9 \\
2MASS 1750+44AB   & 2006 Jun 20 & source & Kp & 20x4 & 8 \\
                  & 2007 May 17 &  & Kp & 10x12 & 7 \\
                  & 2008 May 13 &  & Kp & 10x12 & 6 \\
                  & 2008 May 30 &  & H       & 5x12 & 6 \\
                  & 2008 May 30 &  & J       & 10x1 & 4 \\
                  & 2009 May 01 &  & Kp & 5x12 & 9 \\
2MASS 1847+55AB   & 2006 May 21 & source & Kp & 5x6 & 6 \\
                  & 2007 May 14 &  & Kp & 1.452x1 & 9 \\
                  & 2008 May 20 &  & Kp & 5x12 & 9 \\
                  & 2008 May 20 &  & H       & 5x5 & 6 \\
                  & 2008 May 20 &  & J       & 10x1 & 19 \\
                  & 2009 May 04 &  & Kp & 5x12 & 9 \\
2MASS 2140+16AB   & 2006 May 21 & USNO-B1.0 1064-0594380 & Kp & 5x1 & 12\\
                  & 2006 Nov 27 &  & Kp & 10x12 & 9 \\
                  & 2007 May 14 &  & Kp & 7x5 & 9 \\
                  & 2007 Dec 02 &  & Kp & 10x12 & 9 \\
                  & 2008 May 15 &  & Kp & 10x12 & 9 \\
                  & 2008 May 30 &  & H       & 5x12 & 4 \\
                  & 2008 May 30 &  & J       & 10x1 & 4 \\
                  & 2008 Dec 19 &  & Kp & 5x12 & 9 \\
                  & 2009 Jun 11 &  & Kp & 5x12 & 8 \\
2MASS 2206-20AB   & 2006 May 21 & source & Kp & 5x6 & 9 \\
                  & 2006 Nov 27 &  & Kp & 10x12 & 9 \\
                  & 2007 May 17 &  & Kp & 10x3 & 8 \\
                  & 2007 Dec 02 &  & Kp & 10x12 & 2 \\
                  & 2008 May 30 &  & Kp & 5x12 & 9 \\
                  & 2008 May 30 &  & H       & 5x6 & 3 \\
                  & 2008 May 30 &  & J       & 10x1 & 6 \\
                  & 2009 Jun 11 &  & Kp & 2.5x12 & 9 \\
GJ 569BC          & 2009 Jun 11 & GJ569A & Kp & 0.5x30 & 10 \\
HD 130948BC       & 2007 May 11 & HD 130948A & Kp & 2x30 & 12 \\
                  & 2007 May 11 &  & H       & 2x60 & 12 \\
                  & 2007 May 11 &  & J       & 4x15 & 12 \\
                  & 2008 Apr 28 &  & Kp & 0.1452x1 & 12 \\
                  & 2009 May 09 &  & H       & 1x15 & 7 \\
LHS 2397aAB       & 2006 Nov 27 & source & Kp & 15x10 & 3 \\
                  & 2007 Dec 02 &  & Kp & 8x15 & 3 \\
                  & 2007 Dec 02 &  & J       & 10x15 & 6 \\
                  & 2008 May 30 &  & Kp & 3x30 & 8 \\
                  & 2008 Dec 18 &  & Kp & 2x30 & 8 \\
                  & 2008 Dec 18 &  & H       & 1.5x30 & 6 \\
                  & 2008 Dec 18 &  & J       & 2x30 & 6 \\
                  & 2009 Jun 10 &  & Kp & 2x30 & 9 \\
LP 349-25AB       & 2006 Nov 27 & source & Kp & 1x30 & 5 \\
                  & 2006 Nov 27 &  & H       & 1x30 & 5 \\
                  & 2006 Nov 27 &  & J       & 1.5x30 & 3 \\
                  & 2007 Dec 02 &  & Kp & 5x6 & 9 \\
                  & 2008 May 30 &  & Kp & 1.452x20 & 7 \\
                  & 2008 Dec 19 &  & Kp & 2x20 & 6 \\
                  & 2008 Dec 19 &  & J       & 1.5x20 & 5 \\
                  & 2009 Jun 11 &  & Kp & 0.5x20 & 12 \\
LP 415-20AB       & 2006 Nov 27 & source & Kp & 8x12 & 6 \\
                  & 2007 Dec 02 &  & Kp & 6x12 & 9 \\
                  & 2008 Dec 18 &  & Kp & 6x12 & 9 \\
                  & 2008 Dec 18 &  & H       & 5x12 & 9 \\
\enddata
\label{tab:astrolog}
\end{deluxetable}
\end{singlespace}

\begin{singlespace}
\begin{deluxetable}{lcccc}
\tabletypesize{\scriptsize}
\tablecolumns{5}
\tablewidth{0pc}
\tablecaption{Log of NIRSPAO-LGS K-band Observations}
\tablehead{
\colhead{Target} &  \colhead{Date of} & \colhead{A0V Star} &
\colhead{Exposure Time} & \colhead{No. of} \\
\colhead{Name} &\colhead{Observation (UT)} &  \colhead{Standard} & \colhead{(sec x coadds)} &
\colhead{Frames}
}
\startdata
2MASS J07464256+2000321AB & 2006 Dec 16 & HIP 41798 & 1200x1 & 4 \\
                          & 2007 Dec 04 & HIP 41798 & 1200x1 &6 \\
                          & 2008 Dec 19 & HIP 41798 & 1200x1 & 6 \\
2MASS J14263161+1557012AB & 2007 Jun 08 & HIP 73087 & 1200x1 & 4 \\ 
                          & 2008 Jun 01 & HIP 73087 & 1200x1 & 4 \\
                          & 2009 Jun 12 & HIP 73087 & 1200x1 & 4 \\
2MASS J17501291+4424043AB & 2008 May 31 & HIP 87045 & 1200x1 & 4 \\
                          & 2009 Jun 12 & HIP 87045 & 1200x1 & 6 \\
2MASS J18470342+5522433AB & 2007 Jun 08 & HIP 93713 & 1200x1 & 4 \\
                          & 2008 Jun 01 & HIP 93713 & 1200x1 & 5 \\
                          & 2009 Jun 13 & HIP 93713 & 1200x1 & 3 \\
2MASS J21402931+1625183AB & 2007 Jun 09 & HIP 108060 & 1200x1 & 4 \\
                          & 2008 May 31 & HIP 108060 & 1800x1 & 3 \\
                          & 2009 Jun 13 & HIP 108060 & 1800x1 & 2 \\
2MASS J22062280-2047058AB & 2007 Jun 09 & HIP 116750 & 1200x1 & 3 \\
                          & 2008 Jun 01 & HIP 109689 & 1200x1 & 4 \\
                          & 2009 Jun 12 & HIP 109689 & 1200x1 & 4 \\
GJ 569b AB                & 2007 Jun 09 & HIP 73087 & 900x1 & 2 \\
                          & 2009 Jun 13 & HIP 73087 & 900x1 & 4 \\
HD 130948BC               & 2007 Jun 09 & HIP 73087 & 1200x1 & 4 \\
LHS 2397aAB               & 2007 Dec 04 & HIP 58188 & 1800x1 & 2 \\
                          & 2008 May 31 & HIP 61318 & 1800x1 & 3 \\
                          & 2008 Dec 19 & HIP 58188 & 1800x1 & 3 \\
                          & 2009 Jun 12 & HIP 61318 & 1800x1 & 2 \\
LP 349-25AB               & 2006 Dec 16 & HIP 5132 & 600x1 & 4 \\
                          & 2007 Dec 04 & HIP 5132 & 900x1 &1 \\
                          & 2008 Dec 19 & HIP 5132 & 1200x1 & 4 \\
                          & 2009 Jun 12 & HIP 5132 & 1200x1 & 4 \\
LP 415-20AB               & 2008 Dec 19 & HIP 24555 & 1200x1 & 4 \\
\enddata
\label{tab:bd_rvlog}
\end{deluxetable}
\end{singlespace}

\begin{singlespace}
\begin{deluxetable}{lcccccc}
\rotate
\tabletypesize{\scriptsize}
\tablecolumns{7}
\tablewidth{0pc}
\tablecaption{NIRC2 LGS AO Results}
\tablehead{
\colhead{Target} & \colhead{Date of} &  \colhead{Filter} &
\colhead{Separation} & \colhead{Separation} & \colhead{Position Angle} & \colhead{Flux Ratio} \\
\colhead{Name} & \colhead{Observation (UT)} & \colhead{} & \colhead{(pixels)\tablenotemark{a}} &\colhead{(arcseconds)\tablenotemark{b}} & \colhead{(degrees)\tablenotemark{c}} &\colhead{(Ab/Aa)\tablenotemark{d}}
}
\startdata
2MASS 0518-28 & 2006 Nov 27 & Kp & $<$ 6.40 & $<$ 0.064 & --- & --- \\ 
              & 2007 Dec 02 & Kp & $<$ 5.55 & $<$ 0.055 & --- & --- \\
              & 2008 Dec 18 & Kp & $<$ 6.55 & $<$ 0.065 & --- & --- \\
2MASS 0652+47 & 2006 Nov 27 & Kp & $<$ 3.05 & $<$ 0.030 & --- & --- \\
              & 2007 Dec 02 & Kp & $<$ 3.73 & $<$ 0.037 & --- & --- \\
2MASS 0700+31 & 2008 Dec 18 & Kp & 19.999 $\pm$ (0.035 $\pm$ 0.013) & 0.1993 $\pm$ 0.0004 & 279.21 $\pm$ (0.14 $\pm$ 0.02) [0.14] & 3.48 $\pm$ (0.10 $\pm$ 0.03) \\
2MASS 0746+20 & 2006 Nov 27 & Kp & 29.924 $\pm$ (0.058 $\pm$ 0.013) & 0.2981 $\pm$ 0.0006 & 233.93 $\pm$ (0.08 $\pm$ 0.03) [0.08] & 1.39 $\pm$ (0.02 $\pm$ 0.04)\\
              & 2007 Nov 30 & Kp & 33.533 $\pm$ (0.028 $\pm$ 0.057) & 0.3341 $\pm$ 0.0007 & 223.54 $\pm$ (0.04 $\pm$ 0.22) [0.23] & 1.39 $\pm$ (0.01 $\pm$ 0.03) \\
              & 2007 Nov 30 & J       & 33.501 $\pm$ (0.165 $\pm$ 0.072) & 0.334 $\pm$ 0.002 & 223.49 $\pm$ (0.36 $\pm$ 0.26) [0.45] & 1.60 $\pm$ (0.14 $\pm$ 0.04)\\
              & 2008 Dec 18 & Kp & 35.240 $\pm$ (0.032 $\pm$ 0.002) & 0.3511 $\pm$ 0.0004 & 214.31 $\pm$ (0.06 $\pm$ 0.02) [0.07] & 1.39 $\pm$ (0.01 $\pm$ 0.03) \\
              & 2008 Dec 18 & H       & 35.268 $\pm$ (0.036 $\pm$ 0.039) & 0.3514 $\pm$ 0.0006 & 214.38 $\pm$ (0.10 $\pm$ 0.07) [0.13] & 1.50 $\pm$ (0.01 $\pm$ 0.04) \\
2MASS 0850+10 & 2007 Dec 01 & Kp & 8.927 $\pm$ (0.197 $\pm$ 0.215) & 0.089 $\pm$ 0.003 & 158.71 $\pm$ (0.93 $\pm$ 0.13) [0.93] & 1.81 $\pm$ (0.19 $\pm$ 0.04) \\
              & 2008 Dec 18 & Kp & 7.611 $\pm$ (0.086 $\pm$ 0.073) & 0.076 $\pm$ 0.001  &165.87 $\pm$ (0.36 $\pm$ 0.12) [0.37]& 2.12 $\pm$ (0.10 $\pm$ 0.03) \\
2MASS 0920+35 & 2006 Nov 27 & Kp & 6.583 $\pm$ (0.194 $\pm$ 0.406) & 0.066 $\pm$ 0.004 & 247.14 $\pm$ (2.04 $\pm$ 0.06) [2.04] & 1.36 $\pm$ (0.08 $\pm$ 0.04) \\
              & 2007 Dec 01 & Kp & 7.561 $\pm$ (0.292 $\pm$ 0.232) & 0.075 $\pm$ 0.004 & 244.91 $\pm$ (3.23 $\pm$ 0.27) [3.24] & 1.19 $\pm$ (0.07 $\pm$ 0.04) \\
              & 2007 Dec 01 & J       & 6.623 $\pm$ (1.50 $\pm$ 0.66) & 0.066 $\pm$ 0.016 & 247.7 $\pm$ (1.6 $\pm$ 0.2) [1.6] & 1.04 $\pm$ (0.27 $\pm$ 0.05) \\
              & 2008 May 30 & Kp & 6.622 $\pm$ (0.079 $\pm$ 0.057) & 0.066 $\pm$ 0.001 & 249.94 $\pm$ (0.53 $\pm$ 0.09) [0.54] &1.75 $\pm$ (0.09 $\pm$ 0.03) \\
              & 2008 Oct 20 & H       & 4.714 $\pm$ 0.145 & 0.047 $\pm$ 0.001 & 252.3 $\pm$ 3.0 & 1.20 $\pm$ 0.07 \\
              & 2008 Dec 18 & H       & 3.753 $\pm$ 0.335 & 0.037 $\pm$ 0.003 & 247.6 $\pm$ 1.8 & 1.07 $\pm$ 0.05 \\
              & 2009 Jun 10 & H       & $<$ 2.63 & $<$ 0.0262 & --- & --- \\
2MASS 1017+13 & 2006 Nov 27 & Kp & 8.777 $\pm$ (2.403 $\pm$ 0.295) & 0.087 $\pm$ 0.024 & 83.11 $\pm$ (4.98 $\pm$ 0.06) [4.98] & 1.27 $\pm$ (0.63 $\pm$ 0.04) \\
2MASS 1021-03 & 2008 Dec 18 & Kp & 14.923 $\pm$ (0.032 $\pm$ 0.026) & 0.1487 $\pm$ 0.0004 & 204.13 $\pm$ (0.13 $\pm$ 0.02) [0.13] & 2.52 $\pm$ (0.03 $\pm$ 0.03) \\
2MASS 1047+40 & 2006 Jun 21 & Kp & 3.178 $\pm$ (0.169 $\pm$ 0.153) & 0.032 $\pm$ 0.002 & 126.77 $\pm$ (4.44 $\pm$ 0.05) [4.44] & 1.52 $\pm$ (0.26 $\pm$ 0.02) \\
              & 2006 Nov 27 & Kp & $<$ 4.68 & $<$ 0.047 & --- & --- \\
              & 2007 Dec 02 & Kp & $<$ 4.68 & $<$ 0.047 & --- & --- \\
2MASS 1426+15 & 2006 Jun 19 & Kp & 26.565 $\pm$ (0.054 $\pm$ 0.018) & 0.265 $\pm$ 0.001 & 343.07 $\pm$ (0.47 $\pm$ 0.04) [0.47] & 1.81 $\pm$ (0.10 $\pm$ 0.02) \\
              & 2008 May 30 & Kp & 30.562 $\pm$ (0.043 $\pm$ 0.015) & 0.3045 $\pm$ 0.0005 & 343.55 $\pm$ (0.06 $\pm$ 0.03) [0.07] & 1.82 $\pm$ (0.02 $\pm$ 0.03) \\
              & 2008 May 30 & H       & 30.479 $\pm$ (0.107 $\pm$ 0.071) & 0.304 $\pm$ 0.001 & 343.53 $\pm$ (0.28 $\pm$ 0.22) [0.36] & 2.02 $\pm$ (0.05 $\pm$ 0.02) \\
              & 2009 May 02 & Kp & 32.389 $\pm$ (0.046 $\pm$ 0.015) & 0.3227 $\pm$ 0.0005 & 343.69 $\pm$ (0.06 $\pm$ 0.05) [0.08] & 1.84 $\pm$ (0.02 $\pm$ 0.04) \\
              & 2009 May 02 & H       & 32.375 $\pm$ (0.038 $\pm$ 0.017) & 0.3226 $\pm$ 0.0006 & 343.84 $\pm$ (0.06 $\pm$ 0.05) [0.08] & 1.91 $\pm$ (0.02 $\pm$ 0.04) \\
2MASS 1534-29 & 2006 Jun 19 & J       & 18.649 $\pm$ 0.125 & 0.186 $\pm$ 0.001 & 15.57 $\pm$ 0.29 & 1.20 $\pm$ 0.04 \\
              & 2008 May 30 & Kp & 9.571 $\pm$ 0.121 & 0.095 $\pm$ 0.001 & 21.53 $\pm$ 0.84 & 1.23 $\pm$ 0.13 \\
              & 2008 May 30 & H       & 9.549 $\pm$ 0.131 & 0.095 $\pm$ 0.001 & 21.69 $\pm$ 0.82 & 1.38 $\pm$ 0.09 \\
              & 2009 May 04 & H       & 3.919 $\pm$ 0.118 & 0.039 $\pm$ 0.001 & 38.52 $\pm$ 3.25 & 1.28 $\pm$ 0.11 \\
2MASS 1600+17 & 2007 May 20 & Kp & $<$ 4.02 & $<$ 0.040 & --- & --- \\
              & 2008 May 30 & Kp & $<$ 3.90 & $<$ 0.039 & --- & --- \\
2MASS 1728+39 & 2007 May 20 & Kp & 20.496 $\pm$ (0.138 $\pm$ 0.030) & 0.204 $\pm$ 0.001 & 85.08 $\pm$ (0.21 $\pm$ 0.14) [0.25]& 1.83 $\pm$ (0.03 $\pm$ 0.02) \\ 
              & 2008 May 30 & Kp & 20.790 $\pm$ (0.589 $\pm$ 0.026) & 0.207 $\pm$ 0.006 & 101.33 $\pm$ (0.13 $\pm$ 0.03) [0.14] & 1.97 $\pm$ (0.14 $\pm$ 0.03) \\
              & 2008 May 30 & J       & 21.467 $\pm$ (0.045 $\pm$ 0.165) & 0.214 $\pm$ 0.002 & 101.85 $\pm$ (0.12 $\pm$ 0.25) [0.28] & 1.34 $\pm$ (0.02 $\pm$ 0.02) \\
              & 2009 May 03 & Kp & 21.854 $\pm$ (0.019 $\pm$ 0.110) & 0.218 $\pm$ 0.001 & 105.85 $\pm$ (0.48 $\pm$ 0.15) [0.50] & 1.74 $\pm$ (0.02 $\pm$ 0.02) \\
              & 2009 Jun 11 & H       & 21.868 $\pm$ 0.034 & 0.218 $\pm$ 0.0004 & 106.41 $\pm$ 0.08 & 1.52 $\pm$ 0.01 \\
2MASS 1750+44 & 2006 Jun 19 & Kp & 15.392 $\pm$ (0.443 $\pm$ 0.050) & 0.153 $\pm$ 0.004 & 33.68 $\pm$ (2.47 $\pm$ 0.07) [2.47] & 1.94 $\pm$ (0.13 $\pm$ 0.02) \\
              & 2007 May 17 & Kp & 17.330 $\pm$ (0.161 $\pm$ 0.022) & 0.173 $\pm$ 0.002 & 42.37 $\pm$ (0.28 $\pm$ 0.02) [0.28] & 1.93 $\pm$ (0.02 $\pm$ 0.04) \\
              & 2008 May 13 & Kp & 18.556 $\pm$ (0.020 $\pm$ 0.057) & 0.1849 $\pm$ 0.0006 & 52.29 $\pm$ (0.05 $\pm$ 0.06) [0.08] & 1.84 $\pm$ (0.02 $\pm$ 0.03) \\
              & 2008 May 30 & J       & 19.321 $\pm$ (0.173 $\pm$ 0.201) & 0.192 $\pm$ 0.003 & 53.78 $\pm$ (0.32 $\pm$ 0.16) [0.35] & 2.41 $\pm$ (0.03 $\pm$ 0.02) \\
              & 2008 May 30 & H       & 18.575 $\pm$ (0.115 $\pm$ 0.394) & 0.185 $\pm$ 0.004 & 52.64 $\pm$ (0.91 $\pm$ 1.77) [1.99] & 2.04 $\pm$ (0.11 $\pm$ 0.19) \\
              & 2009 May 01 & Kp & 20.2779 $\pm$ (0.035 $\pm$ 0.011) & 0.2020 $\pm$ 0.0004 & 60.31 $\pm$ (0.06 $\pm$0.02) [0.07] & 1.83 $\pm$ (0.01 $\pm$ 0.02) \\
2MASS 1847+55 & 2006 May 21 & Kp & 15.289 $\pm$ (0.032 $\pm$ 0.076) & 0.1523 $\pm$ 0.0008 & 110.90 $\pm$ (0.03 $\pm$ 0.01) [0.04] & 1.30 $\pm$ (0.003 $\pm$ 0.02) \\
              & 2007 May 14 & Kp & 17.335 $\pm$ (0.039 $\pm$ 0.059) & 0.173 $\pm$ 0.007 & 114.01 $\pm$ (0.08 $\pm$ 0.04) [0.09] & 1.28 $\pm$ (0.01 $\pm$ 0.02) \\
              & 2008 May 20 & Kp & 19.202 $\pm$ (0.147 $\pm$ 0.052) & 0.191 $\pm$ 0.002 & 116.71 $\pm$ (0.44 $\pm$ 0.04) [0.44] & 1.28 $\pm$ (0.04 $\pm$ 0.02) \\
              & 2008 May 20 & J       & 18.892 $\pm$ (0.133 $\pm$ 0.209) & 0.188 $\pm$ 0.003 & 116.61 $\pm$ (0.28 $\pm$ 0.15) [0.32] & 1.25 $\pm$ (0.01 $\pm$ 0.10) \\
              & 2008 May 20 & H       & 19.245 $\pm$ (0.054 $\pm$ 0.389) & 0.192 $\pm$ 0.004 & 116.64 $\pm$ (0.15 $\pm$ 2.71) [2.72] & 1.30 $\pm$ (0.03 $\pm$ 0.19) \\ 
              & 2009 May 04 & Kp & 20.726 $\pm$ (0.057 $\pm$ 0.008) &  0.2065 $\pm$ 0.0006 & 118.74 $\pm$ (0.13 $\pm$ 0.01) [0.14] & 1.28 $\pm$ (0.02 $\pm$ 0.03) \\
2MASS 2101+17 & 2008 May 15 & Kp & 32.405 $\pm$ (0.047 $\pm$ 0.014) & 0.3229 $\pm$ 0.0005 & 94.47 $\pm$ (0.09 $\pm$ 0.04) [0.11] & 1.31 $\pm$ (0.02 $\pm$ 0.03) \\
2MASS 2140+16 & 2006 May 21 & Kp & 10.922 $\pm$ 0.061 & 0.1088 $\pm$ 0.0006 & 202.91 $\pm$ 0.54 $\pm$ 0 & 1.97 $\pm$ 0.04 \\  
              & 2006 Nov 27 & Kp & 10.803 $\pm$ 0.126 & 0.108 $\pm$ 0.001 & 215.02 $\pm$ 1.16 & 1.94 $\pm$ 0.12 \\
              & 2007 May 14 & Kp & 10.816 $\pm$ 0.044 & 0.1078 $\pm$ 0.0004 & 223.50 $\pm$ 0.25 & 1.96 $\pm$ 0.05 \\
              & 2007 Dec 01 & Kp & 10.879 $\pm$ 0.209 & 0.108 $\pm$ 0.002 & 234.02 $\pm$ 0.66 & 1.95 $\pm$ 0.12 \\
              & 2008 May 15 & Kp & 11.067 $\pm$ 0.096 & 0.111 $\pm$ 0.001 & 243.28 $\pm$ 0.56 & 1.96 $\pm$ 0.07 \\
              & 2008 May 30 & J       & 12.021 $\pm$ 0.173 & 0.120 $\pm$ 0.002 & 241.41 $\pm$ 0.45 & 2.39 $\pm$ 0.34 \\
              & 2008 May 30 & H       & 11.491 $\pm$ 0.075 & 0.115 $\pm$ 0.001 & 242.9 $\pm$ 1.6 & 2.35 $\pm$ 0.38 \\
              & 2008 Dec 19 & Kp & 11.311 $\pm$ 0.390 & 0.113 $\pm$ 0.004 &254.68 $\pm$ 0.32 & 1.94 $\pm$ 0.19 \\
              & 2009 Jun 11 & Kp & 11.478 $\pm$ 0.113 & 0.114 $\pm$ 0.001 & 263.34 $\pm$ 0.23 & 1.93 $\pm$ 0.09 \\
2MASS 2152+09 & 2008 May 30 & Kp & 32.797 $\pm$ (0.413 $\pm$ 0.013) & 0.327 $\pm$ 0.004 & 117.75 $\pm$ (1.04 $\pm$ 0.03) [1.04] & 1.05 $\pm$ (0.20 $\pm$ 0.03) \\
2MASS 2206-20 & 2006 May 21 & Kp & 13.068 $\pm$ (0.147 $\pm$ 0.133) & 0.130 $\pm$ 0.002 & 128.99 $\pm$ (0.27 $\pm$ 0.13) [0.27] & 1.04 $\pm$ (0.05 $\pm$ 0.02) \\
              & 2006 Nov 27 & Kp & 12.747 $\pm$ (0.223 $\pm$ 0.165) & 0.127 $\pm$ 0.003 & 138.65 $\pm$ (0.29 $\pm$ 0.04) [0.30] & 1.06 $\pm$ (0.09 $\pm$ 0.04) \\
              & 2007 May 17 & Kp & 12.313 $\pm$ (0.013 $\pm$ 0.035) & 0.1227 $\pm$ 0.0004 & 147.68 $\pm$ (0.12 $\pm$ 0.02) [0.12] & 1.03 $\pm$ (0.01 $\pm$ 0.04) \\
              & 2007 Dec 01 & Kp & 12.199 $\pm$ (0.07 $\pm$ 0.18)  & 0.122 $\pm$ 0.002 & 160.40 $\pm$ (0.09 $\pm$ 0.27) [0.29] & 0.97 $\pm$ (0.09 $\pm$ 0.04) \\
              & 2008 May 30 & Kp & 12.394 $\pm$ (0.084 $\pm$ 0.042) & 0.1235 $\pm$ 0.0009 & 169.58 $\pm$ (0.34 $\pm$ 0.04) [0.34] & 1.11 $\pm$ (0.11 $\pm$ 0.03) \\
              & 2008 May 30 & J       & 11.834 $\pm$ (0.104 $\pm$ 0.404) & 0.118 $\pm$ 0.004 & 170.49 $\pm$ (0.35 $\pm$ 0.47) [0.59] & 1.15 $\pm$ (0.03 $\pm$ 0.02) \\
              & 2008 May 30 & H       & 11.543 $\pm$ (0.838 $\pm$ 0.453) & 0.115 $\pm$ 0.009 & 169.82 $\pm$ (0.92 $\pm$ 0.35) [0.99] & 1.04 $\pm$ (0.18 $\pm$ 0.19) \\
              & 2009 Jun 11 & Kp & 12.588 $\pm$ (0.033 $\pm$ 0.123) & 0.124 $\pm$ 0.001 & 190.52 $\pm$ (0.07 $\pm$ 0.03) [0.09] & 1.05 $\pm$ 0.02 \\
GJ 569B       & 2009 Jun 11 & Kp & 9.953 $\pm$ (0.047 $\pm$ 0.194) & 0.099 $\pm$ 0.002 & 79.04 $\pm$ (0.20 $\pm$ 0.05) [0.20] & 1.58 $\pm$ (0.02 $\pm$ 0.02) \\
HD 130948 BC  & 2006 Jun 18\tablenotemark{e} & Hn3 & 5.401 $\pm$ 0.279 & 0.109 $\pm$ 0.006 & 136.33 $\pm$ 3.68 & --- \\
              & 2007 May 11 & Kp & 10.620 $\pm$ (0.058 $\pm$ 0.101) & 0.1058 $\pm$ 0.001 & 131.63 $\pm$ (0.11 $\pm$ 0.03) [0.12] & 1.21 $\pm$ (0.12 $\pm$ 0.03) \\
              & 2008 Apr 28 & Kp & 5.068 $\pm$ (0.069 $\pm$ 0.122) & 0.0505 $\pm$ 0.001 &  122.82 $\pm$ (4.93 $\pm$ 0.05) [4.93] &  1.15 $\pm$ (0.31 $\pm$ 0.04)  \\
              & 2009 May 09 & H       & 3.775 $\pm$ (0.318 $\pm$ 0.528) & 0.038 $\pm$ 0.006 & 327.1 $\pm$ (5.0 $\pm$ 1.1) [5.1] & 1.20 $\pm$ (0.13 $\pm$ 0.15) \\
LHS 2397a     & 2006 Nov 27 & Kp & 9.672 $\pm$ (4.976 $\pm$ 0.259) & 0.096 $\pm$ 0.050 & 300.01 $\pm$ (9.38 $\pm$ 0.38) [9.39] & 1.77 $\pm$ (0.82 $\pm$ 0.04) \\
              & 2007 Dec 01 & Kp & 14.629 $\pm$ (0.554 $\pm$ 0.155) & 0.146 $\pm$ 0.006 & 19.95 $\pm$ (2.16 $\pm$ 0.13) [2.17] & 10.16 $\pm$ (0.86 $\pm$ 0.04) \\
              & 2008 May 30 & Kp & 15.983 $\pm$ (0.758 $\pm$ 0.034) & 0.159 $\pm$ 0.008 &37.77 $\pm$ (1.68 $\pm$ 0.26) [1.70] & 12.21 $\pm$ (1.49 $\pm$ 0.03) \\
              & 2008 Dec 18 & Kp & 19.813 $\pm$ (0.101 $\pm$ 0.013) & 0.197 $\pm$ 0.001 & 50.27 $\pm$ (0.11 $\pm$ 0.04) [0.12] & 13.2 $\pm$ (1.0 $\pm$ 0.03) \\
              & 2008 Dec 18 & H       & 19.908 $\pm$ (0.276 $\pm$ 0.055) & 0.196 $\pm$ 0.003 & 50.94 $\pm$ (0.37 $\pm$ 0.11) [0.39] & 17.4 $\pm$ (1.2 $\pm$ 0.04) \\
              & 2009 Jun 10 & Kp & 22.026 $\pm$ (0.035 $\pm$ 0.021) & 0.2195 $\pm$ 0.0004 & 59.44 $\pm$ (0.08 $\pm$ 0.58) [0.59] & 12.89 $\pm$ (0.39 $\pm$ 0.02) \\
LP 349-25     & 2006 Nov 27 & Kp & 12.603 $\pm$ (0.049 $\pm$ 0.169) & 0.126 $\pm$ 0.002 & 234.88 $\pm$ (0.17 $\pm$ 0.70) [0.72] & 1.38 $\pm$ (0.02 $\pm$ 0.04) \\
              & 2006 Nov 27 & J       & 12.439 $\pm$ (0.213 $\pm$ 0.129) & 0.124 $\pm$ 0.002 & 236.67 $\pm$ (2.53 $\pm$ 0.06) [1.53] & 1.64 $\pm$ (0.04 $\pm$ 0.04) \\
              & 2006 Nov 27 & H       & 12.349 $\pm$ (0.093 $\pm$ 0.150) & 0.123 $\pm$ 0.002 & 235.48 $\pm$ (0.46 $\pm$ 0.06) [0.47] & 1.48 $\pm$ (0.08 $\pm$ 0.04) \\
              & 2007 Dec 01 & Kp & 13.126 $\pm$ (0.400 $\pm$ 0.169) & 0.131 $\pm$ 0.004 & 211.47 $\pm$ (1.61 $\pm$ 0.23) [1.62] & 1.20 $\pm$ (0.10 $\pm$ 0.04) \\ 
              & 2008 May 30 & Kp & 12.518 $\pm$ (0.120 $\pm$ 0.041) & 0.125 $\pm$ 0.001 &197.94 $\pm$ (0.40 $\pm$ 0.02) [0.40] & 1.44 $\pm$ (0.04 $\pm$ 0.03) \\
              & 2008 Dec 19 & Kp & 8.555 $\pm$ (0.080 $\pm$ 0.064) & 0.085 $\pm$ 0.001 & 172.41 $\pm$ (0.81 $\pm$ 0.08) [0.82] & 1.31 $\pm$ (0.08 $\pm$ 0.03) \\
              & 2008 Dec 19 & J       & 8.944 $\pm$ (0.364 $\pm$ 0.095) & 0.089 $\pm$ 0.004 & 173.17 $\pm$ (0.59 $\pm$ 0.12) [0.61] & 1.63 $\pm$ (0.16 $\pm$ 0.05) \\
              & 2009 Jun 11 & Kp & 6.653 $\pm$ (0.056 $\pm$ 0.354) & 0.066 $\pm$ 0.004 & 129.62 $\pm$ (0.22 $\pm$ 0.04) [0.22] & 1.34 $\pm$ (0.05 $\pm$ 0.02) \\
LP 415-20     & 2006 Nov 27 & Kp & 4.616 $\pm$ (0.083 $\pm$ 0.541) & 0.046 $\pm$ 0.005 & 35.11 $\pm$ (2.40 $\pm$ 0.85) [2.55] & 1.77 $\pm$ (0.09 $\pm$ 0.04) \\
              & 2007 Dec 01 & Kp & 9.617 $\pm$ (0.152 $\pm$ 0.206) & 0.096 $\pm$ 0.003 & 52.45 $\pm$ (1.06 $\pm$ 0.12) [1.06] & 2.53 $\pm$ (0.21 $\pm$ 0.04) \\
              & 2008 Dec 18 & Kp & 11.215 $\pm$ (0.444 $\pm$ 0.044) & 0.112 $\pm$ 0.004 & 62.56 $\pm$ (0.97 $\pm$ 0.10) [0.97] & 1.42 $\pm$ (0.12 $\pm$ 0.03) \\
              & 2008 Dec 18 & H       & 11.145 $\pm$ (0.179 $\pm$ 0.066) & 0.111 $\pm$ 0.002 & 62.18 $\pm$ (1.26 $\pm$ 0.04) [1.36] & 1.60 $\pm$ (0.19 $\pm$ 0.04) \\
\enddata
\tablenotetext{a}{The first listed uncertainty is that due to
  the measurement itself, while the second is the systematic uncertainty due to
  imperfect PSF matching.  If only one uncertainty is given,
  the source had a suitable PSF in the field of view.}
\tablenotetext{b}{The uncertainties given are the empirically
  estimated statistical uncertainty, the PSF mismatch uncertainty, and the absolute
  plate scale uncertainty added in quadrature}
\tablenotetext{c}{The first listed uncertainty is that due to
  the measurement itself, while the second is the systematic uncertainty due to
  imperfect PSF matching.  In the brackets is the
  combination of these two uncertainties along with the
  absolute uncertainty of the columns with respect to
  north. If only one uncertainty is given, the source had a
  suitable PSF in the field of view.}
\tablenotetext{d}{The first listed uncertainty is systematic uncertainty due to
  the measurement itself, while the second is that due to
  imperfect PSF matching.  If only one uncertainty is given,
  the source had a suitable PSF in the field of view.}
\tablenotetext{e}{Data from the OSIRIS imager, which has a
  plate scale of 0$\farcs$02/pixel.  This camera has not been
  fully characterized for distortion.  However, the
  uncertainties on these measurements are such that they
  should account for distortion on this camera}
\label{tab:bd_astro}
\end{deluxetable}
\end{singlespace}

\begin{singlespace}
\begin{deluxetable}{lcccccc}
\tabletypesize{\scriptsize}
\rotate
\tablecolumns{7}
\tablewidth{0pc}
\tablecaption{Radial Velocity Measurements}
\tablehead{
\colhead{Target} &  \colhead{Date of} & \colhead{Average SNR}
& \colhead{Average SNR} & \colhead{Rad. Velocity} &
\colhead{Rad. Velocity} & \colhead{$\Delta$RV}\\
\colhead{Name} &\colhead{Observation (UT)} &  \colhead{Primary
  (A)} & \colhead{Secondary (B)} & \colhead{Primary (km/s)} & \colhead{Secondary (km/s)} & \colhead{(km/s)}
}
\startdata
2MASS 0746+20AB & 2006 Dec 16 & 52 & 44 & 55.60 $\pm$ 0.68 & 52.94 $\pm$ 0.68 & -2.66 $\pm$ 0.96 \\
                & 2007 Dec 04 & 72 & 59 & 55.18 $\pm$ 0.60 & 52.37 $\pm$ 1.12 & -2.81 $\pm$ 1.27 \\
                & 2008 Dec 19 & 66 & 56 & 56.06 $\pm$ 0.85 & 54.05 $\pm$ 2.30 & -2.01 $\pm$ 2.45\\
2MASS 1426+15AB & 2007 Jun 08 & 44 & 33 & 12.54 $\pm$ 0.43 & 14.41 $\pm$ 1.27 & 1.87 $\pm$ 1.34 \\
                & 2008 Jun 01 & 50 & 36 & 12.67 $\pm$ 0.36 & 15.39 $\pm$ 1.40 & 2.72 $\pm$ 1.45 \\
                & 2009 Jun 12 & 41 & 29 & 12.78 $\pm$ 0.49 & 15.00 $\pm$ 0.68 & 2.22 $\pm$ 0.84\\
2MASS 1750+44AB & 2008 May 31 & 48 & 36 & -17.52 $\pm$ 0.39 & -15.89 $\pm$ 0.54 & 1.63 $\pm$ 0.67 \\
                & 2009 Jun 12 & 41 & 31 & -17.09 $\pm$ 0.53 & -15.25 $\pm$ 1.31 & 1.84 $\pm$ 1.41 \\
2MASS 1847+55AB & 2007 Jun 08 & 69 & 60 & -23.88 $\pm$ 0.32 &-20.46 $\pm$ 0.29 & 3.42 $\pm$ 0.43 \\
                & 2008 Jun 01 & 69 & 60 & -24.15 $\pm$ 0.21 & -20.09 $\pm$ 0.46 & 4.06 $\pm$ 0.51 \\
                & 2009 Jun 13 & 39 & 36 & -24.68 $\pm$ 0.60 & -19.63 $\pm$ 1.00 & 5.05 $\pm$ 1.17 \\
2MASS 2140+16AB & 2007 Jun 09 & 43 & 28 & 13.90 $\pm$ 0.30 & 11.07 $\pm$ 1.21 & -2.83 $\pm$ 1.25 \\
                & 2008 May 31 & 58 & 40 & 13.62 $\pm$ 0.27 & 12.26 $\pm$ 1.62 & -1.36 $\pm$ 1.68\\
                & 2009 Jun 13 & 38 & 26 & 13.47 $\pm$ 0.28 & 10.97 $\pm$ 2.00 & 2.50 $\pm$ 2.01 \\ 
2MASS 2206-20AB & 2007 Jun 09 & 47 & 39 & 13.66 $\pm$ 0.36 & 13.28 $\pm$ 0.48 & -0.38 $\pm$ 0.66 \\
                & 2008 Jun 01 & 54 & 48 & 13.14 $\pm$ 0.39 & 13.46 $\pm$ 0.51 & 0.32 $\pm$ 0.64 \\
                & 2009 Jun 12 & 47 & 44 & 13.37 $\pm$ 0.24 & 12.75 $\pm$ 0.37 & -0.62 $\pm$ 0.44 \\
GJ 569b AB      & 2007 Jun 09 & 89 & 82 & -10.49 $\pm$ 0.20 & -4.90 $\pm$ 0.50 & 5.59 $\pm$ 0.54 \\
                & 2009 Jun 13 & 86 & 67 & -8.97 $\pm$ 0.36 & -6.83 $\pm$ 0.27 & 2.14 $\pm$ 0.45 \\
HD 130948Bc     & 2007 Jun 09 & 44 & 33 & 4.57 $\pm$ 2.61 & -0.72 $\pm$ 1.05 & -5.29 $\pm$ 2.81 \\
LHS 2397aAB     & 2007 Dec 04 & 68 & 27 & 34.43 $\pm$ 0.86 & 34.84 $\pm$ 2.24 & 0.41 $\pm$ 2.40 \\
                & 2008 May 31 & 114 & 44 & 33.85 $\pm$ 0.27 & 36.30 $\pm$ 0.86 & 2.45 $\pm$ 0.90 \\
                & 2008 Dec 19 & 85 & 31 & 33.79 $\pm$ 0.37 & 35.30 $\pm$ 2.49 & 1.51 $\pm$ 2.52 \\
                & 2009 Jun 12 & 103 & 33 & 33.51 $\pm$ 0.66 & 34.27 $\pm$ 2.02 & 0.76 $\pm$ 1.22 \\
LP 349-25AB     & 2006 Dec 16 & 58 & 45 & -11.91 $\pm$ 1.33 & -6.57 $\pm$ 2.50 & 5.34 $\pm$ 2.83 \\
                & 2007 Dec 04 & 63 & 58 & -11.11 $\pm$ 3.00 & -5.50 $\pm$ 3.02 & 5.67 $\pm$ 2.12 \\
                & 2008 Dec 19 & 105 & 84 &-9.89 $\pm$ 1.51 & -6.78 $\pm$ 1.84 & 3.11 $\pm$ 0.98 \\
                & 2009 Jun 12 & 114 & 98 & -8.16 $\pm$ 0.49 & -7.27 $\pm$ 1.35 & 0.89 $\pm$ 1.44 \\
LP 415-20A     & 2008 Dec 19 & 42 & 32 & 41.13 $\pm$ 0.91 & 40.41 $\pm$ 1.06 & -0.72 $\pm$ 1.40 \\
\enddata
\tablecomments{Velocities are in the heliocentric reference frame}
\label{tab:bd_rvdata}
\end{deluxetable}
\end{singlespace}

\begin{singlespace}
\begin{deluxetable}{lccccccccccc}
\rotate
\tabletypesize{\scriptsize}
\setlength{\tabcolsep}{0.9mm}
\tablecolumns{12}
\tablewidth{0pc}
\tablecaption{Astrometric Orbital Parameters}
\tablehead{
\colhead{Target} &  \colhead{Fixed Dist.} & \colhead{Fit Dist.} & \colhead{Total System} &
\colhead{Period} & \colhead{Semi-Major} &
\colhead{Eccentricity} & \colhead{T$_{o}$} &
\colhead{Inc.} & \colhead{$\Omega$} &
\colhead{$\omega$} & \colhead{Best Fit}\\
\colhead{Name} &  \colhead{(pc)} & \colhead{(pc)}&\colhead{Mass (M$_{\odot}$)\tablenotemark{a}} &
\colhead{(years)} & \colhead{Axis (mas)} &
\colhead{} & \colhead{(years)} &
\colhead{(degrees)} & \colhead{(degrees)} &
\colhead{(degrees)} & \colhead{Reduced $\chi^{2}$}
}
\startdata
2MASS 0746+20AB & 12.21 $\pm$ 0.05\tablenotemark{b}& --- & 0.151 $\pm$ 0.003 & 12.71 $\pm$ 0.07 & 237.3$^{+1.5}_{-0.4}$ & 0.487 $\pm$ 0.003 & 2002.83 $\pm$ 0.01 & 138.2 $\pm$ 0.5 & 28.4 $\pm$ 0.5 & 354.4 $\pm$ 0.9 & 0.88\\
2MASS 0850+10AB & 38.1 $\pm$ 7.3\tablenotemark{c} &--- & 0.2 $\pm$ 0.2 & 24$^{+69}_{-6}$ & 126$^{100}_{-32}$ & 0.64 $\pm$ 0.26 & 2016$^{+9}_{-24}$ & 65 $\pm$ 12 & 96 $\pm$ 27 & 236$^{+117}_{-171}$ & 2.85\\
2MASS 0920+35AB & 24.3 $\pm$ 5.0\tablenotemark{d}& ---& 0.11 $\pm$ 0.11 & 6.7$^{+3.3}_{3.4}$ & 69 $\pm$ 24& 0.21$^{+0.65}_{-0.21}$ & 2003.43 $\pm$ 1.15 & 88.6 $\pm$ 2.4 & 69.0 a$\pm$ 1.5 & 317$^{+43}_{-300}$ & 0.92 \\
2MASS 1426+15AB &--- &34 $\pm$ 13 & 0.11$^{+0.08}_{-0.11}$ & 1985$^{+2141}_{-1945}$ & 2273 $\pm$ 1560 & 0.85$^{+0.10}_{-0.41}$ & 1998 $\pm$ 24 & 88.3 $\pm$ 0.8 & 344.8 $\pm$ 0.4 & 282$^{+78}_{-210}$ & 1.89\\
2MASS 1534-29AB& 13.59 $\pm$ 0.22\tablenotemark{e}&--- & 0.060 $\pm$ 0.004 & 23.1 $\pm$ 4.0 & 234 $\pm$ 30 & 0.10 $\pm$ 0.09 & 2006.4 $\pm$ 3.0 & 85.6 $\pm$ 0.4 & 13.4 $\pm$ 0.3 & 25$^{+154}_{-25}$ & 1.57\\
2MASS 1728+39AB & 24.1 $\pm$ 2.1\tablenotemark{c} &---& 0.15$^{+0.25}_{-0.04}$ & 31.3 $\pm$ 12.7 & 220 $\pm$ 26 & 0.28$^{+0.35}_{-0.28}$ & 2017 $^{+4}_{-22}$ & 62 $\pm$ 7 &  118$^{+11}_{-9}$ & 94 $\pm$ 15 & 2.60\\
2MASS 1750+44AB &--- &37.6 $\pm$ 12.3 & 0.20 $\pm$ 0.12 & 317 $\pm$ 240 & 728 $\pm$ 375 & 0.71 $\pm$ 0.18& 2004.3 $\pm$ 1.8 & 44 $\pm$ 10 & 99 $\pm$ 6 & 267 $\pm$ 26 & 1.63 \\
2MASS 1847+55AB &--- &29.8 $\pm$ 7.1 & 0.18$^{+0.35}_{-0.13}$ & 44.2 $\pm$ 18.7 & 237 $\pm$ 36 & 0.1$^{+0.5}_{-0.1}$ &2020$^{+6}_{-28}$ & 79$^{+4}_{-2}$ & 125 $\pm$ 3 & 68 $\pm$ 30 & 0.63 \\ 
2MASS 2140+16AB &--- &25 $\pm$ 10 & 0.10 $\pm$ 0.08 & 20.1$^{+5.3}_{-1.6}$ & 141$^{9}_{-6}$ & 0.26 $\pm$ 0.06 & 2012.0$^{+0.5}_{-2.0}$ & 46.2$^{2.5}_{-8.7}$ & 104 $\pm$ 7 & 223$^{+10}_{-47}$ & 0.50\\
2MASS 2206-20AB & 26.67 $\pm$ 2.63\tablenotemark{f}&--- & 0.16 $\pm$ 0.05 & 23.78 $\pm$ 0.19 & 168.0 $\pm$ 1.5 & 0.000$^{+0.002}_{-0.000}$ & 2000.0$^{+1.9}_{-3.2}$ & 44.3 $\pm$ 0.7 & 74.8 $\pm$ 1.0 & 326$^{+28}_{-52}$ & 2.32\\                                         
GJ569B ab & 9.81 $\pm$ 0.16\tablenotemark{g} &--- &0.126 $\pm$ 0.007 & 2.370 $\pm$ 0.002 & 90.8 $\pm$ 0.8 & 0.310 $\pm$ 0.006 & 2003.150 $\pm$ 0.005 & 33.6 $\pm$ 1.3 & 144.8 $\pm$ 1.9 & 77.4 $\pm$ 1.7 & 1.43 \\
HD 130948BC & 18.18 $\pm$ 0.08\tablenotemark{g} &---& 0.109 $\pm$ 0.002 & 9.83 $\pm$ 0.16 & 120.4 $\pm$ 1.4 & 0.16 $\pm$ 0.01 & 2008.6 $\pm$ 0.2 & 95.7 $\pm$ 0.2 & 313.3 $\pm$ 0.2 & 253.3 $\pm$ 3.9 & 2.12\\
LHS 2397a AB & 14.3 $\pm$ 0.4\tablenotemark{h} &--- &0.144 $\pm$ 0.013 & 14.26 $\pm$ 0.10 & 215.8 $\pm$ 1.5 & 0.348 $\pm$ 0.006 & 2006.29 $\pm$ 0.04 & 40.9 $\pm$ 1.2 & 78.0 $\pm$ 1.5 & 217.7 $\pm$ 2.6 & 1.47\\
LP 349-25AB & 13.19 $\pm$ 0.28\tablenotemark{i} &---& 0.121 $\pm$ 0.009 & 7.31 $\pm$ 0.37 & 141 $\pm$ 7 & 0.08 $\pm$ 0.02 & 2002.5 $\pm$ 0.8 & 118.7 $\pm$ 1.5 & 213.8 $\pm$ 1.1 & 109$^{+37}_{-22}$ & 2.15\\
LP 415-20AB &--- &21 $\pm$ 5 &0.09 $\pm$ 0.06 & 11.5 $\pm$ 1.2 & 108 $\pm$ 24 & 0.9 $\pm$ 0.1 & 2006.5 $\pm$ 0.2 & 55 $\pm$ 12 & 200 $\pm$ 40 & 73 $\pm$ 50 & 1.47\\
\enddata
\tablenotetext{a}{Derived from period and semi-major axis}
\tablenotetext{b}{Distance from parallax measurement by Dahn
  et al. (2002)}
\tablenotetext{c}{Distance from parallax measurement by Vrba
  et al. (2004)}
\tablenotetext{d}{No parallax measurement or radial velocity
  data exists - spectrophotometric distance used here}
\tablenotetext{e}{Distance from parallax measurement by Tinney
  et al. (2003)}
\tablenotetext{f}{Distance from parallax measurement by Costa
  et al. (2006)}
\tablenotetext{g}{Distance from Hipparcos parallax for high
  mass tertiary companion}
\tablenotetext{h}{Distance from parallax measurement by Monet
  et al. (1992)}
\tablenotetext{i}{Distance from parallax measurement by
  Gatewood et al. (2009)}
\label{tab:astro_orb}
\end{deluxetable}
\end{singlespace}

\begin{singlespace}
\begin{deluxetable}{lcccccccc}
\rotate
\tabletypesize{\scriptsize}
\tablecolumns{9}
\tablewidth{0pc}
\tablecaption{Absolute Orbital Parameters}
\tablehead{
\colhead{} & \multicolumn{3}{c}{Fit Parameters} & \colhead{} & \multicolumn{4}{c}{Derived Properties}  \\
\cline{2-4}
\cline{6-9}\\
\colhead{Target} &  \colhead{K$_{Primary}$} & \colhead{Center
  of Mass} & \colhead{Best Fit} & \colhead{} & \colhead{K$_{Secondary}$} &
 \colhead{Mass Ratio} &
\colhead{M$_{Primary}$} & \colhead{M$_{Secondary}$} 
 \\
\colhead{Name} &  \colhead{(km/s)} & \colhead{Velocity (km/s)}
& \colhead{Reduced $\chi^{2}$}& \colhead{} &\colhead{(km/s)} &
\colhead{(M$_{Primary}$ / M$_{Secondary}$)} &
\colhead{(M$_{\odot}$)} & \colhead{(M$_{\odot}$)} 
}
\startdata
2MASS 0746+20AB & 1.0$^{+3.0}_{-0.1}$ &  54.7 $\pm$ 0.8 & 0.44& & 4.1$^{+0.1}_{-3.1}$ & 4.0$^{+0.1}_{-3.8}$ & 0.12$^{+0.01}_{-0.09}$ & 0.03$^{+0.09}_{-0.01}$ \\
2MASS 2140+16AB & 0.8 $\pm$ 0.3 & 13.0 $\pm$ 0.2 & 0.9& & 3.1 $\pm$ 1.1 &  4.0$^{+0}_{-0.1}$ & 0.08 $\pm$ 0.06 & 0.02$^{+0.08}_{-0.02}$\tablenotemark{a} \\
2MASS 2206-20AB & 0.8 $\pm$ 0.2 & 13.3 $\pm$0.2 & 2.2& & 3.1 $\pm$ 0.4 &  4.0$^{+0.0}_{-0.2}$ & 0.13 $\pm$ 0.05 & 0.03$^{+0.07}_{-0.02}$\tablenotemark{a} \\
GJ 569b AB & 2.7 $\pm$ 0.3 & -8.0 $\pm$ 0.2\tablenotemark{b} & 0.56 & & 3.8 $\pm$ 0.4 &  1.4 $\pm$ 0.3 & 0.073 $\pm$ 0.008 & 0.053 $\pm$ 0.006 \\
LHS 2397a AB & 1.7 $\pm$ 1.2 & 34.6 $\pm$ 1.4 & 0.41 & &2.6 $\pm$ 1.4 &  1.5$^{+7.1}_{-1.4}$ & 0.09 $\pm$ 0.05 & 0.06 $\pm$ 0.05 \\
LP 349-25 AB & 4.5 $\pm$ 0.9 & -8.0 $\pm$ 0.5 & 0.8 & &2.2 $\pm$ 0.9 & 0.5 $\pm$ 0.3 & 0.04 $\pm$ 0.02 & 0.08 $\pm$ 0.02 \\
\enddata
\tablecomments{Using our absolute radial velocities in
  conjunction with the parameters from our relative
  orbital solutions, we fit for K$_{Primary}$ and $\gamma$.
  We then use those values to find K$_{Secondary}$ and the
  mass ratio.  We combine the mass ratio and the total system
  mass from the relative orbits to find component masses.}
\tablenotetext{a}{Upper uncertainty set using the uncertainty
  in M$_{Primary}$ and M$_{Tot}$}
\tablenotetext{b}{Set to our value}
\label{tab:spec_orb}
\end{deluxetable}
\end{singlespace}

\begin{singlespace}
%\begin{landscape}
\begin{deluxetable}{lcccccccccccc}
\tabletypesize{\scriptsize}
\setlength{\tabcolsep}{1.0mm}
\rotate
\tablecolumns{13}
\tablewidth{0pc}
\tablecaption{Photometric Measurements}
\tablehead{
\colhead{Target} &  \colhead{M$_{F625W}$} & \colhead{M$_{F775W}$}
& \colhead{M$_{814W}$} & \colhead{M$_{850LP}$} &
\colhead{M$_{F1042}$} & \colhead{M$_{J}$} & \colhead{M$_{H}$}
& \colhead{M$_{Kp}$} 
& \colhead{L$_{Bol}$} & \colhead{T$_{Eff}$} & \colhead{Rad.} &\colhead{Phot.}\\
\colhead{Name} &\colhead{} &  \colhead{} & \colhead{} &
\colhead{} & \colhead{} & \colhead{} &\colhead{} & \colhead{} & \colhead{(Log L/L$_{\odot}$)} &
\colhead{(K)} & \colhead{(R$_{Jup}$)} &\colhead{Ref}
}
\startdata
2MASS 0746+20A & 18.36 $\pm$ 0.05 & 15.55 $\pm$ 0.05 & 14.98 $\pm$ 0.15 & 13.81 $\pm$ 0.05 & --- & 11.85 $\pm$ 0.04 & 11.13 $\pm$ 0.02 & 10.62 $\pm$ 0.02 &-3.64 $\pm$ 0.02 & 2205 $\pm$ 50 & 0.99 $\pm$ 0.03 & 1\\
2MASS 0746+20B & 18.86 $\pm$ 0.06 & 16.23 $\pm$ 0.07 & 15.98 $\pm$ 0.18 & 14.57 $\pm$ 0.06 & ---& 12.36 $\pm$ 0.10 & 11.57 $\pm$ 0.03 & 10.98 $\pm$ 0.02 & -3.77 $\pm$ 0.02 & 2060 $\pm$ 70 & 0.97 $\pm$ 0.06 &1\\
2MASS 0850+10A & 20.93 $\pm$ 0.50 & 18.41 $\pm$ 0.48 & 17.39 $\pm$ 0.44 & 16.17 $\pm$ 0.48 & --- & --- & --- & 11.99 $\pm$ 0.42 & -4.22 $\pm$ 0.18 & 1590 $\pm$ 290 & 1.0 $\pm$ 0.3 &2\\
2MASS 0850+10B & 22.24 $\pm$ 0.57 & 19.57 $\pm$ 0.50 & 18.86 $\pm$ 0.45 & 17.03 $\pm$ 0.49 & --- & --- & --- & 12.80 $\pm$0.43 & -4.47 $\pm$ 0.18 & 1380 $\pm$ 250 & 1.0 $\pm$ 0.3 &2\\  
2MASS 0920+35A & --- & --- & 17.90 $\pm$ 0.48 & --- & --- & 14.43 $\pm$ 0.47 & 13.40 $\pm$ 0.45 & 12.65 $\pm$ 0.45 & -4.47 $\pm$ 0.19 & 1375 $\pm$ 250 & 1.0 $\pm$ 0.3 &3\\ 
2MASS 0920+35B & --- & --- & 18.78 $\pm$ 0.49 & --- & --- & 14.47 $\pm$ 0.56 & 13.60 $\pm$ 0.46 & 12.97 $\pm$ 0.46 & -4.54 $\pm$ 0.20 & 1320 $\pm$ 250 & 1.0 $\pm$ 0.3 &3\\ 
2MASS 1426+15A & 16.98 $\pm$ 0.87 & 13.92 $\pm$ 0.86 & 13.49 $\pm$ 0.85 & 12.23 $\pm$ 0.86 & 11.33 $\pm$ 0.85 & 10.69 $\pm$ 0.83 & 10.00 $\pm$ 0.83 & 9.55 $\pm$ 0.83  & -3.19 $\pm$ 0.34 & 2400 $\pm$ 70 & 1.37$^{+0.54}_{-0.59}$ & 4\\
2MASS 1426+15B & 17.98 $\pm$ 0.87 & 15.16 $\pm$ 0.87 & 14.89 $\pm$ 0.85 & 13.29 $\pm$ 0.86 & 12.63 $\pm$ 0.85 & 11.46 $\pm$ 0.83 & 10.70 $\pm$ 0.83 & 10.20 $\pm$ 0.83 & -3.48 $\pm$ 0.34 & 2240 $\pm$ 70 & 1.12$^{+0.48}_{-0.50}$ &4\\
2MASS 1534-29A & --- & --- & 19.57 $\pm$ 0.04 & --- & 15.74 $\pm$ 0.12 & 14.61 $\pm$ 0.10 & 14.79 $\pm$ 0.11 & 14.84 $\pm$ 0.12  & -4.97 $\pm$ 0.10 & 1130 $\pm$ 50 & 0.80 $\pm$ 0.03 & 5\\
2MASS 1534-29B & --- & --- & 19.87 $\pm$ 0.05 & --- & 15.94 $\pm$ 0.24 & 14.77 $\pm$ 0.10 & 15.14 $\pm$ 0.13 & 15.03 $\pm$ 0.13  & -5.05 $\pm$ 0.10 & 1097 $\pm$ 50 & 0.80 $\pm$ 0.03 &5\\
2MASS 1728+39A & --- & --- & 18.35 $\pm$ 0.25 & --- & 15.89 $\pm$ 0.21 & 14.68 $\pm$ 0.20 &  13.40 $\pm$ 0.20 & 12.47 $\pm$ 0.20 & -4.38 $\pm$ 0.10& 1450 $\pm$ 230 & 1.0 $\pm$ 0.3 &3\\
2MASS 1728+39B & --- & --- & 19.00 $\pm$ 0.28 & --- & 15.64 $\pm$ 0.22 & 15.00 $\pm$ 0.20 & 13.85 $\pm$ 0.20 & 13.13 $\pm$ 0.20 & -4.60 $\pm$ 0.10 & 1280 $\pm$ 200 & 1.0 $\pm$ 0.3 & 3\\
2MASS 1750+44A & --- & --- & --- & --- & --- & 10.30 $\pm$ 0.71 & 9.72 $\pm$ 0.71 & 9.36 $\pm$ 0.71 & -3.08 $\pm$ 0.29 & 2200 $\pm$ 230 & 1.88$^{+0.72}_{-0.73}$ &6\\ 
2MASS 1750+44B & --- & --- & --- & --- & --- & 11.26 $\pm$ 0.71 & 10.49 $\pm$ 0.72 & 10.03 $\pm$ 0.71 & -3.40 $\pm$ 0.29 & 2020 $\pm$ 215 & 1.62$^{+0.78}_{-0.65}$ &6\\ 
2MASS 1847+55A & --- & --- & --- & --- & --- & 10.19 $\pm$ 0.52 & 9.52 $\pm$ 0.52 & 9.16 $\pm$ 0.52 & -2.98 $\pm$ 0.22 & 2400 $\pm$ 300 & 1.70$^{+0.26}_{-0.28}$ & 6 \\
2MASS 1847+55B & --- & --- & --- & --- & --- & 10.43 $\pm$ 0.53 & 9.81 $\pm$ 0.55 & 9.43 $\pm$ 0.52 & -3.11 $\pm$ 0.22 & 2100 $\pm$ 230 & 1.99 $\pm$ 0.59 &6\\
2MASS 2140+16A & --- & --- & 14.05 $\pm$ 0.89 & --- & 11.79 $\pm$ 0.89 & 11.33 $\pm$ 0.87 & 10.66 $\pm$ 0.87 & 10.28 $\pm$ 0.87 & -3.48 $\pm$ 0.35 & 2300 $\pm$ 80 & 1.13$^{+0.47}_{-0.44}$&3\\
2MASS 2140+16B & --- & --- & 15.56 $\pm$ 0.89 & --- & 13.17 $\pm$ 0.89 & 12.28 $\pm$ 0.88 & 11.59 $\pm$ 0.89 & 11.02 $\pm$ 0.87  & -3.83 $\pm$ 0.35 & 2075 $\pm$ 70 & 0.92$^{+0.39}_{-0.36}$&3\\
2MASS 2206-20A & --- & --- & 13.59 $\pm$ 0.21 & --- & 11.81 $\pm$ 0.21 & 10.92 $\pm$ 0.21 & 10.28 $\pm$ 0.22 & 9.91 $\pm$ 0.21 & -3.32 $\pm$ 0.10 & 2350 $\pm$ 80 & 1.27$^{+0.15}_{-0.14}$&3\\
2MASS 2206-20B & --- & --- & 13.67 $\pm$ 0.21 & --- & 11.83 $\pm$ 0.21 & 11.07 $\pm$ 0.22 & 10.33 $\pm$ 0.24 & 9.98 $\pm$ 0.21 & -3.35 $\pm$ 0.10 & 2250 $\pm$ 80 & 1.30$^{+0.15}_{-0.18}$&3\\
GJ 569Ba & --- & --- & --- & --- & --- & 11.18 $\pm$ 0.08 & 10.47 $\pm$ 0.05 & 9.90 $\pm$ 0.06 & -3.33 $\pm$ 0.07 & 2000 $\pm$ 210 & 1.69 $\pm$ 0.09 & 7 \\
GJ 569Bb & --- & --- & --- & --- & --- & 11.69 $\pm$ 0.08 & 11.08 $\pm$ 0.06 & 10.43 $\pm$ 0.07 & -3.56 $\pm$ 0.07 & 2000 $\pm$ 215 & 1.28 $\pm$ 0.07 & 7 \\
HD 130948B & --- & --- & --- &--- & --- & 12.51 $\pm$ 0.06 & 11.74 $\pm$ 0.10 & 10.96 $\pm$ 0.03 & -3.84 $\pm$ 0.06 & 1840 $\pm$ 65 & 1.09 $\pm$ 0.03 &8 \\
HD 130948C & --- & --- & --- & --- & --- & 12.82 $\pm$ 0.07 & 12.03 $\pm$ 0.11 & 11.16 $\pm$ 0.03 & -3.92 $\pm$ 0.06 & 1790 $\pm$ 65 & 1.02 $\pm$ 0.03 &8\\
LHS 2397aA & --- & --- & 14.29 $\pm$ 0.07 & --- & --- & 11.33 $\pm$ 0.06 & 10.52 $\pm$ 0.07 & 10.04 $\pm$ 0.07 & -3.37 $\pm$ 0.07 & 2180$^{+70}_{-100}$ & 1.28 $\pm$ 0.15 &9 \\
LHS 2397aB & --- & --- & 18.71 $\pm$ 0.18 & --- & --- & 14.45 $\pm$ 0.10 & 13.62 $\pm$ 0.10 & 12.82 $\pm$ 0.07 & -4.50 $\pm$ 0.07 & 1350 $\pm$ 210 & 1.0 $\pm$ 0.3 &9\\
LP 349-25A & --- & --- & --- & --- & --- & 10.53 $\pm$ 0.05 & 9.93 $\pm$ 0.06 & 9.58 $\pm$ 0.06  & -3.19 $\pm$ 0.06 & 2200 $\pm$ 210 &1.70$^{+0.08}_{-0.09}$ &6\\
LP 349-25B & --- & --- & --- & --- & --- & 11.07 $\pm$ 0.07 & 10.35 $\pm$ 0.09 & 9.88 $\pm$ 0.09 & -3.34 $\pm$ 0.07 & 2050 $\pm$ 210 &1.68$^{+0.09}_{-0.08}$ &6\\
LP 415-20A & --- & --- & --- & --- & --- & 11.48 $\pm$ 0.52 & 10.98 $\pm$ 0.52 & 10.64 $\pm$ 0.52  & -3.57 $\pm$ 0.22 & 2300 $\pm$ 230 & 1.00$^{+0.24}_{-0.29}$&6\\
LP 415-20B & --- & --- & --- & --- & --- & 12.32 $\pm$ 0.54 & 11.49 $\pm$ 0.54 & 11.02 $\pm$ 0.53  & -3.80 $\pm$ 0.22 & 2000 $\pm$ 230 & 1.00$^{+0.30}_{-0.25}$&6\\
\enddata
\tablecomments{References for photometric measurements:
  (1) Optical from Bouy et al. (2004), NIR from this work; (2)
  F814W from Bouy et al. (2003), all others from this work;
  (3) Optical from Bouy et al. (2003), NIR from this work; (4)
F814W and F1042M from Bouy et al. (2003), all others from this
work; (5) F814W and J from Liu et al. (2008), F1042M from
Burgasser et al. (2003), all others from this work; (6) All
photometry from this work; (7)
Photometry from Lane et al. (2001) and Simon et al. (2006); (8) Photometry from Dupuy
et al. (2009a); (9) Optical from Freed et al. (2004), J from
Dupuy et al. (2009b), all others from this work}
\label{tab:phot}
\end{deluxetable}
%\end{landscape}
\end{singlespace}

\begin{singlespace}
\begin{deluxetable}{lccccccccc}
\rotate
\tabletypesize{\scriptsize}
\setlength{\tabcolsep}{1.0mm}
\tablecolumns{10}
\tablewidth{0pc}
\tablecaption{Evolutionary Model Predictions}
\tablehead{
\colhead{Target} &  
\colhead{M$_{Primary}$} & \colhead{M$_{Secondary}$} &
\colhead{M$_{Total}$} & \colhead{M$_{Primary}$} &
\colhead{M$_{Secondary}$} & \colhead{M$_{Total}$}&  \colhead{M$_{Primary}$} &
\colhead{M$_{Secondary}$} & \colhead{M$_{Total}$}\\
\colhead{Name} & 
\colhead{Tucson (M$_{\odot}$)} & \colhead{(Tucscon (M$_{\odot}$)} &
\colhead{Tucscon (M$_{\odot}$)} & \colhead{DUSTY (M$_{\odot}$)} &
\colhead{DUSTY (M$_{\odot}$)} & \colhead{DUSTY (M$_{\odot}$)}
& \colhead{COND (M$_{\odot}$)} &
\colhead{COND (M$_{\odot}$)} & \colhead{COND (M$_{\odot}$)}
}
\startdata
2MASS 0746+20AB & 0.050 $\pm$ 0.01 & 0.050 $\pm$ 0.01 & 0.10 $\pm$ 0.01 & 0.07 $\pm$ 0.01& 0.06$\pm$0.01 & 0.13 $\pm$ 0.01 & --- & --- & ---\\
2MASS 0850+10AB & 0.04 $\pm$ 0.04 & 0.04 $\pm$ 0.04 & 0.08 $\pm$ 0.06 & 0.04$^{+0.04}_{-0.03}$ & 0.04$^{+0.04}_{-0.03}$ & 0.08$^{+0.06}_{-0.04}$ & 0.03$^{+0.05}_{-0.03}$ & 0.03$^{+0.05}_{-0.03}$& 0.06$^{+0.07}_{-0.04}$\\
2MASS 0920+35AB & 0.04 $\pm$ 0.04& 0.03$^{+0.05}_{-0.03}$ & 0.07$^{+0.06}_{-0.05}$ & 0.04$^{+0.04}_{-0.03}$ & 0.03$^{+0.05}_{-0.03}$ & 0.07$^{+0.06}_{-0.04}$ & 0.03$^{+0.05}_{-0.03}$& 0.03$^{+0.05}_{-0.03}$& 0.06$^{+0.07}_{-0.04}$\\
2MASS 1426+15AB & 0.03$^{+0.04}_{-0.02}$ & 0.04$^{+0.04}_{-0.03}$ & 0.07$^{+0.06}_{-0.04}$ & 0.04$^{+0.04}_{-0.02}$ & 0.05 $\pm$ 0.03 & 0.09$^{+0.05}_{-0.04}$ & --- & --- & --- \\
2MASS 1534-29AB & 0.06 $\pm$ 0.02 & 0.06$^{+0.03}_{-0.02}$ & 0.12$^{+0.04}_{-0.03}$  &--- & --- & ---& 0.04$^{+0.04}_{-0.01}$ & 0.04$^{+0.04}_{-0.01}$ & 0.08$^{+0.06}_{-0.01}$ \\
2MASS 1728+39AB & 0.04 $\pm$ 0.04 & 0.03$^{+0.05}_{-0.03}$ & 0.07$^{+0.06}_{-0.05}$ & 0.04 $\pm$ 0.04& 0.04 $\pm$ 0.04& 0.08 $\pm$ 0.06& 0.03$^{+0.05}_{-0.03}$& 0.03$^{+0.04}_{-0.03}$ & 0.06$^{+0.06}_{-0.04}$\\
2MASS 1750+44AB & 0.01$^{+0.02}_{-0.01}$ & 0.01$^{+0.02}_{-0.01}$& 0.02$^{+0.03}_{-0.01}$ & 0.02 $\pm$ 0.02 & 0.02 $\pm$ 0.02 & 0.04$\pm$ 0.03& --- & --- & --- \\
2MASS 1847+55AB & 0.02$^{+0.03}_{-0.02}$ & 0.01 $\pm$ 0.01 & 0.03$^{+0.03}_{-0.02}$ & 0.03$^{+0.05}_{-0.02}$ & 0.01$^{+0.02}_{-0.01}$ & 0.04$^{+0.05}_{-0.02}$ & --- & --- & --- \\
2MASS 2140+16AB & 0.04$^{+0.05}_{-0.03}$ & 0.07$^{+0.03}_{-0.05}$ & 0.11 $\pm$ 0.06 & 0.06 $\pm$ 0.04 & 0.07$^{+0.03}_{-0.04}$ & 0.13$^{+0.05}_{-0.06}$ & --- & --- & --- \\
2MASS 2206-20AB & 0.032 $\pm$ 0.010 & 0.026$^{+0.007}_{-0.010}$ & 0.058$^{+0.012}_{-0.014}$ & 0.047$^{+0.016}_{-0.012}$ & 0.037$^{+0.011}_{-0.009}$ & 0.084$^{+0.019}_{-0.015}$ & --- & --- & --- \\
GJ569B ab       & 0.01 $\pm$ 0.01 & 0.02 $\pm$ 0.02 & 0.03$\pm$ 0.02 & 0.02$^{+0.01}_{-0.02}$  & 0.03$^{+0.03}_{-0.02}$ & 0.05$\pm$ 0.03 & --- & --- & --- \\
HD 130948BC     & 0.030 $\pm$ 0.010 & 0.032 $\pm$ 0.010 & 0.062 $\pm$ 0.014 & 0.035 $\pm$ 0.010 & 0.037$^{+0.013}_{-0.010}$ & 0.072$^{+0.016}_{-0.014}$ & --- & --- & --- \\
LHS2397a AB     & 0.02 $\pm$ 0.01 & 0.04 $\pm$ 0.04& 0.06 $\pm$ 0.04 & 0.03 $\pm$ 0.01 & 0.04 $\pm$ 0.04 & 0.07 $\pm$ 0.04 & --- & 0.03$^{+0.04}_{-0.03}$ & 0.06$^{+0.04}_{-0.03}$\tablenotemark{a}\\
LP 349-25AB     & 0.01$^{+0.02}_{-0.01}$ & 0.01 $\pm$ 0.01 & 0.02$^{+0.02}_{-0.01}$ & 0.02 $\pm$ 0.02 & 0.02 $\pm$ 0.02 & 0.04 $\pm$ 0.03& --- & --- & --- \\
LP 415-20AB     & 0.06 $\pm$ 0.04 & 0.05$^{+0.04}_{-0.03}$ & 0.11$^{+0.06}_{-0.05}$ & 0.08$^{+0.02}_{-0.05}$ & 0.06$^{+0.03}_{-0.04}$ & 0.14$^{+0.04}_{-0.06}$ & --- & --- & --- \\
\enddata
\tablenotetext{a}{Total mass found by adding DUSTY prediction
  for primary to COND prediction for secondary}
\label{tab:models}
\end{deluxetable}
\end{singlespace}

%%figures

\begin{figure}% 
\centering 
\subfloat[][]{\label{fig:2mass07464_orb}\includegraphics[angle=0,width=\linewidth]{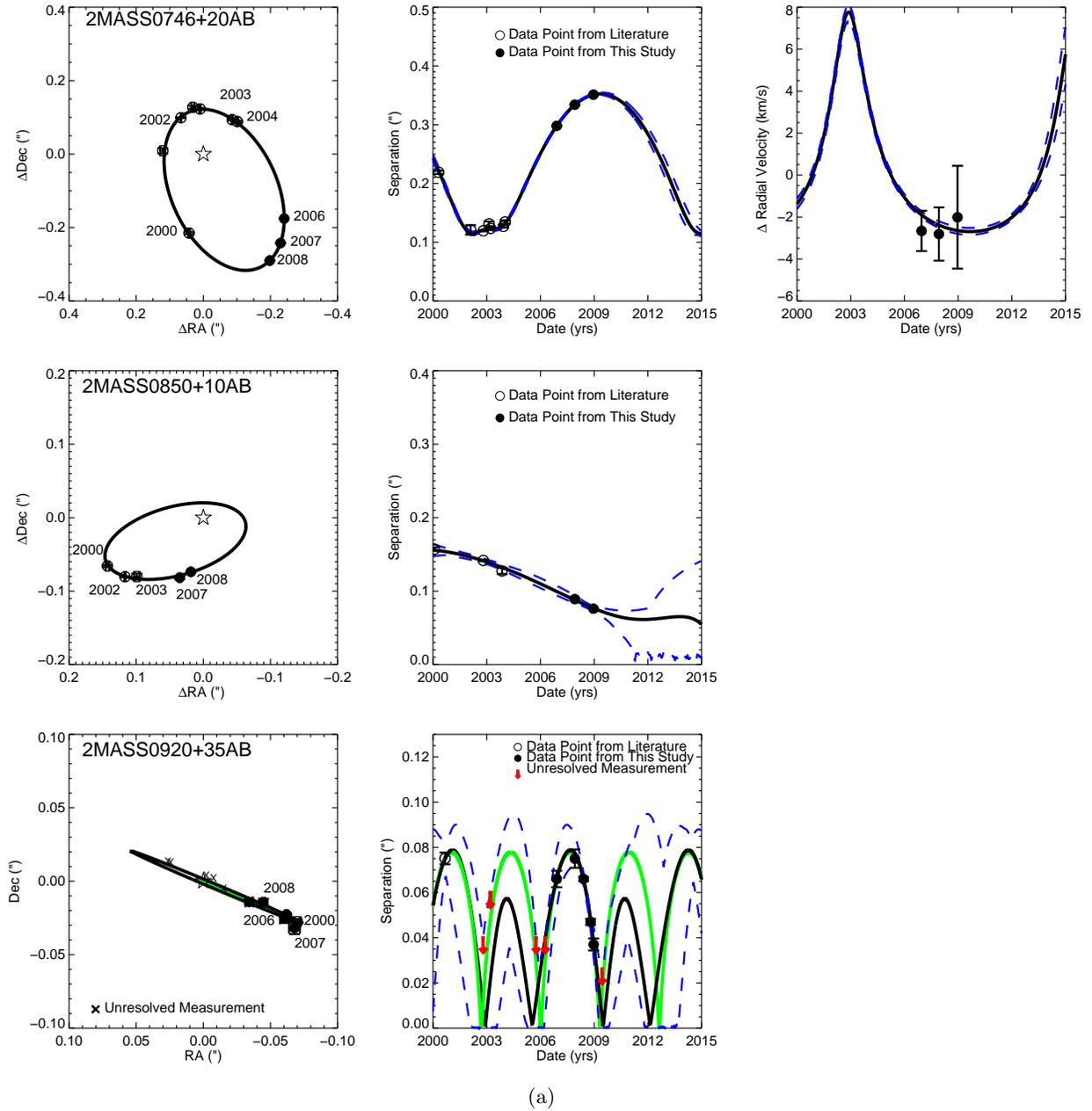}}
\caption{Best
  fit relative orbit for 2MASS0746+20AB (top), 2MASS 0850+10
  AB (middle) and 2MASS 0920+35AB (bottom).  The left panel shows
the relative astrometry data points overplotted with the best
fit orbit.  The middle panel shows separation of the
components as a function of time overplotted with the best fit
orbit.  Finally, the right hand panel shows the relative
radial velocity measurements as a function of time overplotted
with the best fit orbit.  The blue dotted lines represent
the 1$\sigma$ allowed range of separations and relative radial
velocities at a given time.  Astrometric data from the literature
is from Reid et al. (2001), Bouy et al. (2004), and Buoy et al
(2008 - unresolved data points for 2MASS 0920+35AB).  For 2MASS
0920+35 (bottom), the black line shows the best fit
  orbital solution (period $\sim$6.7 years), while the green
  line shows the other 
  allowed solution which has a very short period ($\sim$3.3
  years) and a high eccentricity.  The unresolved measurements
from Bouy et al. (2008) are used to throw out solutions that
do not lead to the binary being unresolved on those dates (Xs
and arrows).}% 
\label{fig:orbfig}% 
\end{figure} 

\begin{figure}% 
\ContinuedFloat 
\centering 
\subfloat[][]{\label{fig:2mass14263_orb}\includegraphics[angle=0,width=\linewidth]{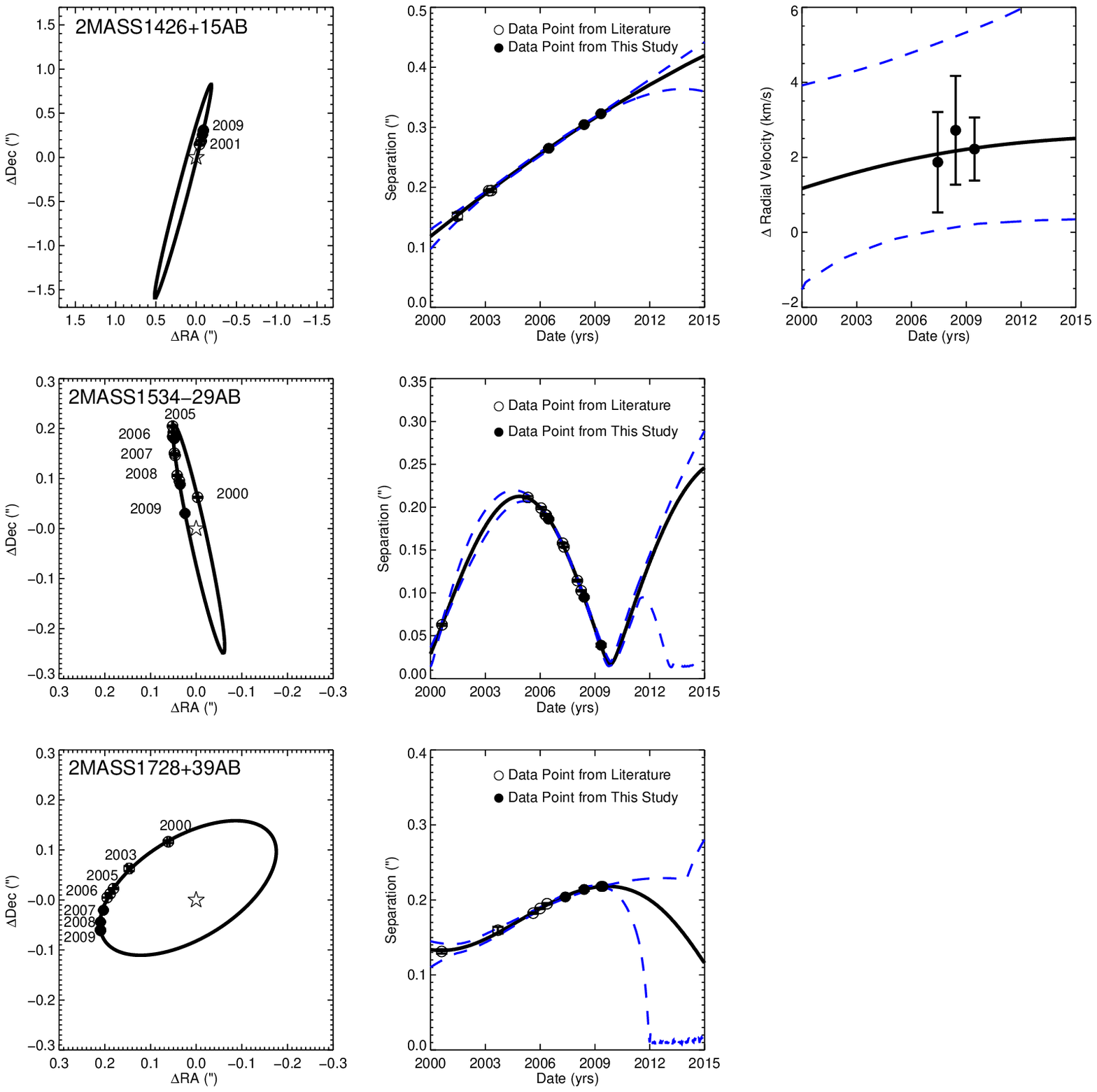}}
\caption[]{The same as Figure \ref{fig:2mass07464_orb} for 2MASS1426+15AB (top),
  2MASS 1534-29AB (middle), and 2MASS1728+39AB.  Astrometric data from the literature
is from Close et al. (2002), Bouy et al. (2003), Burgasser et
al. (2003), Bouy et al. (2008), and Liu et al. (2008).}% 
\label{fig:orbfig}% 
\end{figure} 

\begin{figure}% 
\ContinuedFloat 
\centering 
\subfloat[][]{\label{fig:2mass17501_orb}\includegraphics[angle=0,width=\linewidth]{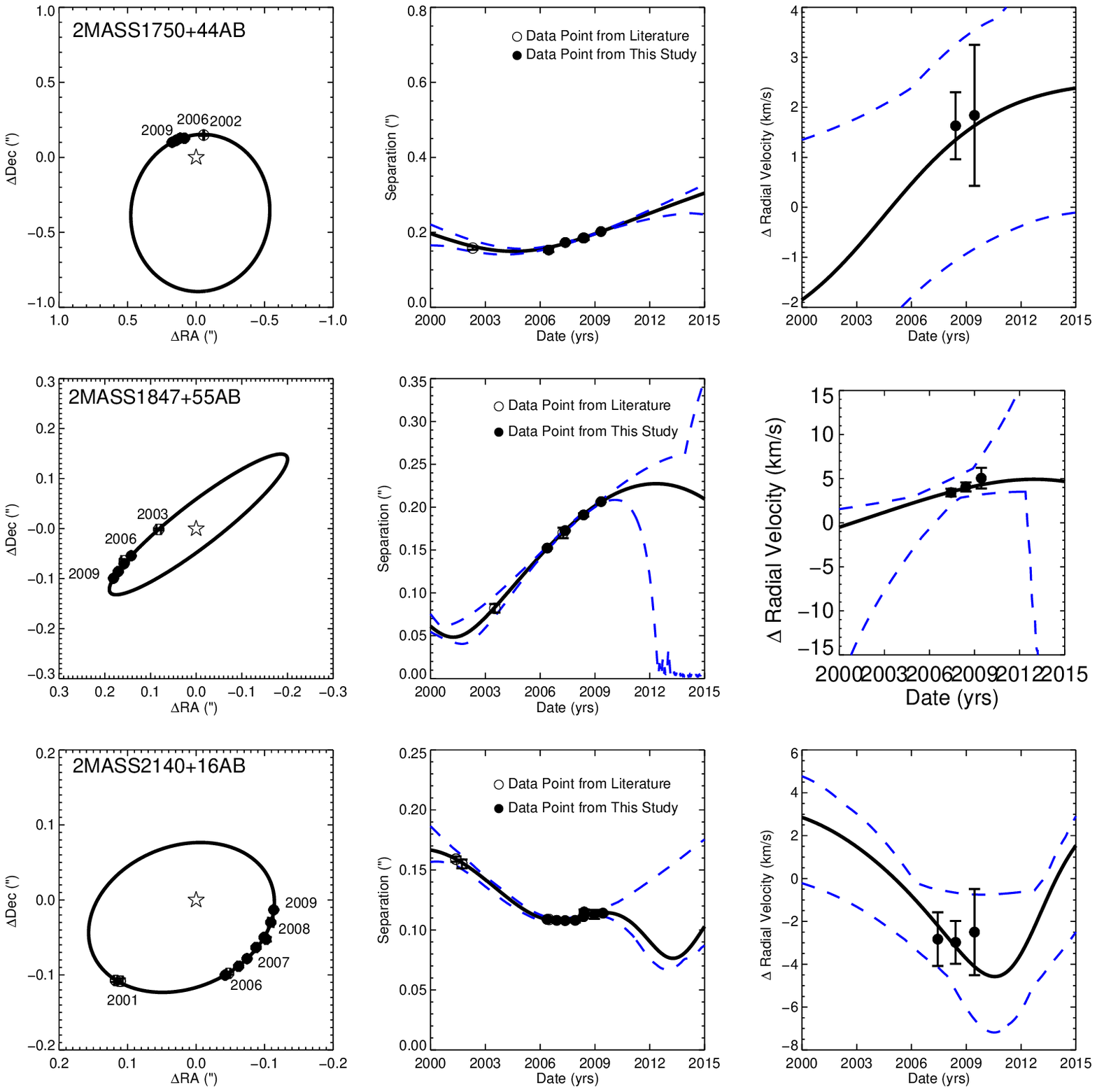}}
\caption[]{The same as Figure \ref{fig:2mass07464_orb} for 2MASS1750+44AB (top),
  2MASS1847+55AB (middle), and 2MASS2140+16AB.  Astrometric
  data from the literature 
is from Bouy et al. (2003), Close et al. (2003), Siegler et
al. (2003), Siegler et al. (2005), and Bouy et al. (2008)}% 
\label{fig:orbfig}% 
\end{figure} 

\begin{figure}% 
\ContinuedFloat 
\centering 
\subfloat[][]{\label{fig:2mass22062_orb}\includegraphics[angle=0,width=\linewidth]{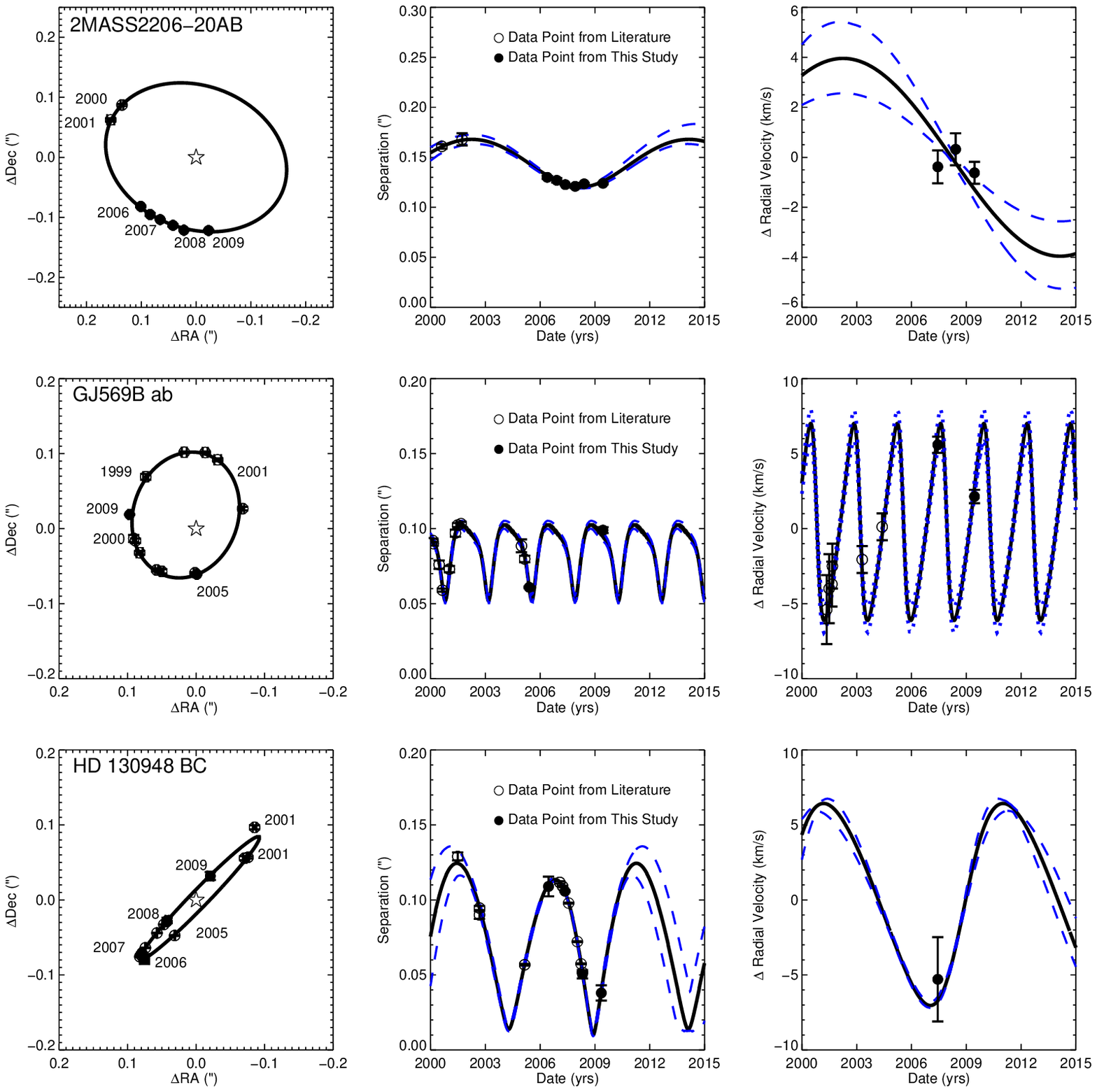}}
\caption[]{The same as Figure \ref{fig:2mass07464_orb} for 2MASS2206-20AB (top), GJ 569Bab
  (middle) and HD 130948BC (bottom).  Astrometric and radial
  velocity data from the literature 
is from Close et al. (2002), Potter et al. (2002), Bouy et
al. (2003), Zapatero Osorio et al. (2004), Simon et
al. (2006), and Dupuy et al. (2009a).}% 
\label{fig:orbfig}% 
\end{figure} 

\begin{figure}% 
\ContinuedFloat 
\centering 
\subfloat[][]{\label{fig:lhs2397a_orb}\includegraphics[angle=0,width=\linewidth]{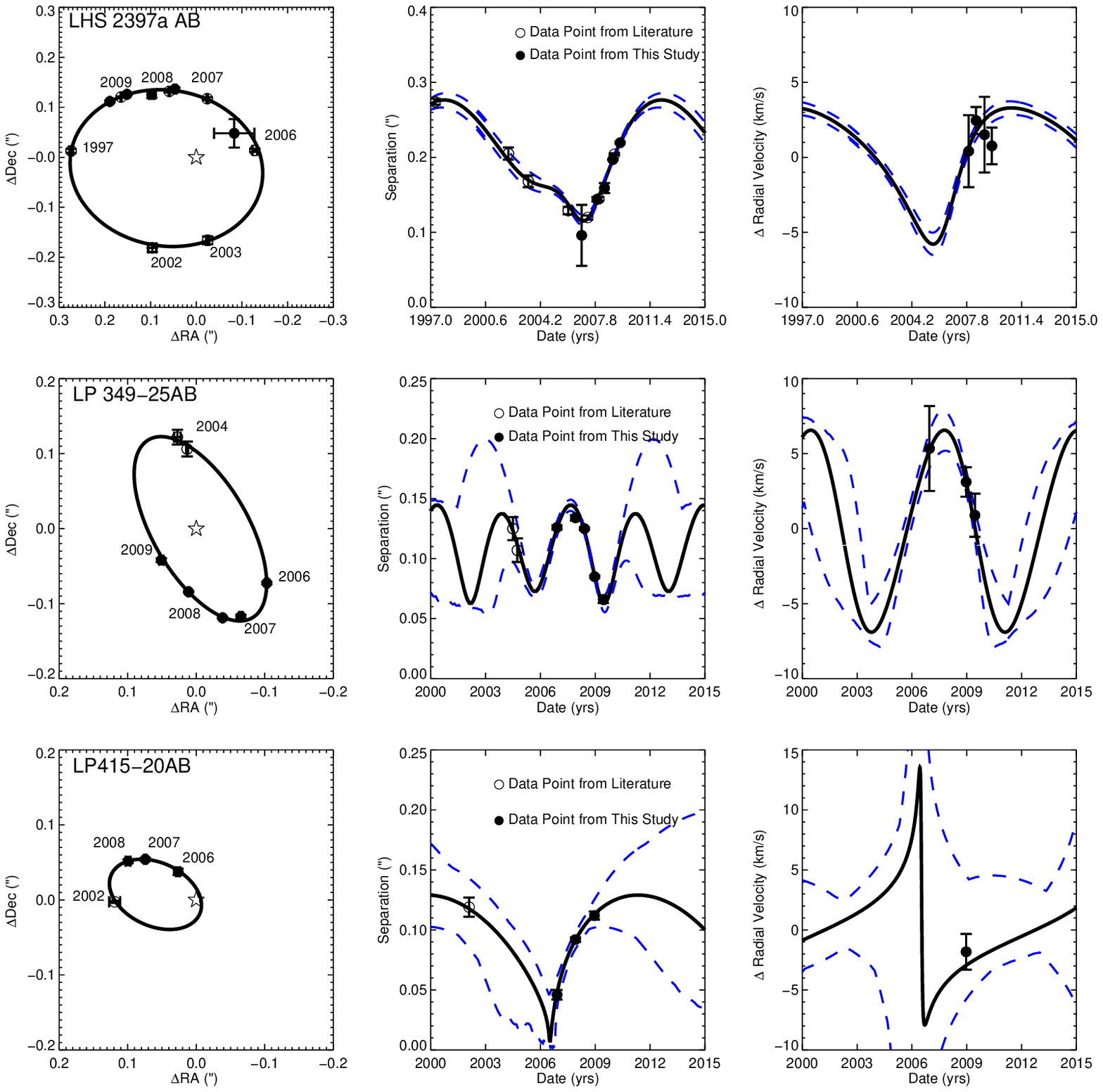}}
\caption[]{The same as Figure \ref{fig:2mass07464_orb} for LHS 2397a AB (top), LP 349-25AB
  (middle) and LP 415-20AB (bottom).  Astrometric data from the literature
taken from Freed et al. (2003), Forveille et al. (2005),
Siegler et al. (2005), and Dupuy et al. (2009b). }% 
\label{fig:orbfig}% 
\end{figure}

\clearpage

\begin{figure}% 
\centering 
\subfloat[][]{\label{fig:2mass0746_astrhist}\includegraphics[width=1.0\textwidth]{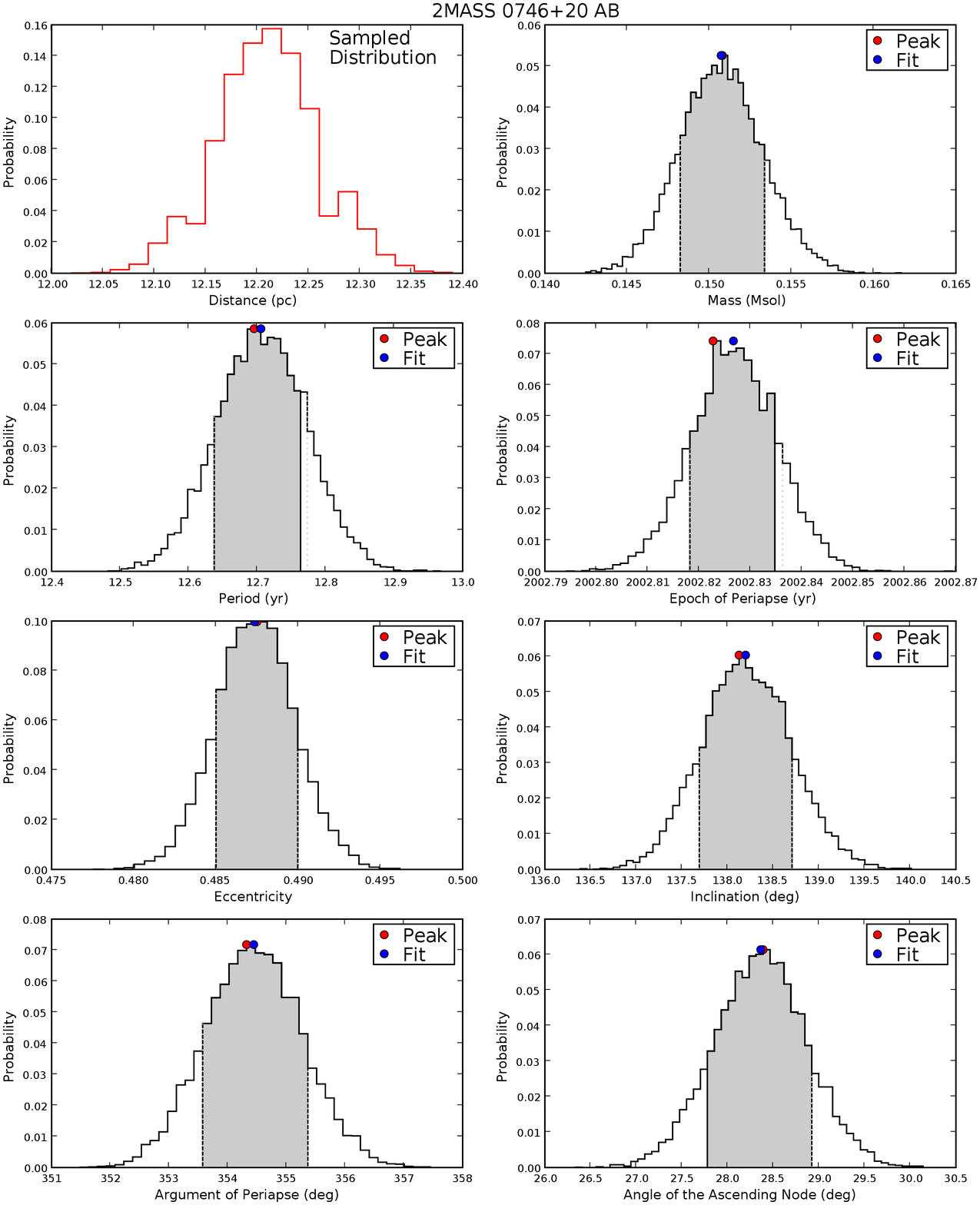}}
\caption{One-dimensional PDFs for the relative orbit (total
  system mass) of 2MASS 0746+20AB.  This is an example of a
  typical system with a well-measured mass and a distance
  sample from a parallax measurement.}% 
\label{fig:astrhist}% 
\end{figure} 

\begin{figure}% 
\ContinuedFloat 
\centering 
\subfloat[][]{\label{fig:2mass09201_astrhist}\includegraphics[width=1.0\textwidth]{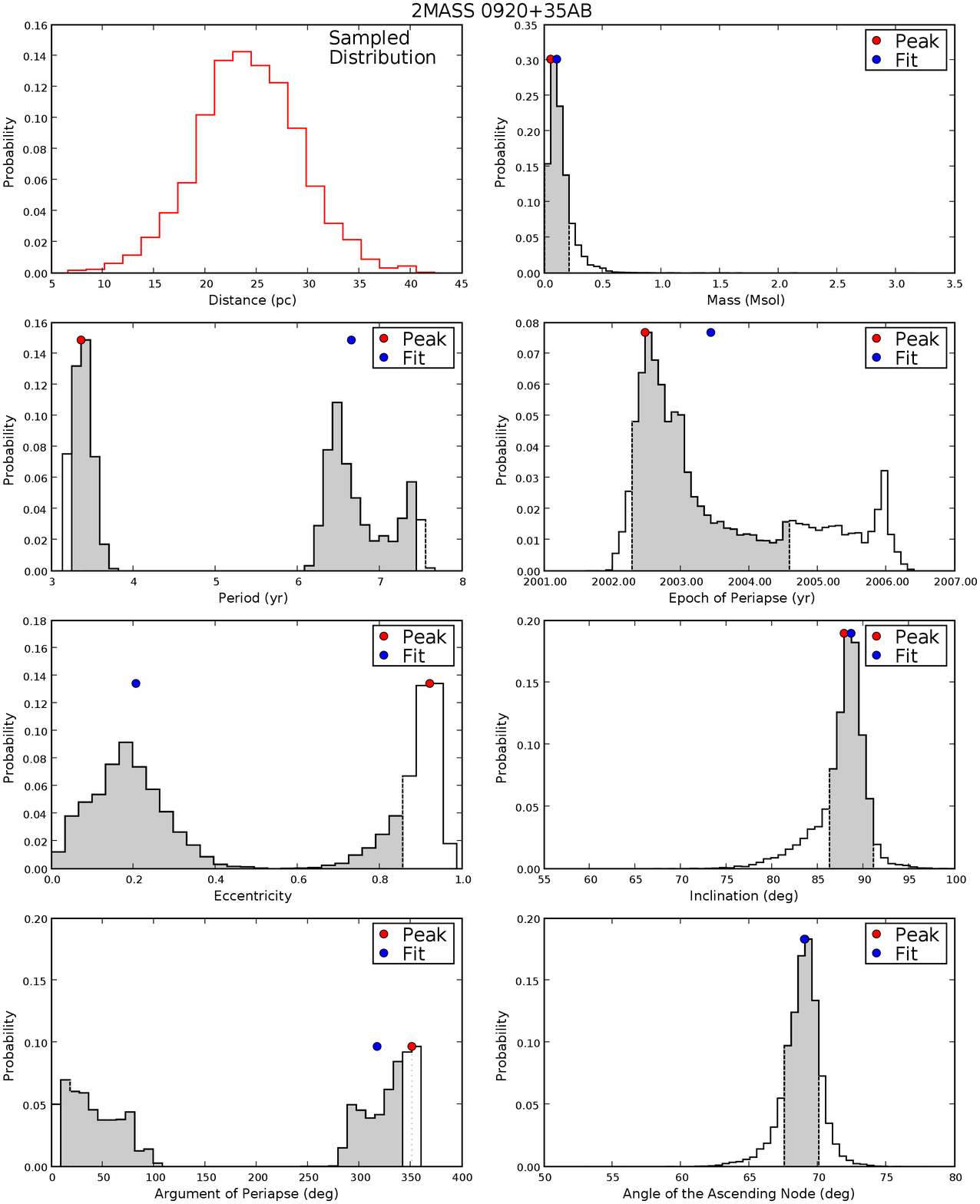}}
\caption[]{One-dimensional PDFs for the relative orbit (total
  system mass) of 2MASS 0920+35AB.  A set of solutions exists
  with a period of $\sim$3.5 years and very high
  eccentricities, making the distributions of period, e and
  $\omega$ strongly bifurcated.  We obtain the uncertainties
  on each parameter as in Ghez et al. (2008), where the
  distribution of each parameter is marginalized against all
  others and confidence limits are determined by integrating
  the resulting one‐dimensional distribution out to a
  probability of 34$\%$ on each side of the best fitting value.}% 
\label{fig:astrhist}% 
\end{figure} 

\begin{figure}% 
\ContinuedFloat 
\centering 
\subfloat[][]{\label{fig:2mass21402_astrhist}\includegraphics[width=1.0\textwidth]{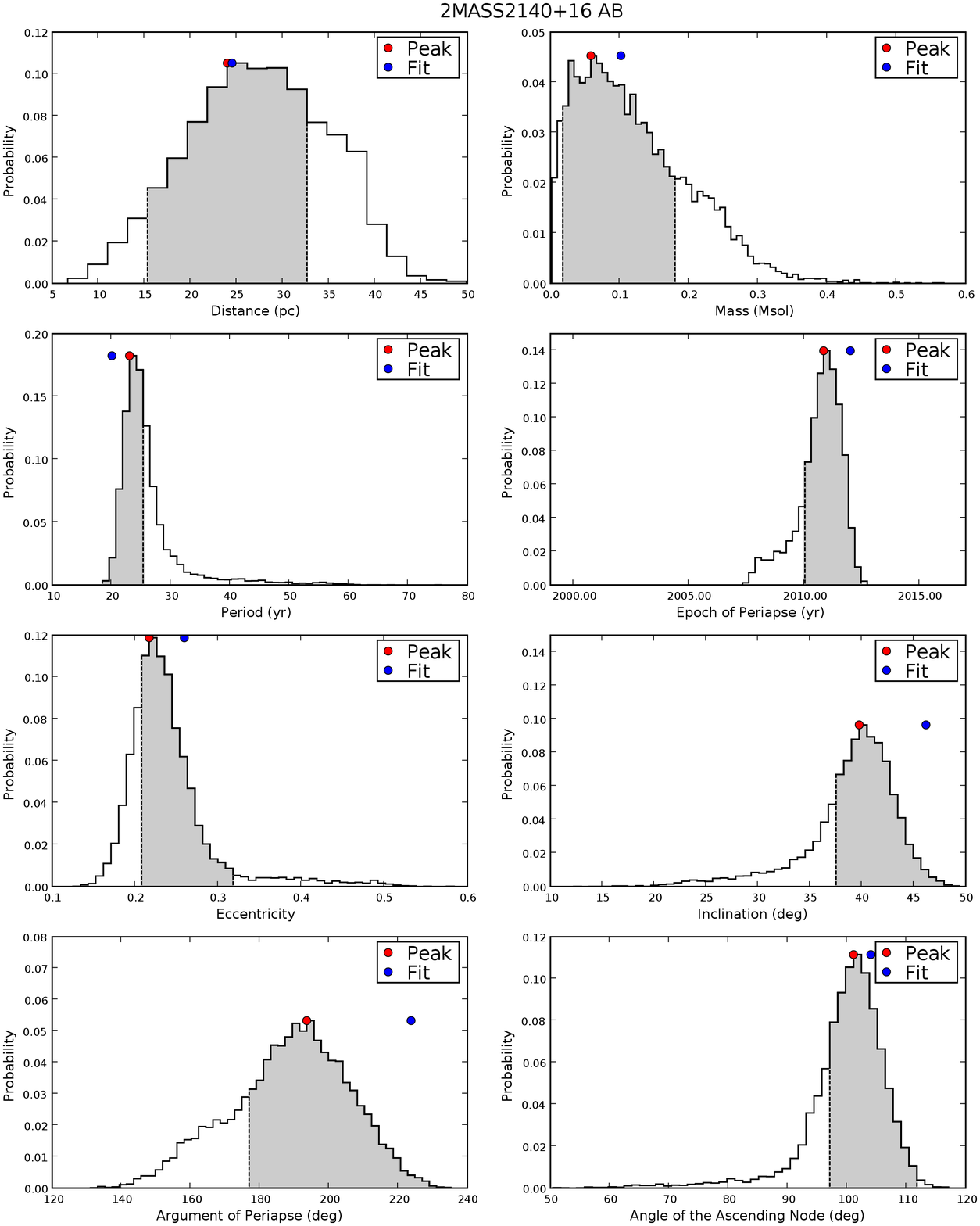}}
\caption[]{One-dimensional PDFs for the relative orbit (total
  system mass) of 2MASS 2140+16AB.  This is an example of a
  system for which we fit for distance using our relative
  radial velocities.  An extended version of this figure is
  shown online.}% 
\label{fig:astrhist}% 
\end{figure}

\clearpage

\begin{figure*}
\epsscale{1.0}
\plotone{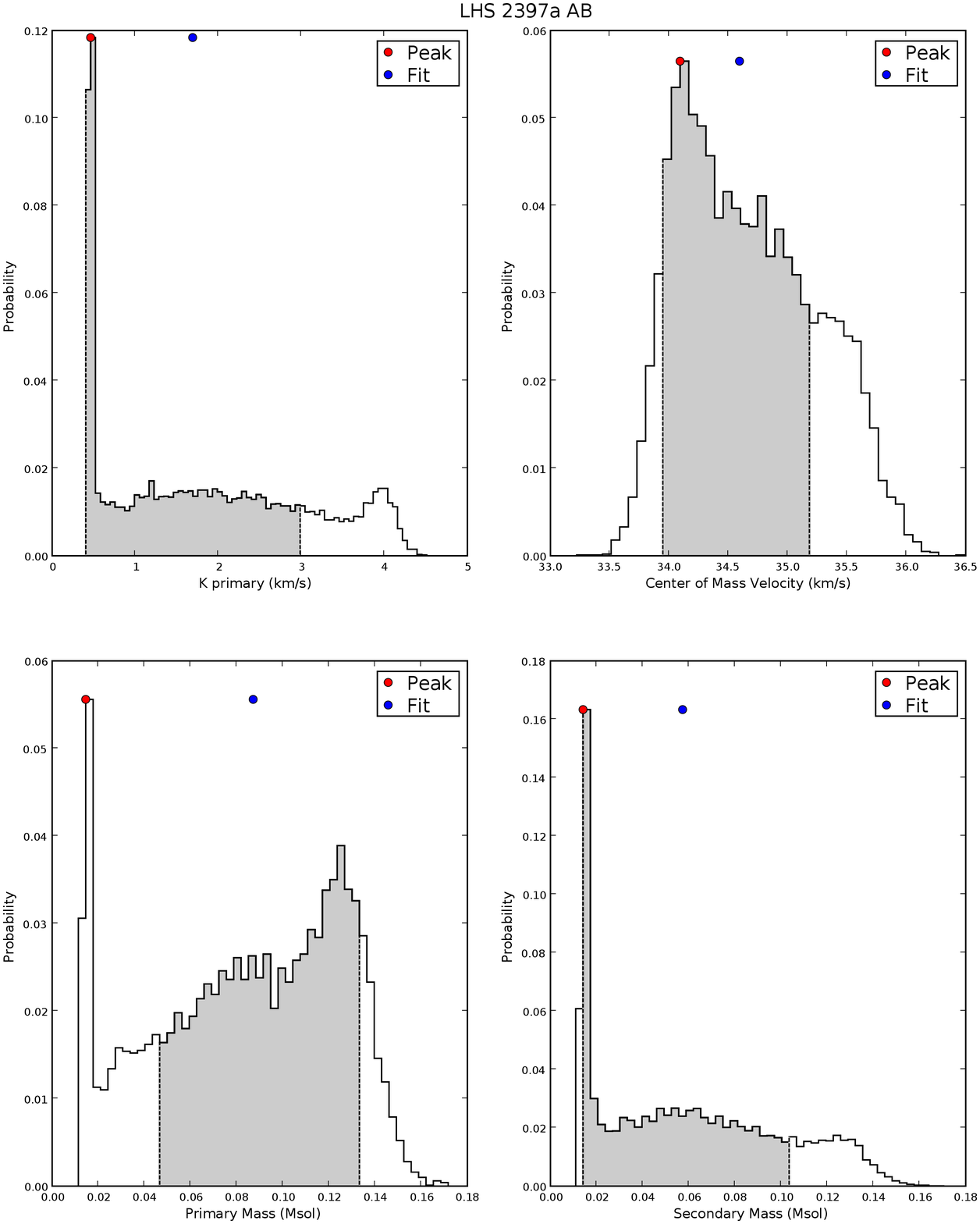}
\caption{One-dimensional PDFs for the absolute orbit
  of LHS 2397a AB.  Fit parameters are K$_{Primary}$ and
  $\gamma$ (top panels).  The distributions for parameters in common
  between this orbit and the relative orbit, namely P, e,
  T$_{o}$, and $\omega$, are shown above in Figure
  \ref{fig:2mass0746_astrhist} online.  From K$_{Primary}$ and $\gamma$,
  K$_{Secondary}$ is calculated, giving the mass ratio, which
  we use in conjunction with the total system mass to derive
  component masses (bottom panels).  PDFs for the other 5
  systems with absolute orbit derivations are shown online.}
\label{fig:lhs2397a_spechist}
\end{figure*}

\clearpage

\clearpage

\clearpage

\clearpage

\clearpage

\begin{thebibliography}{}

\bibitem[Allard et al.(1997)]{allard97} Allard, F., Hauschildt, P.~H., Alexander, D.~R., \& Starrfield, S.\ 1997, \araa, 35, 137 

\bibitem[Allard et al.(2001)]{allard01} Allard, F., Hauschildt, P.~H., Alexander, D.~R., Tamanai, A., \& Schweitzer, A.\ 2001, \apj, 556, 357 

\bibitem[Bailey et al. (2010)]{bailey09} Bailey, J.I., III, et al. 2010, in prep

\bibitem[Baraffe et al.(2009)]{2009ApJ...702L..27B} Baraffe, I., Chabrier, G., \& Gallardo, J.\ 2009, \apjl, 702, L27 

\bibitem[Berger et al.(2009)]{berger09} Berger, E., et al.\ 2009, \apj, 695, 310 

\bibitem[Blake et al.(2007)]{blake07} Blake, C.~H., Charbonneau, D., White, R.~J., Marley, M.~S., \& Saumon, D.\ 2007, \apj, 666, 1198 

\bibitem[Bouy et al.(2003)]{bouy03} Bouy, H., Brandner, W., Mart{\'{\i}}n, E.~L., Delfosse, X., Allard, F., \& Basri, G.\ 2003, \aj, 126, 1526 

\bibitem[Bouy et al.(2004)]{bouy04} Bouy, H., et al.\ 2004, A\&A, 423, 341 

\bibitem[Bouy et al.(2008)]{bouy08} Bouy, H., et al.\ 2008, A\&A, 481, 757 

\bibitem[Browning(2008)]{browning08} Browning, M.~K.\ 2008, \apj, 676, 1262 

\bibitem[Burgasser et al.(2003)]{burg03} Burgasser, A.~J., Kirkpatrick, J.~D., Reid, I.~N., Brown, M.~E., Miskey, C.~L., \& Gizis, J.~E.\ 2003, \apj, 586, 512 

\bibitem[Burgasser et al.(2006)]{burg06} Burgasser, A.~J., Kirkpatrick, J.~D., Cruz, K.~L., Reid, I.~N., Leggett, S.~K., Liebert, J., Burrows, A., \& Brown, M.~E.\ 2006, \apjs, 166, 585 

\bibitem[Burgasser et al.(2007)]{burg07} Burgasser, A.~J., Reid, I.~N., Siegler, N., Close, L., Allen, P., Lowrance, P., \& Gizis, J.\ 2007, Protostars and Planets V, 427 

\bibitem[Burrows et al.(1997)]{burrows97} Burrows, A., et al.\ 1997, \apj, 491, 856 

\bibitem[Burrows et al.(2006)]{burrows06} Burrows, A., Sudarsky, D., \& Hubeny, I.\ 2006, \apj, 640, 1063 

\bibitem[Chabrier \& Baraffe(1997)]{cb97} Chabrier, G., \& Baraffe, I.\ 1997, A\&A, 327, 1039 

\bibitem[Chabrier \& Baraffe(2000)]{cb00} Chabrier, G., \& Baraffe, I.\ 2000, \araa, 38, 337 

\bibitem[Chabrier et al.(2000)]{chabrier00} Chabrier, G., Baraffe, I., Allard, F., \& Hauschildt, P.\ 2000, \apj, 542, 464 

\bibitem[Chabrier \& Kuker(2006)]{chabrier06} Chabrier, G. \& Kuker, M.\ 2006, \aap, 446, 1027 

\bibitem[Close et al.(2003)]{close03} Close, L.~M., Siegler, N., Freed, M., \& Biller, B.\ 2003, \apj, 587, 407 

\bibitem[Close et al.(2002)]{close02} Close, L.~M., Siegler, N., Potter, D., Brandner, W., \& Liebert, J.\ 2002, \apjl, 567, L53 

\bibitem[Collins et al.(1998)]{collins98} Collins, G.~W., et al.\ 1998, Science, 281, 1178 

\bibitem[Cruz et al.(2003)]{cruz03} Cruz, K.~L., Reid, I.~N., Liebert, J., Kirkpatrick, J.~D., \& Lowrance, P.~J.\ 2003, \aj, 126, 2421 

\bibitem[Cutri et al.(2003)]{cutri03} Cutri, R.~M., et al.\ 2003, The IRSA 2MASS All-Sky Point Source Catalog, NASA/IPAC Infrared Science Archive.~http://irsa.ipac.caltech.edu/applications/Gator/

\bibitem[Cushing et al.(2008)]{cushing08} Cushing, M.~C., et al.\ 2008, \apj, 678, 1372 

\bibitem[Dahn et al.(2002)]{dahn02} Dahn, C.~C., et al.\ 2002, \aj, 124, 1170 

\bibitem[Diolaiti et al.(2000)]{diolaiti00} Diolaiti, E., Bendinelli, O., Bonaccini, D., Close, L.~M., Currie, D.~G., \& Parmeggiani, G.\ 2000, \procspie, 4007, 879 

\bibitem[Dupuy et al.(2009a)]{dupuy09a} Dupuy, T.~J., Liu, M.~C., \& Ireland, M.~J.\ 2009, \apj, 692, 729 

\bibitem[Dupuy et al.(2009b)]{dupuy09b} Dupuy, T.~J., Liu, M.~C., \& Ireland, M.~J.\ 2009, \apj, 699, 168 

\bibitem[Duquennoy \& Mayor(1991)]{duq91} Duquennoy, A., \& Mayor, M.\ 1991, A\&A, 248, 485 

\bibitem[Fabrycky \& Murray-Clay(2008)]{fab08} Fabrycky, D.~C. \& Murray-Clay, R.~A.\ 2008, arXiv:0812.0011v1 [astro-ph]

\bibitem[Figer et al.(2003)]{figer03} Figer, D.~F., et al.\ 2003, \apj, 599, 1139 

\bibitem[Forveille et al.(2005)]{forveille05} Forveille, T., et al.\ 2005, A\&A, 435, L5 

\bibitem[Freed et al.(2003)]{freed03} Freed, M., Close, L.~M., \& Siegler, N.\ 2003, \apj, 584, 453 

\bibitem[Gaidos(1998)]{gaidos98} Gaidos, E.~J.\ 1998, \pasp, 110, 1259 

\bibitem[Ghez et al.(2008)]{ghez08} Ghez, A.~M., et al.\ 2008, \apj, 689, 1044 

\bibitem[Gizis \& Reid(2006)]{gizis06} Gizis, J.~E., \& Reid, I.~N.\ 2006, \aj, 131, 638 

\bibitem[Gizis et al.(2003)]{gizis03} Gizis, J.~E., Reid, I.~N., Knapp, G.~R., Liebert, J., Kirkpatrick, J.~D., Koerner, D.~W., \& Burgasser, A.~J.\ 2003, \aj, 125, 3302 

\bibitem[Golimowski et al.(2004)]{gol04} Golimowski, D.~A., et al.\ 2004, \aj, 127, 3516 

\bibitem[Go{\'z}dziewski \& Migaszewski(2009)]{2009MNRAS.397L..16G} Go{\'z}dziewski, K., \& Migaszewski, C.\ 2009, \mnras, 397, L16 

\bibitem[Hauschildt et al.(1999)]{hauschildt99} Hauschildt, P.~H., Allard, F., \& Baron, E.\ 1999, \apj, 512, 377 

\bibitem[Helling et al.(2008)]{helling08} Helling, C., et al.\ 2008, \mnras, 391, 1854 

\bibitem[Hilditch(2001)]{hilditch01} Hilditch, R.~W.\ 2001, An Introduction to Close Binary Stars (Cambridge: Cambridge University Press)

\bibitem[Kirkpatrick(2005)]{kirkpatrick05} Kirkpatrick, J.~D.\ 2005, \araa, 43, 195 

\bibitem[Konopacky et al.(2007)]{kono07a} Konopacky, Q.~M., Ghez, A.~M., Duch{\^e}ne, G., McCabe, C., \& Macintosh, B.~A.\ 2007a, \aj, 133, 2008 

\bibitem[Lane et al.(2001)]{lane01} Lane, B.~F., Zapatero Osorio, M.~R., Britton, M.~C., Mart{\'{\i}}n, E.~L., \& Kulkarni, S.~R.\ 2001, \apj, 560, 390 

\bibitem[Leggett et al.(2002)]{leggett02} Leggett, S.~K., et al.\ 2002, \apj, 564, 452 

\bibitem[Liu et al.(2008)]{liu08} Liu, M.~C., Dupuy, T.~J., \& Ireland, M.~J.\ 2008, \apj, 689, 436 

\bibitem[Livingston \& Wallace(1991)]{livingston91} Livingston, W., \& Wallace, L.\ 1991, NSO Technical Report, Tucson: National Solar Observatory, National Optical Astronomy Observatory, 1991,  

\bibitem[Lubow \& Artymowicz(1992)]{lubow92} Lubow, S.~H., \& Artymowicz, P.\ 1992, Workshop on Binaries as Tracers of Star Formation, p.~145 - 154, 145 

\bibitem[Luhman et al.(2003)]{luhman03} Luhman, K.~L., Stauffer, J.~R., Muench, A.~A., Rieke, G.~H., Lada, E.~A., Bouvier, J., \& Lada, C.~J.\ 2003, \apj, 593, 1093 

\bibitem[Marois et al.(2008)]{marois08} Marois, C., Macintosh, B., Barman, T., Zuckerman, B., Song, I., Patience, J., Lafreni{\`e}re, D., \& Doyon, R.\ 2008, Science, 322, 1348 

\bibitem[Mart{\'{\i}}n et al.(2000)]{martin00} Mart{\'{\i}}n, E.~L., Koresko, C.~D., Kulkarni, S.~R., Lane, B.~F., \& Wizinowich, P.~L.\ 2000, \apjl, 529, L37 

\bibitem[McGovern et al.(2004)]{mcgovern04} McGovern, M.~R., Kirkpatrick, J.~D., McLean, I.~S., Burgasser, A.~J., Prato, L., \& Lowrance, P.~J.\ 2004, \apj, 600, 1020 

\bibitem[McLean et al.(2000)]{mclean00} McLean, I.~S., Graham, J.~R., Becklin, E.~E., Figer, D.~F., Larkin, J.~E., Levenson, N.~A., \& Teplitz, H.~I.\ 2000, \procspie, 4008, 1048 

\bibitem[Morales et al.(2008)]{morales08} Morales, J.~C., Ribas, I., \& Jordi, C.\ 2008, \aap, 478, 507 

\bibitem[Potter et al.(2002)]{potter02} Potter, D., Mart{\'{\i}}n, E.~L., Cushing, M.~C., Baudoz, P., Brandner, W., Guyon, O., \& Neuh{\"a}user, R.\ 2002, \apjl, 567, L133 

\bibitem[Reid et al.(2001)]{reid01} Reid, I.~N., Gizis, J.~E., Kirkpatrick, J.~D., \& Koerner, D.~W.\ 2001, \aj, 121, 489 

\bibitem[Reid et al.(2006)]{reid06} Reid, I.~N., Lewitus, E., Allen, P.~R., Cruz, K.~L., \& Burgasser, A.~J.\ 2006, \aj, 132, 891 

\bibitem[Rice et al.(2009)]{rice09} Rice, E.~L., Barman,  T.~S., McLean, I.~S., Prato, L., \& Kirkpatrick, J.D.\ 2009,  \apj, in press, arXiv:0911.3844

\bibitem[Saumon et al.(1995)]{saumon95} Saumon, D., Chabrier, G., \& van Horn, H.~M.\ 1995, \apjs, 99, 713 

\bibitem[Siegler et al.(2005)]{siegler05} Siegler, N., Close, L.~M., Cruz, K.~L., Mart{\'{\i}}n, E.~L., \& Reid, I.~N.\ 2005, \apj, 621, 1023 

\bibitem[Siegler et al.(2003)]{siegler03} Siegler, N., Close, L.~M., Mamajek, E.~E., \& Freed, M.\ 2003, \apj, 598, 1265 

\bibitem[Simon et al.(2006)]{simon06} Simon, M., Bender, C., \& Prato, L.\ 2006, \apj, 644, 1183 

\bibitem[Stassun et al.(2006)]{stassun06} Stassun, K.~G., Mathieu, R.~D., \& Valenti, J.~A.\ 2006, \nat, 440, 311 

\bibitem[Stassun et al.(2007)]{stassun07} Stassun, K.~G., Mathieu, R.~D., \& Valenti, J.~A.\ 2007, \apj, 664, 1154 

\bibitem[van Dam et al.(2006)]{vandam06} van Dam, M.~A., et al.\ 2006, \pasp, 118, 310 

\bibitem[Vrba et al.(2004)]{vrba04} Vrba, F.~J., et al.\ 2004, \aj, 127, 2948 

\bibitem[Wizinowich et al.(2006)]{wiz06} Wizinowich, P.~L., et al.\ 2006, \pasp, 118, 297 

\bibitem[Yelda et al.(2009)]{yelda09} Yelda, S., Lu, J.~R., Ghez, A.~M., Clarkson, W., \& Anderson, J.\ 2009, in prep

\end{thebibliography}
\end{document}